\documentclass[twocolumn,showpacs,amsmath,amssymb,prd]{revtex4}
\usepackage{graphicx}
\usepackage{graphicx}
\usepackage{dcolumn}
\usepackage{bm}
\usepackage{epsf}

\newcommand{\beq}{\begin{equation}}
\newcommand{\eeq}{\end{equation}}
\newcommand{\beqn}{\begin{eqnarray}}
\newcommand{\eeqn}{\end{eqnarray}}

\newcommand{\cB}{{\cal{B}}}

\def\bI{\hbox{$\,I\!\!\!$--}}

\begin{document}

\title{Magnetorotational collapse of very massive stars to black holes in 
full general relativity}

\author{Yuk Tung Liu}

\author{Stuart L. Shapiro}
\altaffiliation{Also at the Department of Astronomy and NCSA, University
of Illinois at Urbana-Champaign, Urbana, IL 61801}

\author{Branson C. Stephens}
\affiliation{Department of Physics, University of Illinois at 
Urbana-Champaign, Urbana, IL 61801, USA} 

\begin{abstract}
We perform axisymmetric simulations of the magnetorotational collapse 
of very massive stars in full general relativity. 
Our simulations are applicable to the collapse of
supermassive stars with masses $M \gtrsim 10^3M_{\odot}$ and to 
very massive Population~III stars. We model our initial
configurations by $n=3$ polytropes, uniformly rotating near the 
mass-shedding limit and at the onset of radial instability to collapse. 
The ratio of magnetic to rotational kinetic energy in these configurations
is chosen to be small (1\% and 10\%). We find that such
magnetic fields do not affect the initial collapse significantly.
The core collapses to a black hole, 
after which black hole excision is employed to continue the evolution 
long enough for the hole to reach a quasi-stationary state. 
We find that the black hole mass is $M_h = 0.95M$ and its spin parameter is 
$J_h/M_h^2 = 0.7$, with the remaining matter forming a torus 
around the black hole.  The subsequent evolution of the torus depends 
on the strength of the magnetic field. We freeze the spacetime metric 
(``Cowling approximation'') and continue to follow the evolution of the 
torus after the black hole has relaxed to quasi-stationary equilibrium.
In the absence of magnetic fields, 
the torus settles down following ejection of a small amount of matter due 
to shock heating. When magnetic fields are present, 
the field lines gradually collimate along the  
hole's rotation axis. MHD shocks and the
magnetorotational instability (MRI) generate MHD turbulence in the 
torus and stochastic accretion onto the central 
black hole. When the magnetic field is strong, a wind is generated in 
the torus, and the torus undergoes radial oscillations that drive
episodic accretion onto the hole. 
These oscillations produce long-wavelength
gravitational waves potentially detectable by the 
Laser Interferometer Space Antenna (LISA). 
The final state of the magnetorotational collapse always 
consists of a central black hole 
surrounded by a collimated magnetic field and a hot, thick accretion 
torus. This system is a viable candidate for the 
central engine of a long-soft gamma-ray burst.
\end{abstract}

\maketitle

\section{Introduction}

Population~III stars born with zero metallicity comprise 
the first generation of stars. It is believed 
that their formation causes the reionization 
of the universe and terminates the ``dark ages''  
(see, e.g., ~\cite{lb01} and references therein). 
The disruption of Pop~III stars following nuclear burning 
may be responsible 
for the small metal abundance observed in later generations 
of stars (e.g. Pop~II stars). 
Simulations of the collapse of primordial molecular clouds suggest that 
Pop~III stars tend to be massive. Masses in the range between 
$100M_{\odot}$ and $1000M_{\odot}$ are not uncommon~\cite{PopIII-mass}. 
Some of these calculations suggest that the initial mass function for Pop~III
stars has a bimodal distribution, with peaks at 
$\sim 100 M_{\odot}$ and $1$--$2 M_{\odot}$~\cite{nu01}. Stars with masses 
between $140 M_{\odot}$ and $260 M_{\odot}$ encounter a pair instability 
and are likely to be completely disrupted by nuclear-powered 
explosions~\cite{hw02}. The recent observation of
the peculiar Type IIn supernova SN2006gy in NGC1260 points to the 
possibility that such a disruption can occur in a massive 
star ($\gtrsim 120M_{\odot}$) even during the current epoch ~\cite{lum-SN}.
For stars with masses above $260 M_{\odot}$, the explosive nuclear burning 
is unable to reverse the implosion and the stars are likely to collapse 
directly to black holes~\cite{hw02}. 

Growing evidence indicates that supermassive black holes (SMBHs) 
with masses in the range  $ 10^6 - 10^{10} M_{\odot}$ exist and 
are the engines that power active galactic nuclei (AGNs) and 
quasars~\cite{rees1,rees}.
There is also ample evidence that SMBHs reside at the centers
of many, and perhaps most, galaxies~\cite{richstone},
including the Milky Way~\cite{genzel}.
The highest redshift of a quasar discovered to 
date is $z_{\rm QSO} = 6.43$, corresponding to
QSO SDSS 1148+5251~\cite{fan03}. Accordingly, if they are the 
energy sources in quasars (QSOs), the first SMBHs
must have formed prior to $z_{\rm QSO} = 6.43$, or within $t = 0.87$ Gyr 
after the Big Bang in the concordance $\Lambda$CDM cosmological model. 
This requirement sets a significant constraint on black
hole seed formation and growth mechanisms in the early universe.
Once formed, black holes grow by a combination of
mergers and gas accretion.

The more massive the initial seed, the less time is required for it 
to grow to SMBH scale and the easier it is to have a SMBH in 
place by $z \geq 6.43$. One possible progenitor
that readily produces a SMBH is a supermassive
star (SMS) with $M \gg 10^3 M_{\odot}$~\cite{rees1,s03}. 
SMSs can form when gaseous 
structures build up sufficient radiation pressure to inhibit 
fragmentation and prevent normal star formation; plausible cosmological 
scenarios have been proposed that can lead to
this situation~\cite{g01}.
Alternatively, the seed black holes that later grow to become
SMBHs may originate from the collapse of Pop III 
stars $\lesssim 10^3 M_{\odot}$~\cite{madau}. 
To achieve the required growth to $\sim 10^9 M_{\odot}$ 
by $z_{\rm QSO} \gtrsim 6.43$, it may be
necessary for gas accretion, if restricted by the Eddington limiting
luminosity, to occur 
at low efficiency of rest-mass to radiation conversion ($\lesssim 0.2$).
Recent relativistic simulations~\cite{accretion_disks,gsm04,dv05} 
show that accretion onto a rotating black hole that has
reached spin equilibrium 
does occur at low efficiency in a magnetized disk with turbulence driven by
the magnetorotational instability (MRI)~\cite{MRI0,MRI,MRIrev}.
Such accretion may enable a Pop~III seed to achieve the necessary  
growth by $z=6.43$~\cite{s05}. But it may be more difficult to use the 
Pop~III seeds to explain the origin of the first generation of the SMBHs
should quasars be detected at redshifts significantly higher than 
$z_{\rm QSO} = 6.43$.

Recent simulations of binary black hole mergers suggest that gravitational 
radiation reaction can induce a large kick velocity ($\gtrsim 1000$~km/s) 
in the remnants following mergers~\cite{GW_kick}. 
These large kick velocities may pose a great hazard for the growth of
black hole seeds to SMBHs by $z \sim 6$~\cite{v07}, but such large
kicks are possible only if the spins of 
the black hole binary companions are appreciable and their 
masses are comparable. Determining the spins 
of the seed black holes formed from collapse and tracking their subsequent
evolution via accretion and minor mergers~\cite{hb03,gsm04}
is therefore important for estimating the kick velocities following
major mergers.

It is very likely that  massive Pop~III {\it and} SMSs
are rotating and have magnetic fields. A SMS 
does not reach sufficiently 
high temperature for nuclear burning to become important before 
the onset of the general relativistic radial instability~\cite{bs99}.
Quasistatic contraction driven by radiative cooling 
will spin up the star to the mass-shedding 
limit, provided that viscosity and/or magnetic fields are sufficient
to maintain uniform rotation~\cite{bs99}. The star will then evolve 
secularly along the 
mass-shedding limit, simultaneously emitting electromagnetic 
radiation, matter, and angular momentum (see, e.g.~\cite{zn,sms, bs99}). 
After reaching 
the onset of radial instability, the star collapses on a 
dynamical timescale. During the collapse, the rotation becomes differential
and the rotational and magnetic 
energies are both amplified. The black hole that forms 
will be rotating and surrounded by a magnetized accretion disk. A qualitatively similar final fate should characterize a massive 
Pop~III star $ \gtrsim 260 M_{\odot}$. 

Shibata and Shapiro performed 
the first full general relativistic (GR) simulation of the collapse of 
a very massive, 
rotating star~\cite{ss02}. They modeled the massive star as a uniformly 
rotating $n=3$ polytrope spinning at the mass-shedding 
limit at the onset of radial instability. A massive star is supported
largely by thermal radiation pressure and is adequately modeled by
an $n=3$ polytrope. They terminated their simulation
soon after the black hole formed because of numerical inaccuracies 
associated with the spacetime 
singularity that inevitably forms inside the black hole.  
They estimated the final state of the system using semi-analytic 
methods (see also~\cite{ss02b,s04}). They concluded that, independent 
of the initial mass $M$ of the progenitor star, the mass of the 
black hole that forms is $M_h \sim 0.9M$ and the hole spin parameter is
$J_h/M_h^2 \sim 0.75$. The remaining gas forms a rotating torus around 
the nascent black hole.

In this paper, we first repeat the full GR 
(axisymmetric) simulation performed by Shibata and 
Shapiro of massive star collapse to the appearance of a black hole. 
We then employ 
the technique of black hole excision~\cite{excision,excision_hydro} 
to continue the evolution. We are able to follow the spacetime evolution for 
another $200M$ by this means. 
By this time, the central black hole and the spacetime metric 
have both settled 
down to a quasi-stationary state. We find that the mass and spin parameter 
of the final black hole are $M_h \approx 0.95 M$
and $J_h/ M_h^2 \approx 0.7$.
These results are close to the semi-analytic estimates 
in~\cite{ss02,ss02b,s04}.
The torus surrounding the black hole continues to 
evolve long after the black hole has settled down. 
In order to study the subsequent evolution of the torus,
we adopt the ``Cowling approximation'' whereby we freeze the metric 
at $t \sim 150M$ after the excision and continue to evolve the system 
for another $2000M$. 
We find that a small amount of material ($\sim 10^{-3}M$) is
ejected from the system due to shock heating, and the torus relaxes 
to a dynamical equilibrium state $\sim 1000M$ after the formation of the 
central black hole.

Next, to study the important role of magnetic fields, we add a small, seed 
poloidal magnetic field to the initial rotating star and follow 
the collapse once again. We consider two 
different strengths of the seed magnetic fields (models S1 and S2). 
The initial magnetic energy $\cal M$ is 1\% of the 
initial rotational kinetic energy $T$ for model S1, and 10\% of 
$T$ for model S2. 
Since $T/|W|=0.009$, we have ${\cal M}/|W| \ll 1$ in both models, where $W$ is 
the gravitational potential energy. Hence in both cases the 
magnetic fields represent small 
perturbations to the dynamics of the initial star.  
During the collapse, the frozen-in poloidal field is amplified as a result of 
compression. The development of differential rotation generates a 
toroidal field due to magnetic winding. However, 
we find that magnetic fields do not affect the collapse 
significantly before the 
formation of the central black hole. The final mass and spin parameter 
of the black hole are about the same as in the unmagnetized case. But 
magnetic fields {\it do} affect the evolution of the torus significantly. 
Magnetic fields intensify the outflow of the ejected material. The 
outflow also lasts longer than in the unmagnetized case. As the torus 
evolves, magnetic fields are collimated along the black hole's 
rotation axis. For model S1, MHD shocks and the 
MRI in the torus create 
turbulence, which leads to stochastic accretion of material 
from the torus to the 
central black hole. For model S2, a strong wind is generated (possibly by 
the magneto-centrifugal mechanism~\cite{magnetocen}) during the period
$\sim 900M$--$1200M$ following central black hole formation. This wind 
induces a radial oscillation of the torus, which leads to episodic
accretion of material to the central black hole, and long-wavelength 
gravitational radiation potentially detectable by the 
Laser Interferometer Space Antenna (LISA).

The final state of the magnetorotational collapse consists of a central black 
hole surrounded by a collimated magnetic field and a massive, hot, accretion  
torus. These features provide the essential ingredients
for generating ultrarelativistic jets at large distance.  
Our simple equation of state (EOS) is a reasonable approximation for the 
collapse of SMSs, but
our omission of neutrino emission and other microphysics 
is certainly not adequate to capture all of the physical processes 
occurring during the collapse of massive Pop~III  
and Pop~I/II stars. Nevertheless, we expect that
the black hole-torus remnant that we find 
will be qualitatively similar to the remnants  
formed from these progenitors if they are rotating rapidly at the onset of
collapse. The reason is that these stars may
also be crudely modeled by $n \approx 3$ polytropes initially and their EOSs may
also be represented by an adiabatic $\Gamma \approx 4/3$ law during collapse  
(see~\cite{s04}).

Our simulations may also help explain the formation of the central engine 
in the collapsar model~\cite{mw99} of long-soft gamma-ray bursts (GRBs). 
In addition, some GRBs observed at very high redshift might be related to 
the gravitational collapse 
of very massive Pop~III stars~\cite{PopIII-GRB}. 
Hence our simulations may also provide 
insights into the formation of GRB central engines arising from these stars.

The reminder of this paper is organized as follows. 
In Sec.~\ref{sec:formalism}, we 
briefly describe the mathematical formulation of the Einstein-Maxwell-MHD 
coupled equations 
and numerical techniques used to solve them. We then describe our 
initial data and computational setup in Sec.~\ref{sec:id}. 
We present our numerical results in Sec.~\ref{sec:results} and 
provide a summary of our simulations in Sec.~\ref{sec:conclusion}. 
Throughout this paper, we
adopt geometrical units in which $G=1=c$, where $G$ and $c$ denote the
gravitational constant and speed of light, respectively.  Cartesian
coordinates are denoted by $x^k=(x, y, z)$. The coordinates are
oriented so that the rotation axis is along the $z$-direction. We
define the coordinate radius $r=\sqrt{x^2+y^2+z^2}$, cylindrical
radius $\varpi=\sqrt{x^2+y^2}$, and azimuthal angle
$\varphi=\tan^{-1}(y/x)$. Coordinate time is denoted by $t$. Greek
indices $\mu, \nu, \cdots$ denote spacetime components ($t, x, y, z$),
small Latin indices $i, j, \cdots$ denote spatial components
($x, y$, and $z$).

\section{Formulation}
\label{sec:formalism}

\subsection{Basic equations and numerical methods}

The formulation and numerical scheme for our GRMHD simulations are
the same as those reported in~\cite{DLSS}, to which the reader 
may refer for details. Here we briefly summarize the 
method and introduce our notation.

We use the 3+1 formulation of general relativity and decompose 
the metric into the following form:
\beq
  ds^2 = -\alpha^2 dt^2  
+ \gamma_{ij} (dx^i + \beta^i dt) (dx^j + \beta^j dt) \ .
\eeq
The fundamental variables for the metric evolution are the spatial
three-metric $\gamma_{ij}$ and extrinsic curvature $K_{ij}$. We adopt
the Baumgarte-Shapiro-Shibata-Nakamura (BSSN)
formalism~\cite{BSSN} to evolve $\gamma_{ij}$ and $K_{ij}$. In this
formalism, the evolution variables are the conformal exponent $\phi
\equiv \ln \gamma/12$, the conformal 3-metric $\tilde
\gamma_{ij}=e^{-4\phi}\gamma_{ij}$, three auxiliary functions
$\tilde{\Gamma}^i \equiv -\tilde \gamma^{ij}{}_{,j}$, the trace of
the extrinsic curvature $K$, and the tracefree part of the conformal extrinsic
curvature $\tilde A_{ij} \equiv e^{-4\phi}(K_{ij}-\gamma_{ij} K/3)$.
Here, $\gamma={\rm det}(\gamma_{ij})$. The full spacetime metric $g_{\mu \nu}$
is related to the three-metric $\gamma_{\mu \nu}$ by $\gamma_{\mu \nu}
= g_{\mu \nu} + n_{\mu} n_{\nu}$, where the future-directed, timelike
unit vector $n^{\mu}$ normal to the time slice can be written in terms
of the lapse $\alpha$ and shift $\beta^i$ as $n^{\mu} = \alpha^{-1}
(1,-\beta^i)$.

The Einstein equations are solved in Cartesian coordinates.
In this paper, we assume both equatorial and axisymmetry so we only 
evolve the region with $x>0$ and $z>0$. We adopt 
the Cartoon method~\cite{cartoon} to impose axisymmetry in the metric 
evolution, and use a cylindrical grid to evolve the MHD and Maxwell 
equations. As for the gauge conditions, 
we adopt the hyperbolic driver conditions as in~\cite{excision_hydro} to 
evolve the lapse $\alpha$ and shift $\beta^i$. 

The fundamental variables in ideal MHD are the rest-mass density 
$\rho_0$, specific internal energy $\epsilon$, pressure $P$, four-velocity 
$u^{\mu}$, and magnetic field $B^{\mu}$ measured by a normal
observer moving with a 4-velocity $n^{\mu}$ (note that $B^{\mu} n_{\mu}=0$). 
The ideal MHD condition is written as $u_{\mu} F^{\mu\nu}=0$, 
where $F^{\mu\nu}$ is the electromagnetic tensor. The tensor 
$F^{\mu\nu}$ and its dual in the ideal MHD approximation are 
given by 
\beqn
&&F^{\mu\nu}=\epsilon^{\mu\nu\alpha\beta}u_{\alpha}b_{\beta}, \label{eqFF}\\
&&F^*_{\mu\nu} \equiv {1 \over 2}\epsilon_{\mu\nu\alpha\beta} F^{\alpha\beta}
=b_{\mu} u_{\nu}- b_{\nu} u_{\mu}, 
\eeqn
where $\epsilon_{\mu\nu\alpha\beta}$ is the Levi-Civita tensor.  
Here we have introduced an auxiliary magnetic 4-vector 
$b^{\mu}=B^{\mu}_{(u)}/\sqrt{4\pi}$, where $B^{\mu}_{(u)}$ is the 
magnetic field measured by an observer comoving with the fluid and 
is related to $B^{\mu}$ by 
\beq
  B^{\mu}_{(u)} = -\frac{(\delta^{\mu}{}_{\nu} + u^{\mu} u_{\nu}) B^{\nu}} 
  {n_{\lambda}u^{\lambda}} \ .
\eeq 

The energy-momentum tensor is written as
\beqn
T_{\mu\nu}=T_{\mu\nu}^{\rm Fluid} + T_{\mu\nu}^{\rm EM}, 
\eeqn
where $T_{\mu\nu}^{\rm Fluid}$ and $T_{\mu\nu}^{\rm EM}$ denote the
fluid and electromagnetic pieces of the stress-energy 
tensor. They are given by 
\beqn
&&T_{\mu\nu}^{\rm Fluid}=
\rho_0 h u_{\mu} u_{\nu} + P g_{\mu\nu}, \\
&&T_{\mu\nu}^{\rm EM}= \frac{1}{4\pi} \left(
F_{\mu\sigma} F^{~\sigma}_{\nu}-{1 \over 4}g_{\mu\nu}
F_{\alpha\beta} F^{\alpha\beta} \right) \nonumber \\
&&~~~~~~=\biggl({1 \over 2}g_{\mu\nu}+u_{\mu}u_{\nu}\biggr)b^2
-b_{\mu}b_{\nu}, 
\eeqn
where $h\equiv 1+\epsilon+P/\rho_0$ is the specific enthalpy, and 
$b^2\equiv b^{\mu}b_{\mu}$. Hence, the total stress-energy tensor becomes
\beq
  T_{\mu\nu}= (\rho_0 h + b^2) u_{\mu} u_{\nu} + \left( P + \frac{b^2}{2}
\right) g_{\mu\nu} - b_{\mu} b_{\nu} \ .
\label{eq:mhdTab}
\eeq

In our numerical implementation of the GRMHD and magnetic 
induction equations, 
we evolve the densitized density $\rho_*$, 
densitized momentum density $\tilde{S}_i$, 
densitized energy density $\tilde{\tau}$, 
and densitized magnetic field $\cB^i$. They are defined as 
\beqn
&&\rho_* \equiv - \sqrt{\gamma}\, \rho_0 n_{\mu} u^{\mu}, 
\label{eq:rhos} \\
&& \tilde{S}_i \equiv -  \sqrt{\gamma}\, T_{\mu \nu}n^{\mu} \gamma^{\nu}_{~i}, \\
&& \tilde{\tau} \equiv  \sqrt{\gamma}\, T_{\mu \nu}n^{\mu} n^{\nu} - \rho_*, 
\label{eq:S0} \\
&& \cB^i \equiv  \sqrt{\gamma}\, B^i. 
\eeqn 
During the evolution, we also need the three-velocity $v^i = u^i/u^t$.

The GRMHD and induction equations are written in conservative form for
variables $\rho_*$, $\tilde{S}_i$, $\tilde{\tau}$, and $\cB^i$ and 
evolved using a
high-resolution shock-capturing (HRSC) scheme. Specifically, we 
use the monotonized central (MC) scheme~\cite{vL77} for data reconstruction 
and the HLL (Harten, Lax and van-Leer) scheme~\cite{HLL} to compute 
the flux. The
magnetic field $\cB^i$ has to satisfy the no monopole constraint
$\partial_i \cB^i=0$. We adopt the flux-interpolated constrained 
transport (flux-CT) scheme~\cite{t00} to impose this constraint. 
This scheme guarantees that no magnetic monopoles will be created in the
computational grid during numerical evolution. At each timestep, the
primitive variables $(\rho_0,P,v^i)$ must be computed from the evolution
variables $(\rho_*,\tilde{\tau},\tilde{S}_i)$. This is done by 
numerically solving the
algebraic equations~(\ref{eq:rhos})--(\ref{eq:S0}) together with an 
EOS $P=P(\rho_0,\epsilon)$. 

As in many hydrodynamic simulations in
astrophysics, we add a tenuous ``atmosphere'' that covers the
computational grid outside the star. The atmospheric rest-mass density
is set to $\approx 10^{-10} \rho_c(0)$ before the black hole
forms, where $\rho_c(0)$ is the initial rest-mass central density of the star.
In the excision evolution where the system consists of a central black
hole and a surrounding torus, the maximum density in the torus is
$\sim 100\rho_c(0)$, and we set the atmosphere density to
$10^{-3} \rho_c(0)$.

The codes used here have been tested in multiple relativistic MHD simulations,
including MHD shocks, nonlinear MHD wave propagation, magnetized Bondi
accretion, and MHD waves induced by linear gravitational waves~\cite{DLSS}.
We have also compared this code with the GRMHD code developed 
independently by Shibata and Sekiguchi~\cite{SS05}
by performing simulations of the evolution of magnetized,
differentially rotating, relativistic, hypermassive neutron
stars~\cite{DLSSS,DLSSS2}, and of magnetorotational collapse of
stellar cores~\cite{SLSS}. We obtain good agreement between
these two independent codes.

\subsection{Equation of state}

In this paper, we adopt the simple $n=3$ ($\Gamma=4/3$) 
polytropic EOS to construct the initial model and 
$P=(\Gamma-1)\rho_0 \epsilon$ ($\Gamma$-law EOS) during the 
evolution. This EOS is a good 
approximation for the pre-collapse core of a massive Pop~III 
star~\cite{bac84} or the bulk of a SMS~\cite{zn,st83}, where pressure is 
dominated by thermal radiation. 
For a Pop~I/II star, which has smaller mass, the pressure 
of the pre-collapse core is dominated by the relativistic degenerate 
electron pressure, which is also well-approximated by a $\Gamma=4/3$ 
EOS. During the collapse, the EOS stiffens when the density 
exceeds nuclear 
density $\rho_{\rm nuc} \approx 2\times 10^{14}~{\rm g}~{\rm cm}^{-3}$. 
However, if the mass of the collapsing core exceeds a critical 
value $M_{\rm crit}$, the black hole horizon appears before the star 
reaches the nuclear density. In this case, the stiffening of the EOS 
has no effect on the collapse. To estimate $M_{\rm crit}$, consider 
the collapse of a uniform density dust sphere (Oppenheimer-Snyder collapse). 
A horizon appears when the areal radius of the sphere reaches 
$R=2M$. At this time, the density is 
$\rho_0 = 3M/(4\pi R^3) \approx 1.7 \times 10^{16} 
(M_{\odot}/M)^2~{\rm g}~{\rm cm}^{-3}$. 
Setting $\rho_0=\rho_{\rm nuc}$, gives $M_{\rm crit} \approx 10M_{\odot}$. 
For a SMS, the mass is much larger than $M_{\rm crit}$. 
For a Pop~III star of mass $M=300M_{\odot}$, the mass of the collapsing 
core is $180M_{\odot}$~\cite{hw02},
which is still much larger than $M_{\rm crit}$. 
Hence the $\Gamma=4/3$ EOS is also a good approximation during the 
entire collapse phase for very massive Pop~III stars. For Pop~I/II stars, 
on the other hand, the core mass is less than $2M_{\odot}$ and a more 
realistic EOS is required in the late stages. 
In addition, neutrino emission and transport 
are also important to the dynamics of the collapse for these stars. 
Neutrino generation and transport also play a 
role in the collapse of Pop~III stars~\cite{hw02,stks07}, 
but are probably not dynamically important for the most massive 
progenitors or for SMSs because of their low temperature and density. 

\subsection{Diagnostics}

During the evolution, we monitor the L2 norm of the Hamiltonian 
and momentum constraints as in~\cite{DLSSS2}. We find the violation
of the constraints is at most a few percent before excision. 
After the excision, 
the constraints can rise to $\approx 10$\%. We terminate the excision evolution 
before the constraints reach $\sim$20\%. 

We also compute the rest mass $M_0$, ADM mass $M$ and angular momentum $J$ 
during the evolution. They are computed by the following volume integrals: 
\beqn
 M_0 &=& \int_V \rho_* d^3x \ , \label{eq:m0} \\
M &=& \int_V \Bigl[e^{5\phi}(\rho + {1\over 16\pi}\tilde A_{ij}\tilde A^{ij}
                             - {1\over 24\pi}K^2) \cr
    & &  \quad - {1\over 16\pi}\tilde\Gamma^{ijk}\tilde\Gamma_{jik}
	     + {1-e^{\phi}\over 16\pi}\tilde R\Bigr] d^3x \ , 
\label{eq:adm_mass_vol} \\
 J &=& \int_V \tilde{S}_{\varphi} d^3x \ , \label{eq:j_vol}
\eeqn
where $\tilde{\Gamma}_{ijk}$ is the Christoffel symbol and $\tilde R$ 
is the Riemann scalar associated with $\tilde{\gamma}_{ij}$.
Note that the above formula for $J$ is only valid in an axisymmetric 
spacetime~\cite{Wald}. 
The rest mass $M_0$ is conserved as a result of the baryon number conservation. 
Angular momentum $J$ is conserved in axisymmetry, as 
gravitational radiation carries no angular momentum. However, 
$M$ is not conserved since gravitational radiation carries 
energy and propagates off the computational grid. 
We find that $M$ remains constant to within 2\%. 
Our finite difference scheme guarantees that $M_0$ and $J$ computed from 
the above volume integrals are conserved to machine precision provided that 
no fluid leaves the computational grid. However, we perform 
several regriddings during the calculation (see Section III) and these 
leave behind a few percent of 
$M_0$ and $J$ in the outermost layers.

During the excision evolution, we compute the rest mass $M_{\rm disk}$ 
and angular momentum $J_{\rm disk}$ of the disk outside the black hole 
by computing integrals~(\ref{eq:m0}) and (\ref{eq:j_vol}) 
over the volume outside the apparent horizon. The irreducible mass 
$M_{\rm irr}$ of the black hole is given by $M_{\rm irr} = \sqrt{A/16\pi}$, 
where $A$ is the surface area of the apparent horizon. Since $J$ is 
conserved, we can compute the black hole's angular momentum $J_h$ by 
\beq
  J_h = J - J_{\rm loss} - J_{\rm disk} \ ,
\eeq
where $J_{\rm loss}$ is the loss of angular momentum as a result 
of regriddings and matter leaving the grid. The black hole's mass 
$M_h$ is then computed from the formula 
\beq
  M_h = \sqrt{M_{\rm irr}^2+(J_h/2M_{\rm irr})^2} \ ,
\eeq
which is exact for a Kerr spacetime, and is in accord with the 
formula derived using the isolated and dynamical 
horizon formalism~\cite{iso-dyn-hor}. 

At $\Delta t \approx 150M$ after the excision evolution, we find that the 
spacetime becomes nearly stationary. In this case, the energy $E$ 
is approximately conserved thereafter, where 
\beq
E =\int \alpha \sqrt{\gamma}\, T^t_{~t} d^3x \ .
\eeq
We can then define the fluxes of rest mass, energy, and angular momentum 
across any closed two-dimensional surface $S$ in a time slice:
\beqn
&&F_M(r)= \oint_S \alpha \rho_0 v^i d^2 \Sigma_i \ ,\\
&&F_E(r)=-\oint_S \alpha T^i_{~t} d^2 \Sigma_i \ ,\\
&&F_J(r)= \oint_S \alpha T^i_{~\varphi} d^2 \Sigma_i \ , 
\eeqn
where 
\beq
  d^2 \Sigma_i = \frac{1}{2} \epsilon_{ijk} dx^j \wedge dx^k \ , 
\eeq
and $\epsilon_{ijk} = n_{\mu} \epsilon^{\mu}{}_{ijk}$ is the Levi-Civita 
tensor associated with the three-metric $\gamma_{ij}$. If $S$ is a sphere 
of radius $r$, the above expressions reduce to 
\beqn
&&F_M(r)= \oint_{r={\rm const}} dA \rho_* v^r r^2 \\
&&F_E(r)=-\oint_{r={\rm const}} dA \alpha \sqrt{\gamma}\, T^r_{~t},\\
&&F_J(r)= \oint_{r={\rm const}} dA \alpha \sqrt{\gamma}\, T^r_{~\varphi},
\eeqn
where $dA=r^2 \sin\theta d\theta d\phi$.
The total energy flux $F_E$ is very close to the rest-mass flux $F_M$
since $F_E$ is primarily composed of the rest-mass energy flow. Thus,
we define another energy flux by subtracting the rest-mass flow:
$F_e=F_E-F_M$. We note that $F_e$ contains kinetic, thermal, 
electromagnetic, and gravitational potential energy fluxes. If $F_e >0$
at sufficiently large radius, an unbound outflow (overcoming gravitational 
binding energy) is present. 

Another method to determine whether a fluid particle is unbound is to 
compute $u_t$. In a stationary spacetime, the value of $u_t$ of a particle 
moving on a geodesic is conserved. If the particle is unbound, the radial 
velocity $v^r>0$ and $-u_t = 1/\sqrt{1-v^2} > 1$ at infinity. Hence $v^r$ 
and $u_t$ are useful diagnostics to determine if the fluid element is 
unbound, provided that the fluid motion is predominantly ballistic and 
pressure and electromagnetic forces can be neglected. This is usually the 
case in the low-density region.

During the excision evolution, the MRI may develop in the torus 
surrounding the central black hole. 
The growth time ($e$-folding time) and wavelength of the fastest-growing
MRI mode can be roughly estimated by the following formulae derived in
linear perturbation theory in Newtonian gravitation~\cite{MRIrev,SLSS}:
\beqn
t_{\rm MRI} & = & 
2 \left|\partial \Omega/\partial \ln \varpi \right|^{-1}
\ ,  \label{eq:tmri} \\
\lambda_{\rm max} & = & {2\pi v_A^z \over \Omega}
\biggl[1-\biggr({\kappa \over 2\Omega}\biggr)^4\biggr]^{-1/2}\ ,
\label{eq:lmri}
\eeqn
where $\Omega$ is angular velocity, $v_A^z=B^z/\sqrt{4\pi \rho_0}$ is 
the $z$-component of the Alfv\'en speed and 
\beq
  \kappa \equiv \left[\frac{1}{\varpi^3}\frac{\partial(\varpi^4\Omega^2)}
{\partial \varpi} \right]^{1/2}
\eeq
is the epicyclic frequency.

\section{Initial Data and Grid setup}
\label{sec:id}

\subsection{Initial Data}
We model the pre-collapse star as a uniformly rotating star 
satisfying the $n=3$ polytropic EOS 
$P=K\rho_0^{4/3}$. We set $K=1$ in our code. As explained in~\cite{cst92}, 
our result can be scaled to arbitrary values of $K$ or, equivalently, the 
ADM mass $M$;
only nondimensional ratios are invariant. 
For example, 
$M \propto K^{3/2}$, $B \propto K^{-3/2}$, $\rho_0 \propto K^{-3}$,... 
etc.

We use the same initial model as in~\cite{ss02}, whereby the star is rotating 
near the mass-shedding limit with $T/|W|=0.009$.
The equatorial radius of the star is 
$R_{\rm eq} = 640M = 10^7 (M/10^4M_{\odot})$km.
This is the configuration 
where the polytrope is on the verge of radial instability against 
gravitational collapse due to general relativity~\cite{bs99}. 
The central density of the
star is $\rho_c=10^3 (M/10^4M_{\odot})^{-2}~{\rm g}~{\rm cm}^{-3}$. 
For a SMS with mass $M \gtrsim 10^4M_{\odot}$, 
this general relativistic instability triggers the collapse, as opposed 
to microphysical processes such as pair instability.

We add a small amount of 
poloidal seed magnetic field to this equilibrium star, employing
a magnetic vector potential of the form
\beq
  A_{\mu} = A_{\varphi} \delta^{\varphi}{}_{\mu} = 
   A_b \varpi^2 \max( \rho_0^{1/6} - \rho_{\rm cut}^{1/6},0 )\, 
   \delta^{\varphi}{}_{\mu} \ ,
\eeq
where $\rho_{\rm cut} = 10^{-5} \rho_c$, and $A_b$ is a constant 
that determines the strength of the initial magnetic field.
The magnetic field is computed by the formula
$B^i = \epsilon^{ijk} \partial_j A_k$. A similar form of the initial 
magnetic field has been used in the study of magnetized accretion 
disks around a stationary black hole~\cite{accretion_disks,dvhk03}, 
the collapse of hypermassive neutron 
stars~\cite{DLSSS,DLSSS2} and the collapse of a magnetized, 
stellar core of a high mass star~\cite{SLSS}. We choose two nonzero 
values of the constants $A_b$ 
so that the values of initial magnetic energy 
\beq
  {\cal M} \equiv \int \sqrt{\gamma}\, n_{\mu} n_{\nu} T^{\mu \nu}_{\rm EM} 
d^3 x
\eeq
are 1\% and 10\% of the initial rotational kinetic energy (corresponding 
to ${\cal M}/|W|=9\times 10^{-5}$ and $9\times 10^{-4}$). Hence adding 
this seed magnetic field causes only a slight perturbation to the star.
We label these two models as S1 and S2, respectively. We also 
study the unmagnetized case (${\cal M}=0$, model S0) to compare with 
the previous result 
reported in~\cite{ss02}. We can also characterize the strength of the 
magnetic field by the volume-averaged ratio of gas pressure to the 
magnetic pressure. Specifically, we define 
$\beta = \langle P \rangle / \langle P_{\rm mag} \rangle$, where 
the magnetic pressure is $P_{\rm mag}=b^2/2$ and 
\[
  \langle q \rangle \equiv  \frac{\int q dV}{V_s} \ .
\]
Here $dV=\sqrt{\gamma} d^3x$ is the proper volume element and 
$V_s=\int_{P>0} dV$ is the volume of the star. This definition of 
$\beta$ is used in~\cite{dvhk03} in the study of magnetized 
accretion disks around central black holes. The value of $\beta$ 
for models S1 and S2 are 3700 and 370, respectively. We also 
define the averaged strength of magnetic field $\bar{B}$ 
by $\bar{B}=\sqrt{8\pi {\cal M}/V_s}$. In cgs units, we find 
\[
  \bar{B} = 3\times 10^8 \left(\frac{M}{10^4M_{\odot}}\right)^{-1}~G
\]
for model S1 and 
\[
  \bar{B} = 10^9 \left(\frac{M}{10^4M_{\odot}}\right)^{-1}~G
\]
for model S2. 

The strength of the magnetic field inside a Pop~III star 
is unknown and is currently not addressed by theoretical models dealing 
with their cosmological formation~\cite{abn02}. Our goal is to determine 
what effects, if any, magnetic fields may have on the eventual collapse 
of the stars to black holes. After all, they are possible progenitors of 
GRBs, and many GRB models require a magnetized disk around a black 
hole.
Here, we choose the strengths of the seed magnetic field to be 
sufficiently large 
for us to perform reliable simulations with limited computational resources 
and still be able to resolve the wavelength of the fastest growing MRI 
mode. We note that 
these magnetic field strengths are still quite small dynamically 
(small ${\cal M}/|W|$ and large $\beta$), and so the magnetic field 
does not affect the dynamics of the collapse (see Sec.~\ref{sec:results}). 
However, the post-collapse 
evolution does depend on the chosen strengths. In Sec~\ref{sec:results}, 
we will discuss how our results may change for even smaller field strengths.

\begin{figure}
\vspace{-4mm}
\begin{center}
\epsfxsize=3.2in
\leavevmode
\epsffile{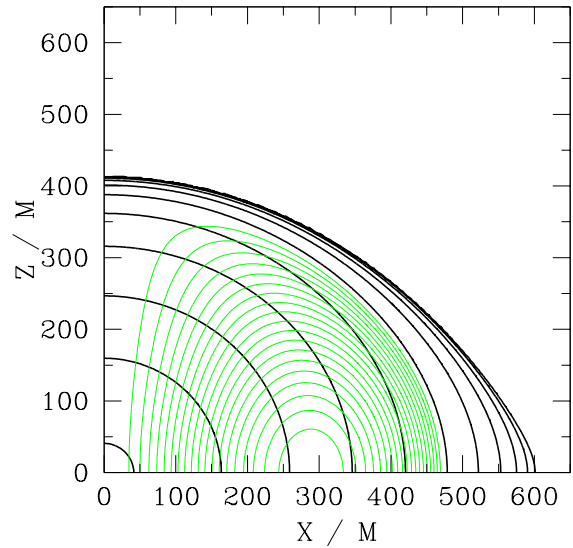}
\vspace{-6mm}
\caption{Initial density contour curves (thick, black) and 
magnetic field lines (thin, green). 
The density contour curves are drawn for 
$\rho_0=10^{-i-0.1}\rho_c$
with $j=0, 1, \cdots, 10$, and the poloidal magnetic field lines (for models S1 and S2 
only), which coincide with contours of $A_{\varphi}$ in axisymmetry,
are for $A_{\varphi}=A_{\varphi,{\rm max}}(j/20)$ with $j=1, 2,
\cdots, 19$ where $\rho_c$ and $A_{\varphi,{\rm max}}$ denote 
the central density and maximum value of $A_{\varphi}$, respectively. 
Note that 
although the magnitudes of the magnetic fields are different for models S1 and 
S2, the field lines have the same profile when normalized as described.}
\label{fig:initial}
\end{center}
\end{figure}

Following~\cite{ss02}, we induce collapse 
by depleting 1\% of the pressure (i.e., $P \rightarrow 0.99P$) 
everywhere inside the star. The parameters of our models are summarized 
in Table~\ref{tab:models}. The density and magnetic field profiles of 
our pre-collapse model are shown in Figure~\ref{fig:initial}. 

\begin{table}
\caption{Model parameters}
\label{tab:models}
\vskip 0.2cm
\begin{tabular}{c|c|c|c}
 Model & ${\cal M}/T$ & $\beta$ & $\bar{B}\times (M/10^4M_{\odot})$ \\
\hline
  S0  &  0 & 0 & 0 \\
  S1  &  0.01 & 3700 & $3\times 10^8$G \\
  S2  &  0.10 & 370 & $10^9$G \\ 
\end{tabular}
\end{table}

\subsection{Grid Setup}

We perform simulations using a cell-centered uniform grid with
size $N \times 3 \times N$ in $x$-$y$-$z$, covering a computational 
domain $\Delta/2 \leq x
\leq L-\Delta/2$, $\Delta/2 \leq z \leq L-\Delta/2$,
and $-\Delta \leq y \leq \Delta$. Here,
$N$ and $L$ are constants and $\Delta = L/N$. The variables in the
$y=\pm \Delta$ planes are computed from the quantities in the $y=0$
plane by imposing axisymmetry. 
Since the characteristic radius of the star decreases by a factor of
$\sim 1000$ during the collapse (from $\sim 600M$ to $\sim M$), using
a fixed uniform grid with sufficient resolution for the entire collapse
phase is computationally prohibitive. In order to save computational
resources and at the same time ensure adequate resolution throughout
the simulation, we adopt a regridding technique similar to the
algorithm described in~\cite{ss02}. When gravity is weak (in the Newtonian
regime), the characteristic radius of the star is proportional to
$1/(1-\alpha_c)$, where $\alpha_c$ is the central lapse. We thus
use a regridding algorithm based on the values of $\alpha_c$. 
During the early stages, the collapse proceeds in a homologous 
manner. We set $N=400$ and $L=929M$ when $\alpha_c > 0.984$. 
Keeping $N$ fixed, we decreases $L$ as the collapse proceeds: 
$L=656M$ when $0.976 \leq \alpha_c < 0.984$, 
$L=459M$ when $0.905 \leq \alpha_c < 0.976$. 
After this stage, the collapse in the core is faster than in 
the outer layers. We increase the grid number $N$ and decrease $\Delta$ 
as follows: $N=900$ and $L=158M$ when $0.7 \leq \alpha_c < 0.905$, 
$N=1400$ and $L=135M$ when $0.3 \leq \alpha_c < 0.7$. In the last 
stage, the star collapses to a black hole. In order to allocate 
our grid more effectively in this last stage, we interpolated the 
data onto a multiple-transition fisheye coordinates~\cite{fisheye} 
when $\alpha_c < 0.3$. 

The multiple transition fisheye coordinates $\bar{x}^i$ are related to the
original coordinates $x^i$ through the following transformation:
\beqn
  x^i &=& \frac{\bar{x}^i}{\bar{r}} r(\bar{r}) ,
\label{eq:fisheyet1}  \\
  r(\bar{r}) &=& a_n \bar{r} + \sum_{i=1}^n \kappa_i \ln
\frac{\cosh [(\bar{r}+\bar{r}_{0i})/s_i]}{\cosh [(\bar{r}-\bar{r}_{0i})/s_i]},
\label{eq:fisheyet2}  \\
 \kappa_i &=& \frac{(a_{i-1}-a_i)s_i}{2\tanh(\bar{r}_{0i}/s_i)} ,
\label{eq:fisheyet3}
\eeqn
where $r=\sqrt{x^2+y^2+z^2}$, $\bar{r} = \sqrt{\bar{x}^2 + \bar{y}^2
+ \bar{z}^2}$, $n$, $a_i$, $\bar{r}_{0i}$ and $s_i$ are constant parameters. 
In the last stage of collapse ($\alpha_c < 0.3$), 
we use a cell-centered uniform grid with $N=600$ in fisheye coordinates 
with parameters $n=3$, $(a_0,a_1,a_2,a_3)=(0.125,0.25,0.5,1)$, 
$(\bar{r}_{01},\bar{r}_{02},\bar{r}_{03})=(31.5M,59.5M,81.4M)$, 
$s_1=s_2=s_3=5.69M$, 
and $\bar{L}=118M$. When transformed back to the original coordinate 
system, the outer boundary is $\approx 60M$ and the resolutions are 
\beq
\Delta \approx \left \{ \begin{array}{ll} 
      0.025M & r \lesssim 4M \\ 
      0.05M  &  4M \lesssim r \lesssim 11M \\ 
      0.1M   &  11M \lesssim r \lesssim 22M \\ 
      0.2M   &  r \gtrsim 22M  \end{array} \right. \ .
\eeq

We find that the total rest mass and angular momentum that are 
discarded as a result of the regriddings are about 1\% and 
5--8\% of their initial values for the models considered.

\section{Results}
\label{sec:results}

\begin{figure}
\begin{center}
\epsfxsize=3in
\leavevmode
\epsffile{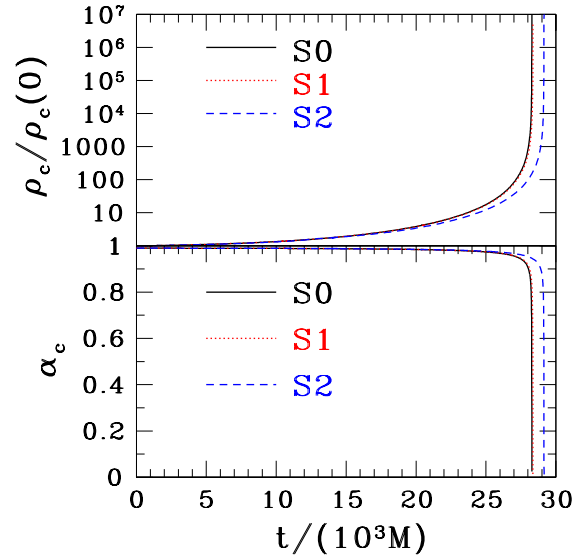}
\caption{Evolution of central rest-mass density (upper panel) and lapse (lower panel) 
for models S0 (black solid lines), S1 (red dotted lines) and S2 (blue dashed 
lines). The central density is normalized by its initial value $\rho_c(0)$.
Note that the results for S0 and S1 are very close and their lines 
almost overlap. The plots terminate soon after the apparent horizons appear.}
\label{fig:rho_alp}
\end{center}
\end{figure}

\begin{figure*}
\vspace{-4mm}
\begin{center}
\epsfxsize=2.15in
\leavevmode
\epsffile{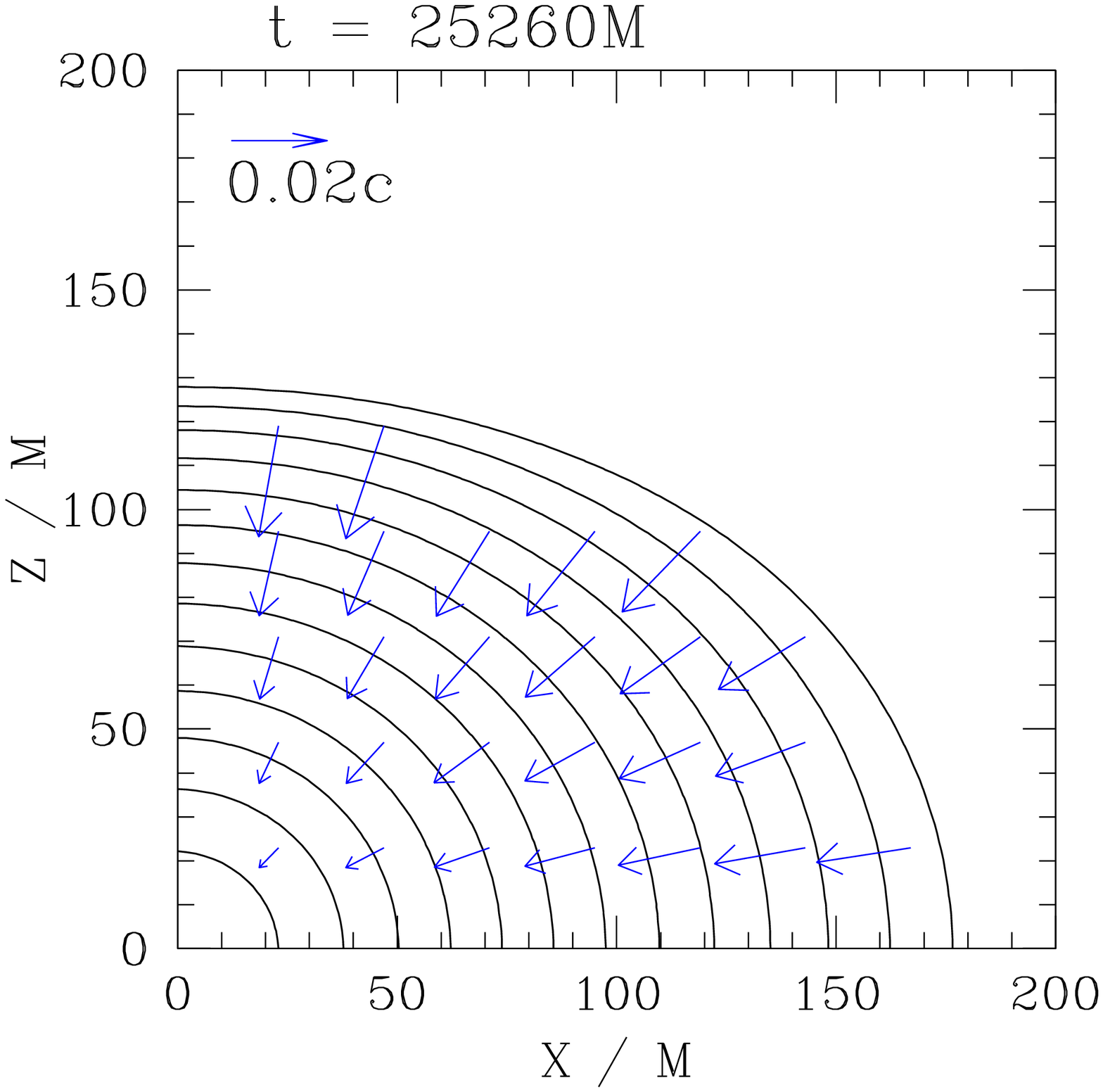}
\epsfxsize=2.15in
\leavevmode
\epsffile{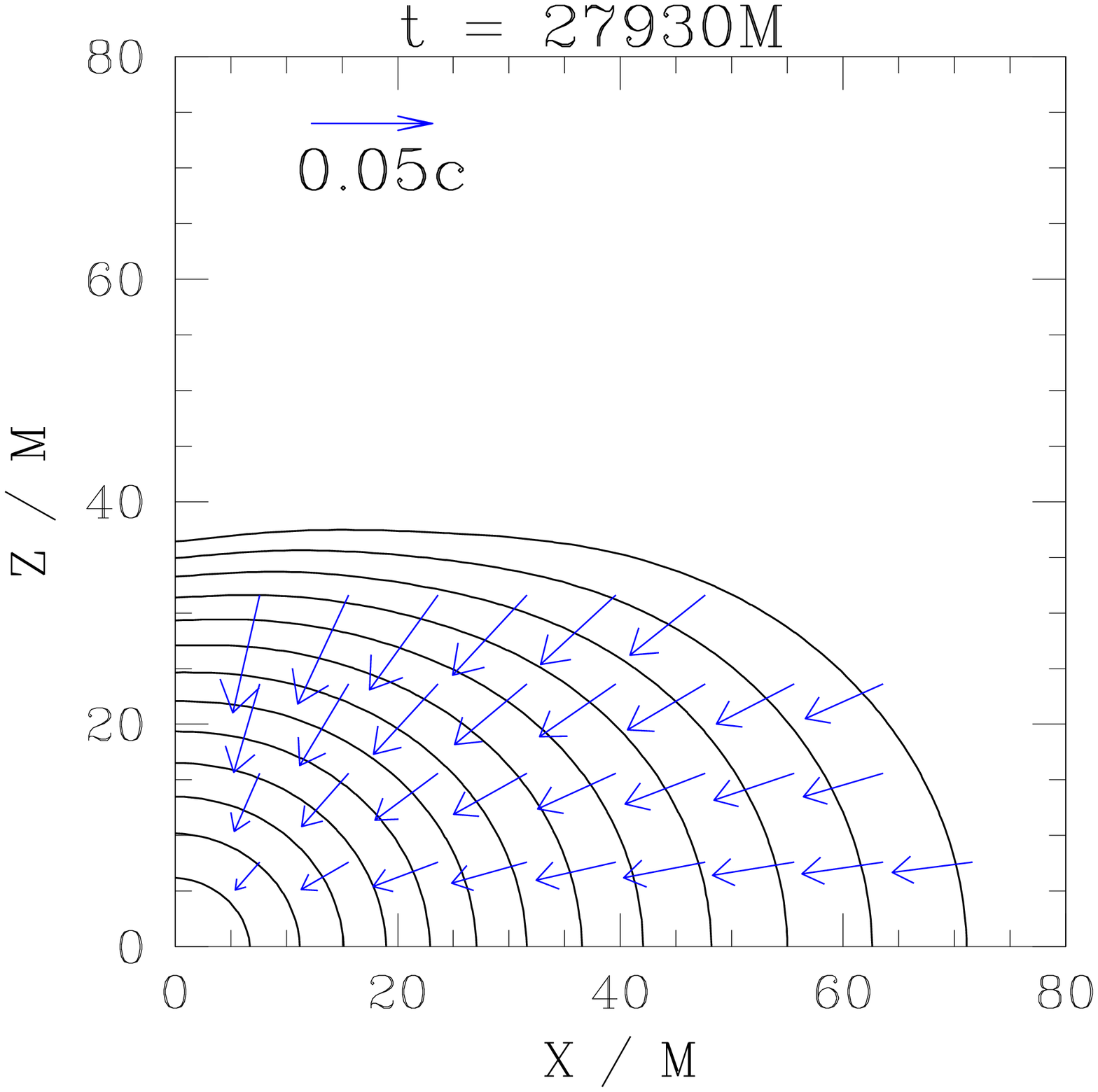}
\epsfxsize=2.15in
\leavevmode
\epsffile{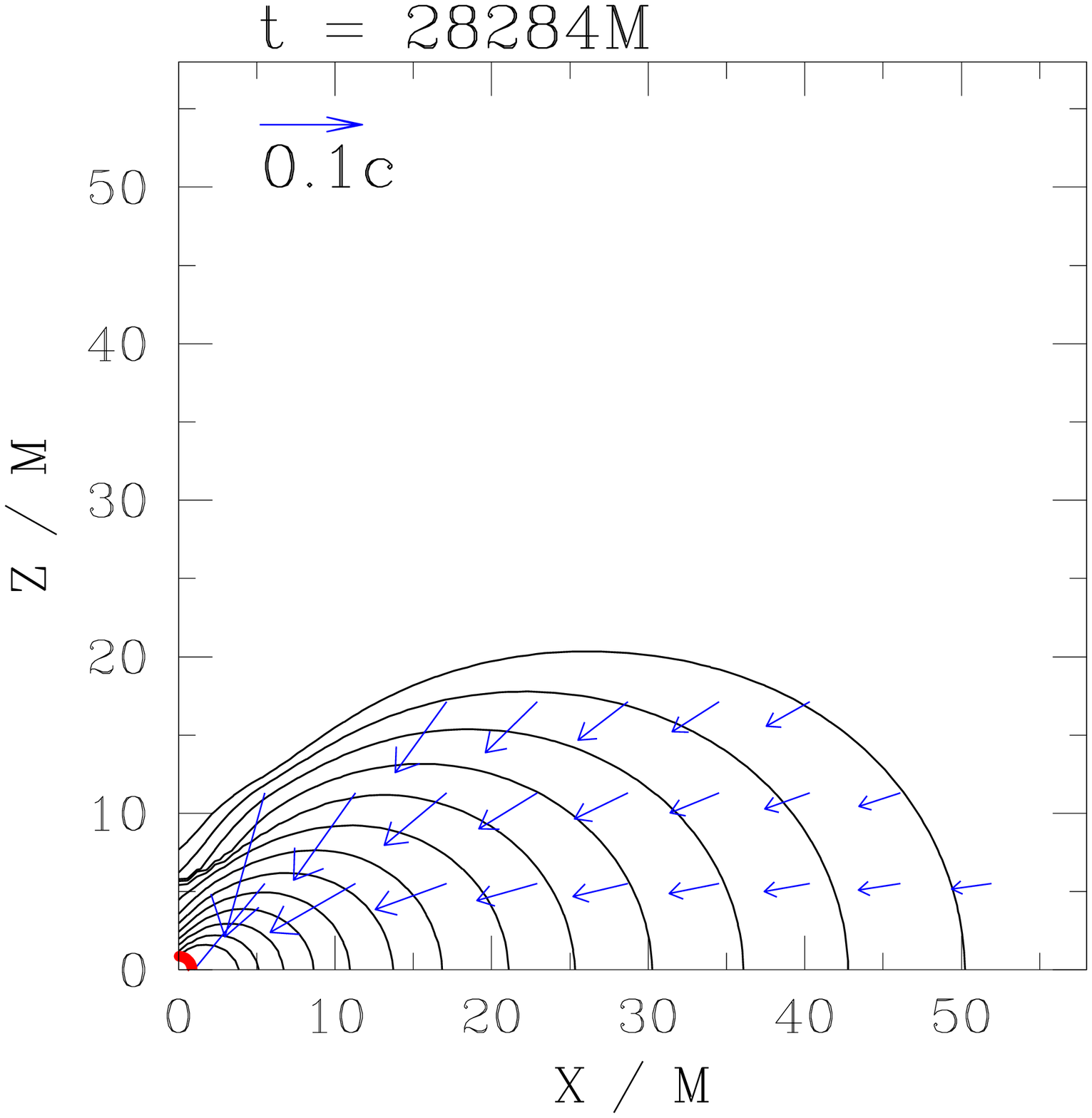}
\caption{Density contour curves and velocity vector fields in the
meridional plane for model S0 (pre-excision). 
The density levels are drawn for $\rho_0 = \rho_{\rm scal}
10^{-0.3j}~(j=0$--12), where $\rho_{\rm scal}=11\rho_c(0)$
at $t=25260M$, $\rho_{\rm scal}=340\rho_c(0)$ at $t=27930$,
and $\rho_{\rm scal}=1000\rho_c(0)$ at $t=28284M$. 
The thick (red) line near the lower left corner in the far right 
graph denotes the apparent horizon. 
Note that the scale is different for each time slice to
show the central region in detail.}
\label{fig:prexconS0}
\end{center}
\end{figure*}

\begin{figure*}
\vspace{-4mm}
\begin{center}
\epsfxsize=2.15in
\leavevmode
\epsffile{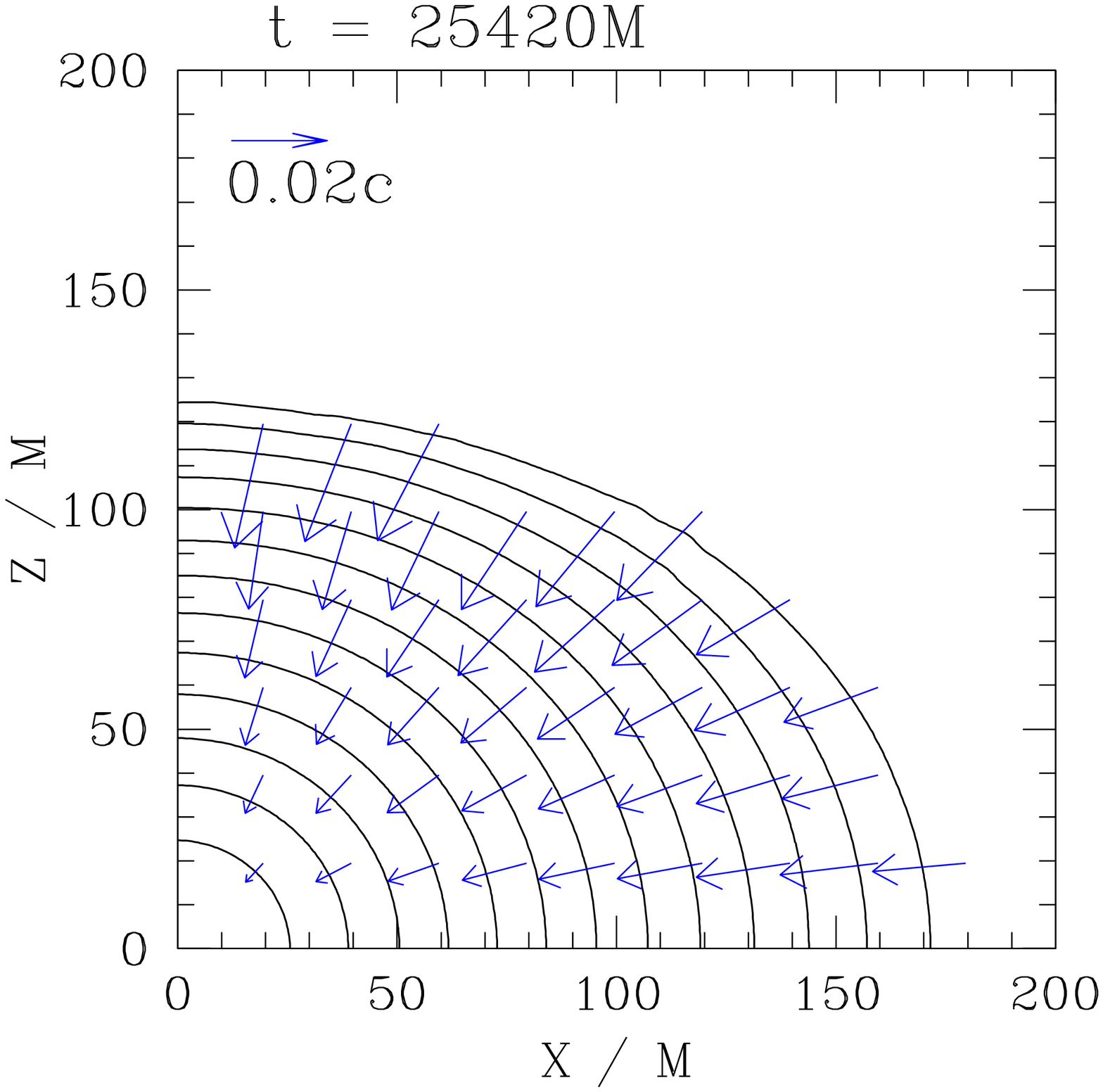}
\epsfxsize=2.15in
\leavevmode
\epsffile{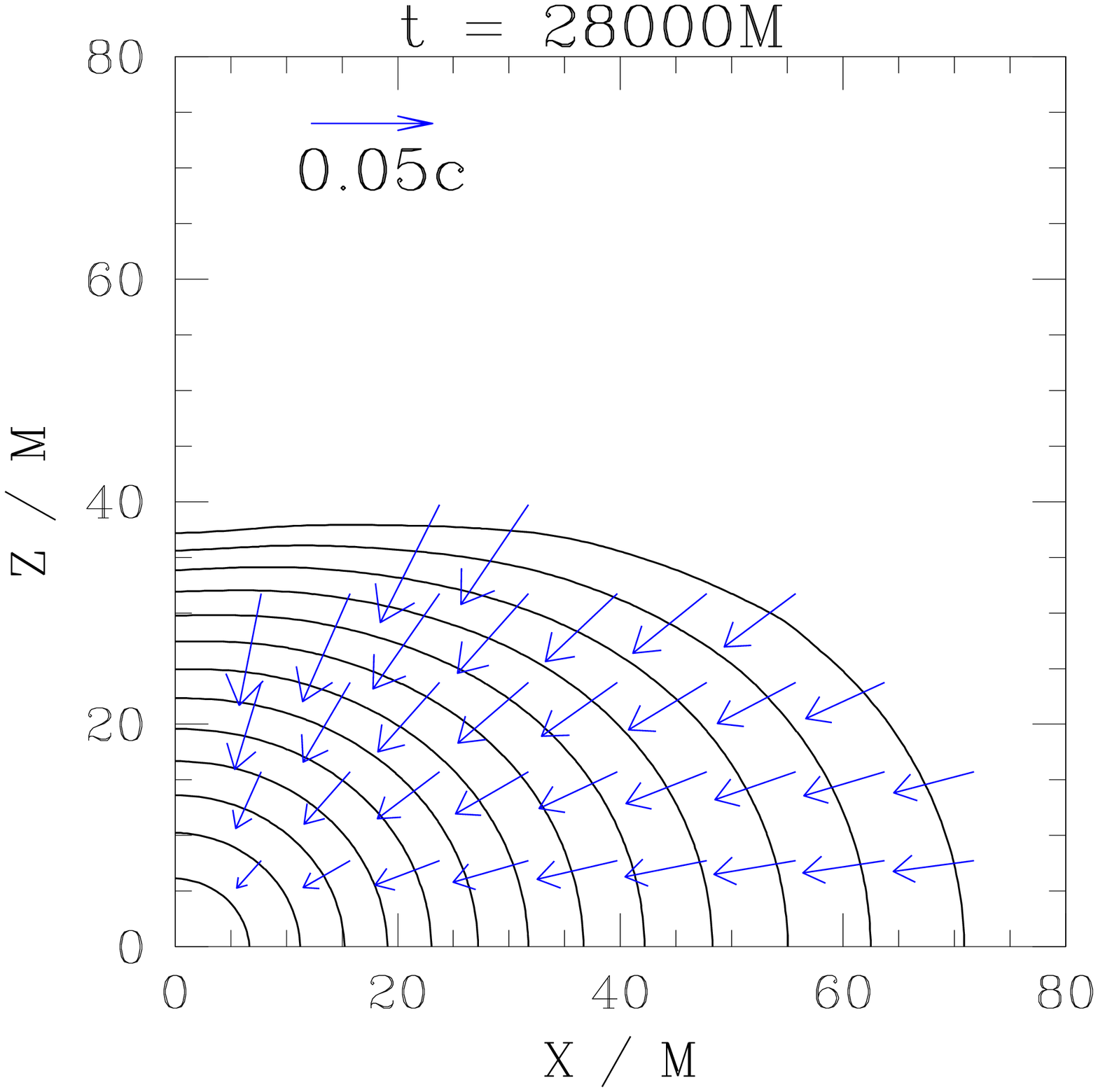}
\epsfxsize=2.15in
\leavevmode
\epsffile{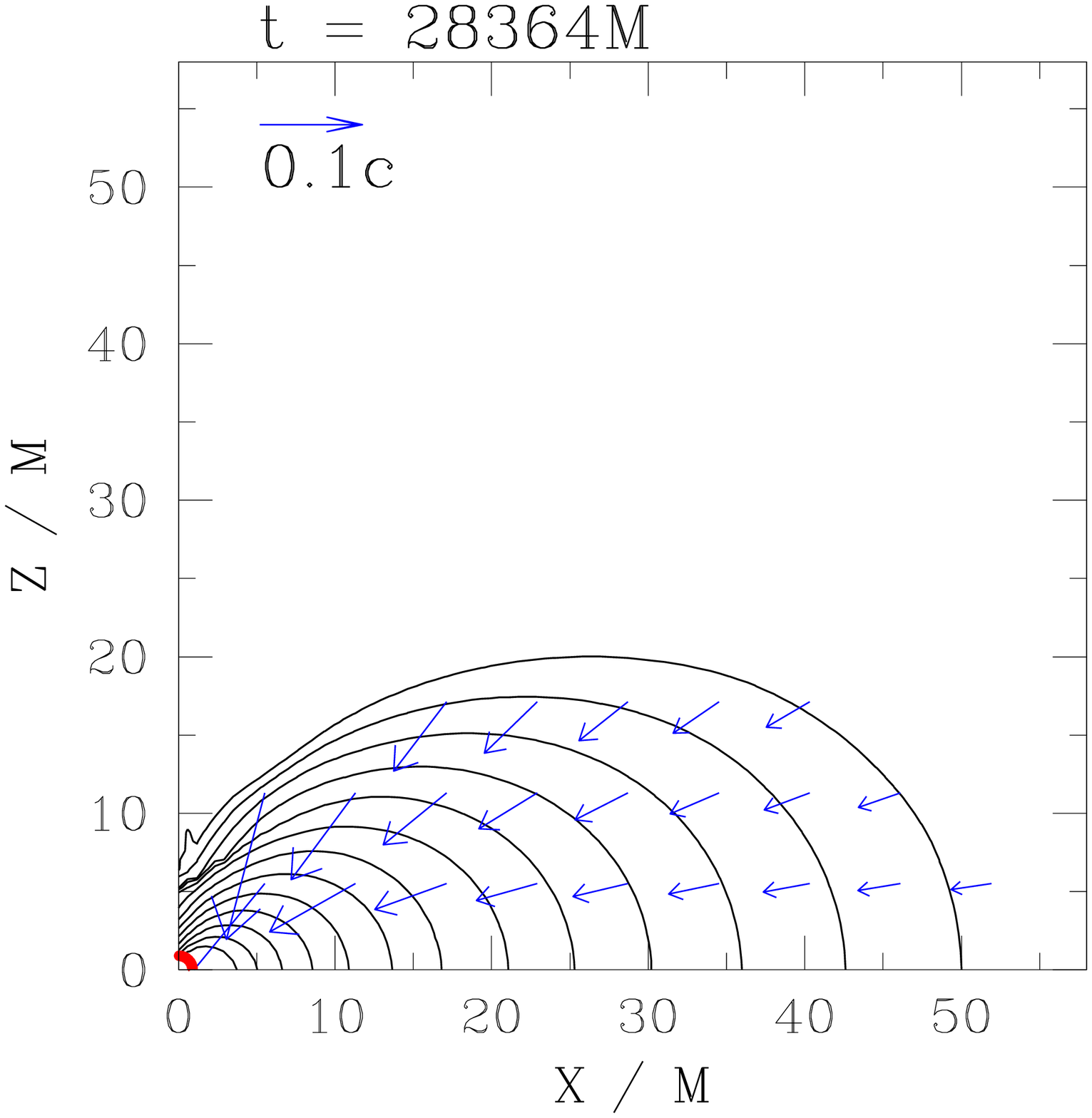}\\
\epsfxsize=2.15in
\leavevmode
\epsffile{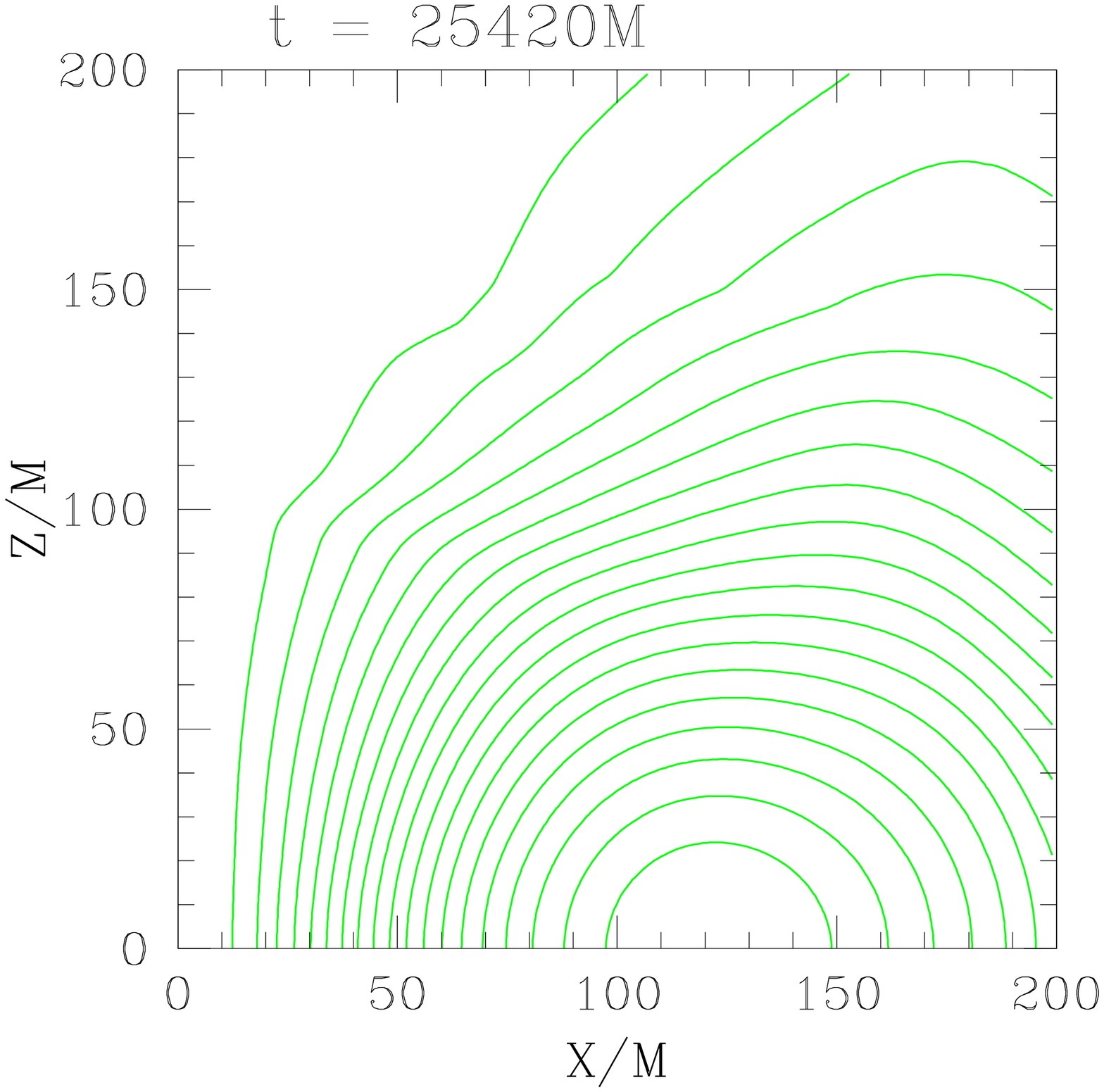}
\epsfxsize=2.15in
\leavevmode
\epsffile{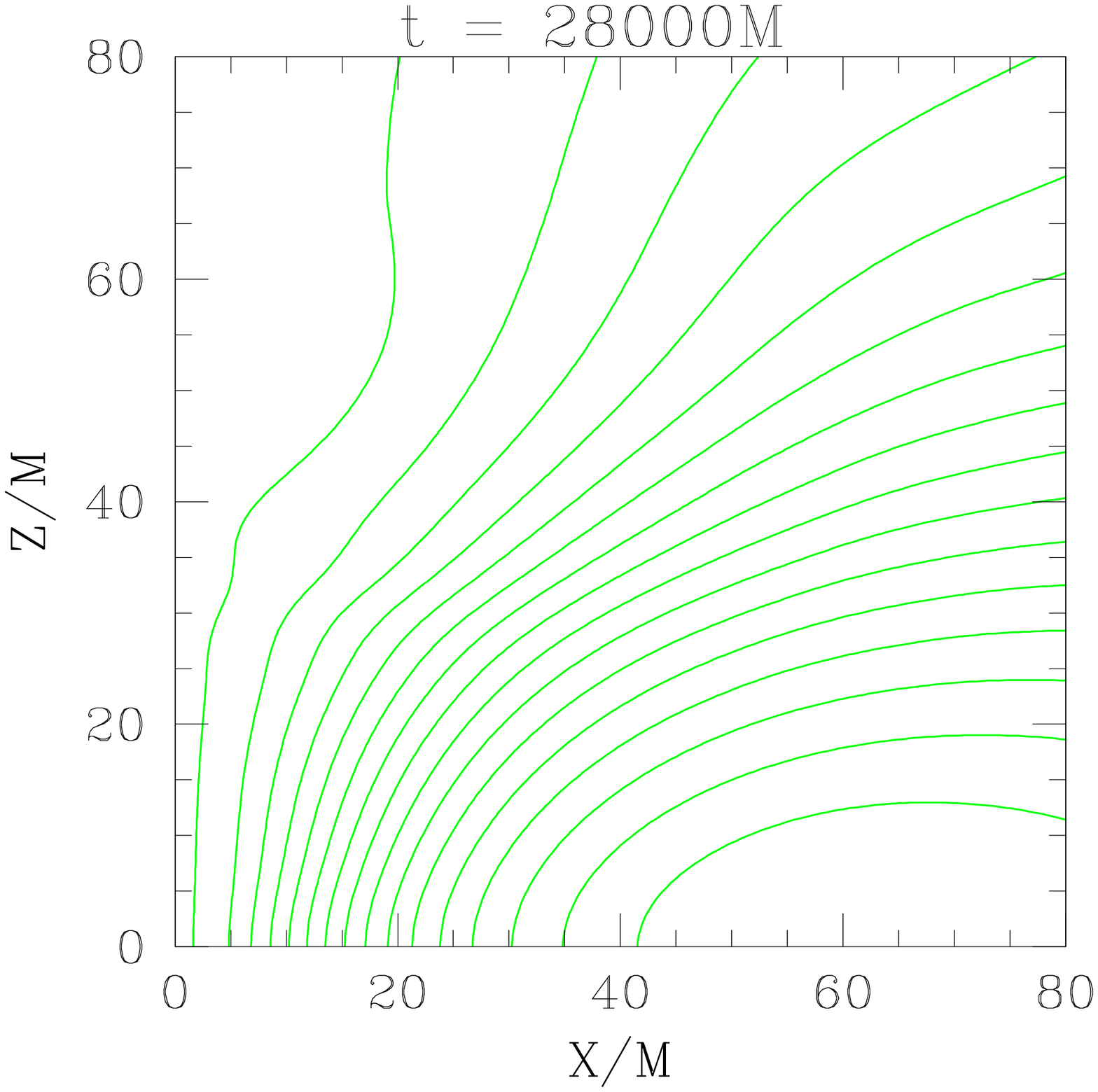}
\epsfxsize=2.15in
\leavevmode
\epsffile{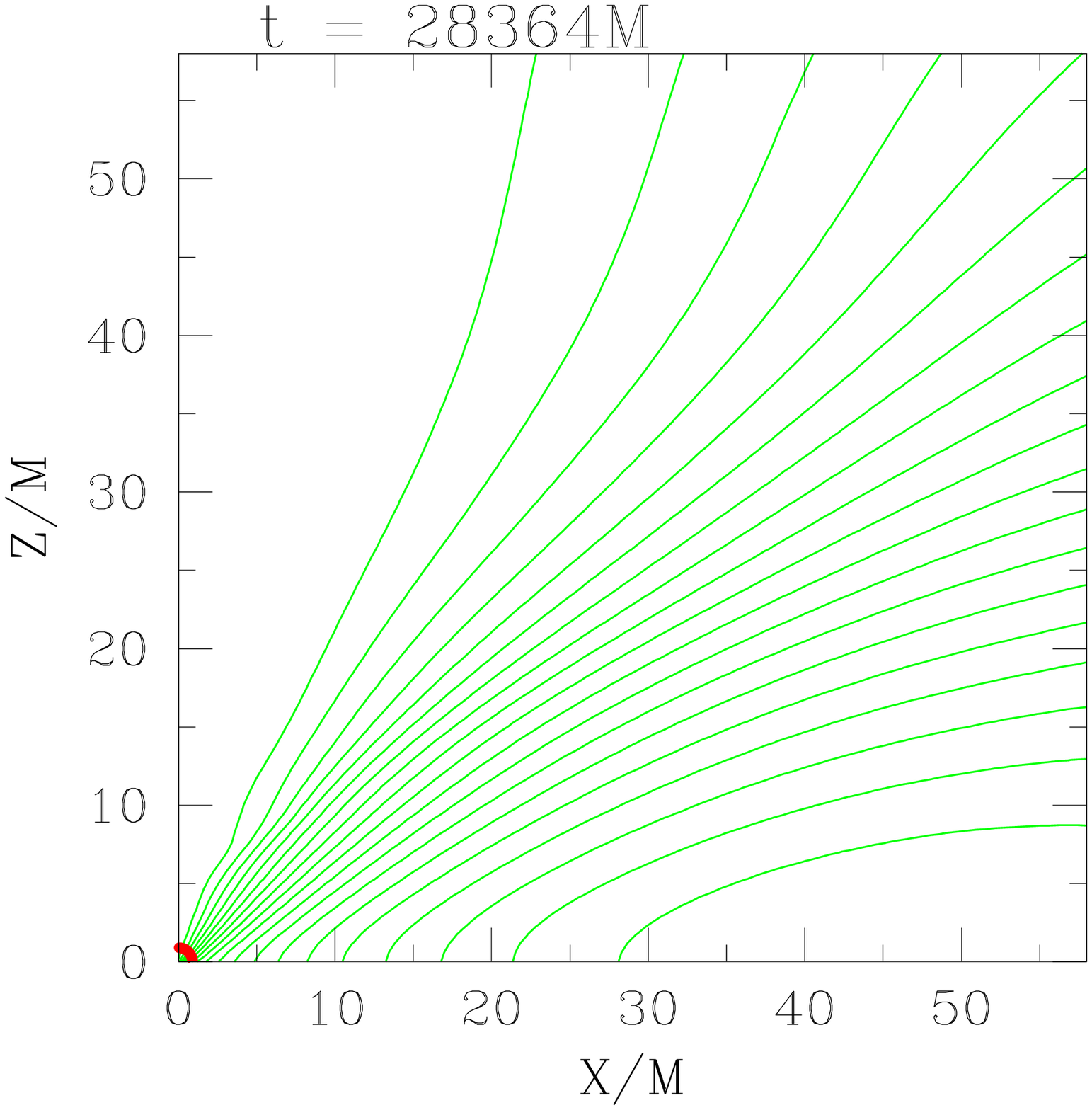}
\caption{Density contour curves and velocity vector fields 
(upper graphs), and magnetic field lines 
(lower graphs) in the meridional plane for model S1. 
The thick (red) lines near the lower left corner in the far right graphs 
denote the apparent
horizon. The density levels are drawn for $\rho_0 = \rho_{\rm scal}
10^{-0.3j}~(j=0$--12), where $\rho_{\rm scal}=11\rho_c(0)$
at $t=25420M$, $\rho_{\rm scal}=340\rho_c(0)$ at $t=28000$,
and $\rho_{\rm scal}=1000\rho_c(0)$ at $t=28364M$.
The poloidal magnetic field lines are drawn as contours of $A_{\varphi}$,
with levels given by $A_{\varphi}= (j/20)A_{\varphi,{\rm max}}$
with $j=1$--19,
where $A_{\varphi,{\rm max}}$ is the maximum value of $A_{\varphi}$
at the given time. Note that the scale is different for each time slice to
show the central region in detail.}
\label{fig:prexconS1}
\end{center}
\end{figure*}

\begin{figure*}
\vspace{-4mm}
\begin{center}
\epsfxsize=2.15in
\leavevmode
\epsffile{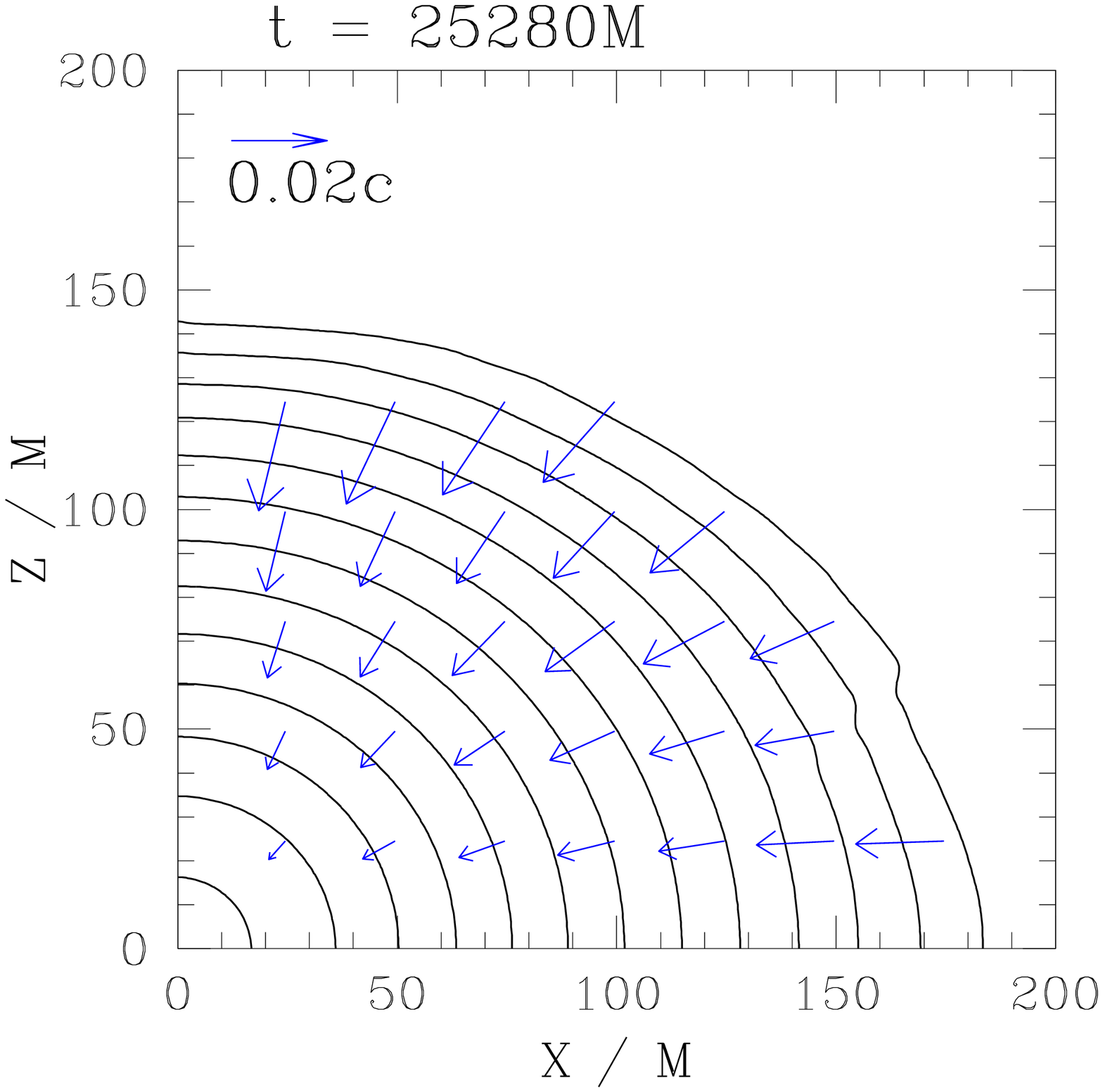}
\epsfxsize=2.15in
\leavevmode
\epsffile{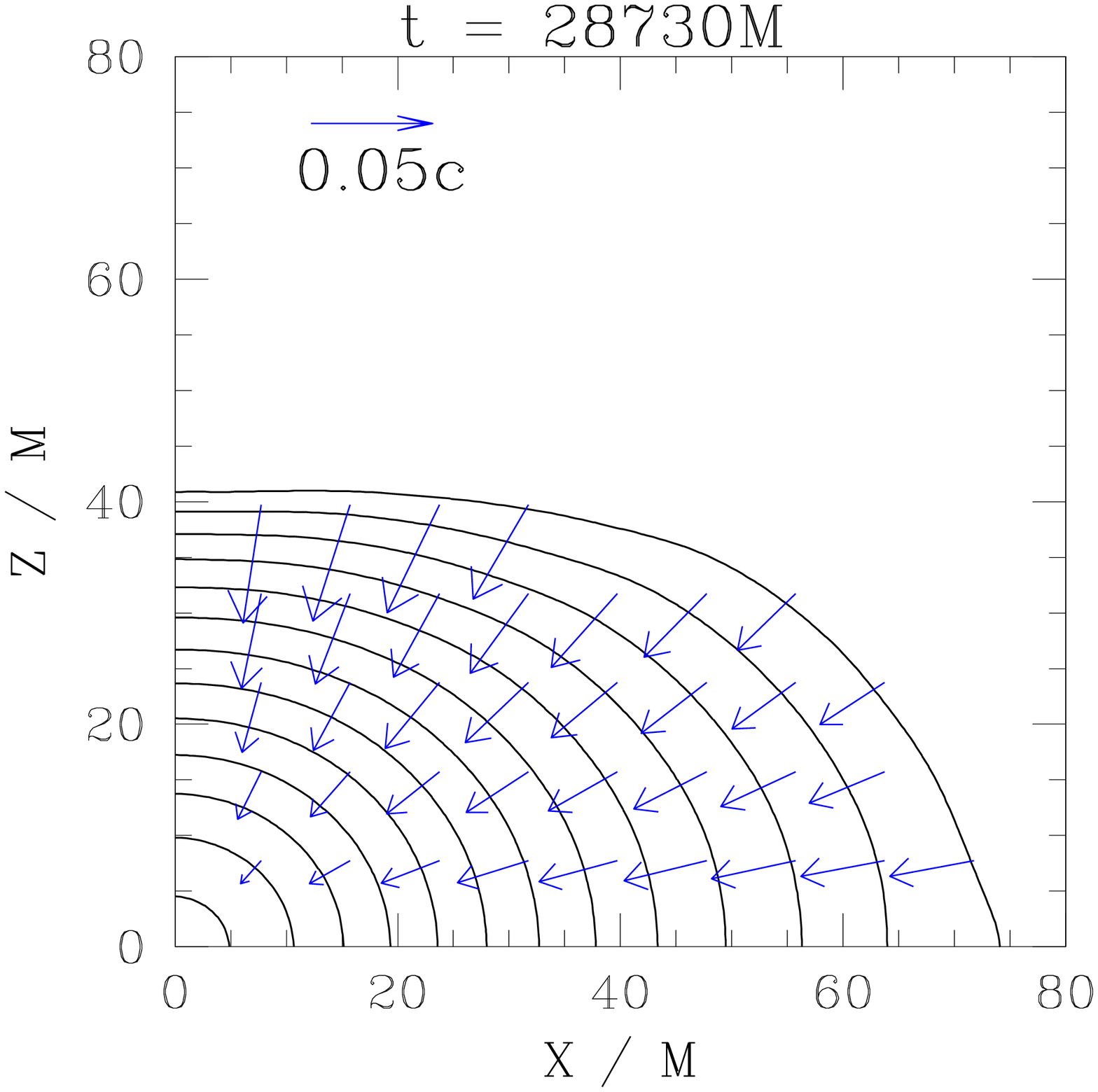}
\epsfxsize=2.15in
\leavevmode
\epsffile{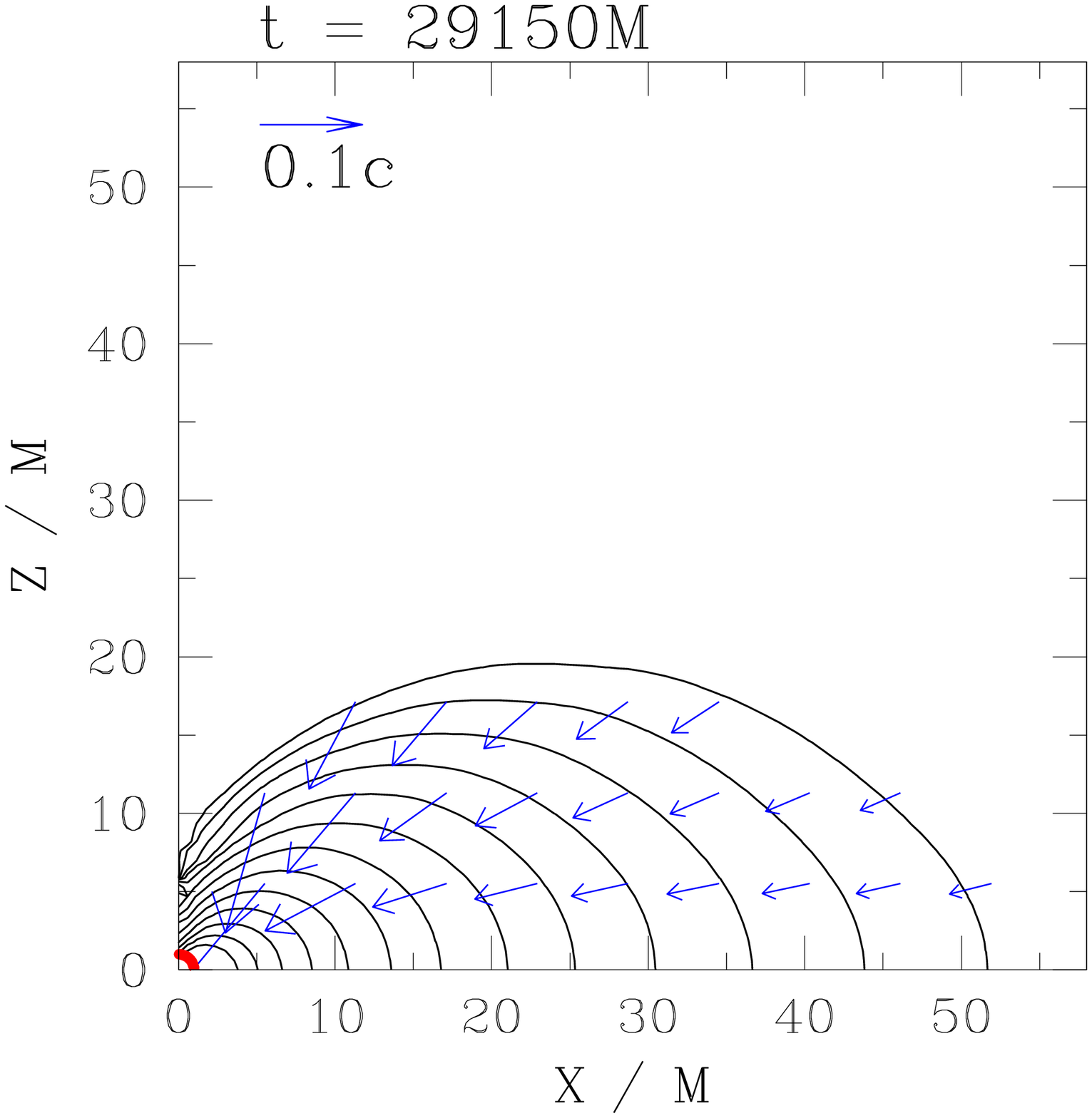}\\
\epsfxsize=2.15in
\leavevmode
\epsffile{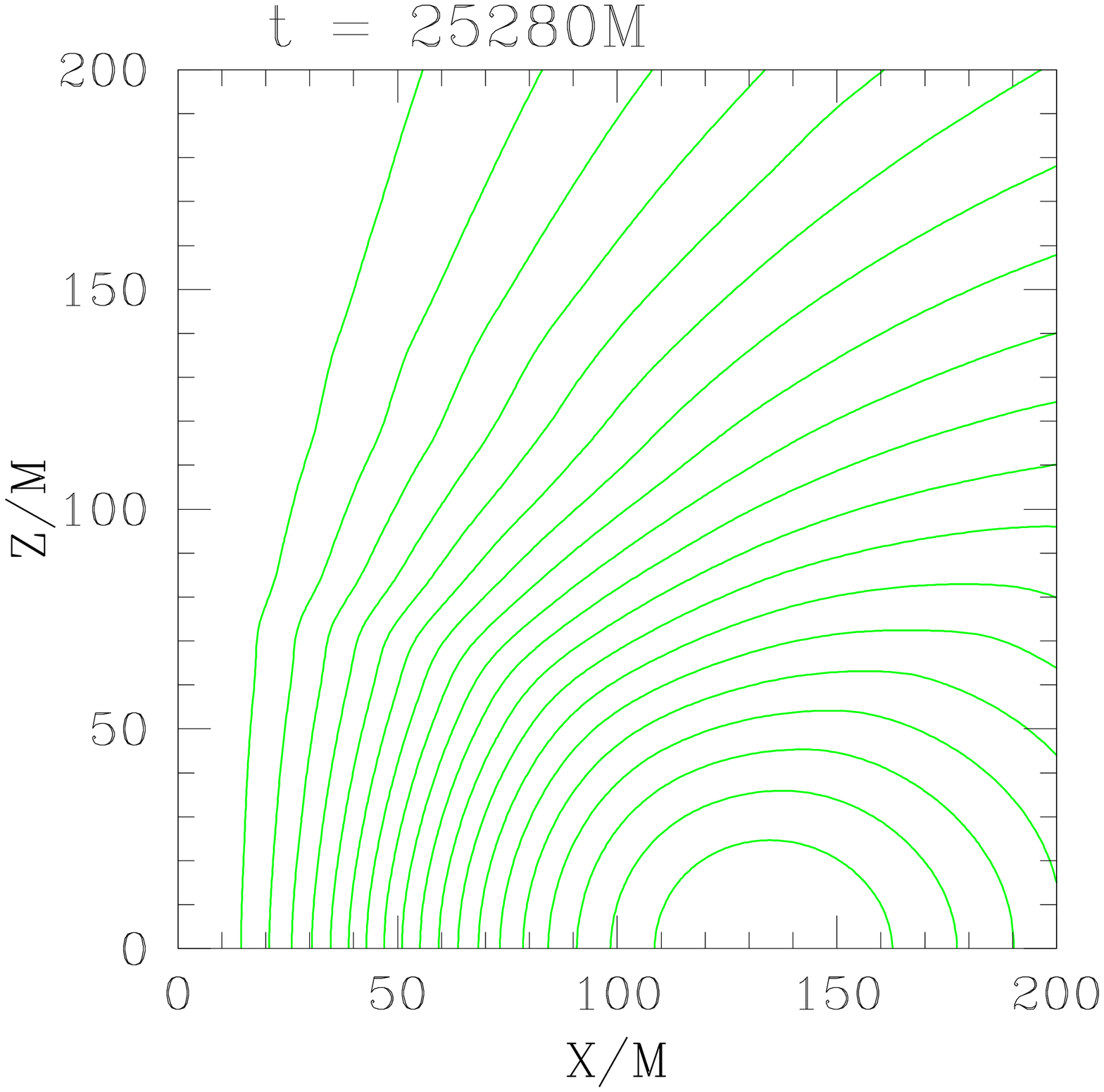}
\epsfxsize=2.15in
\leavevmode
\epsffile{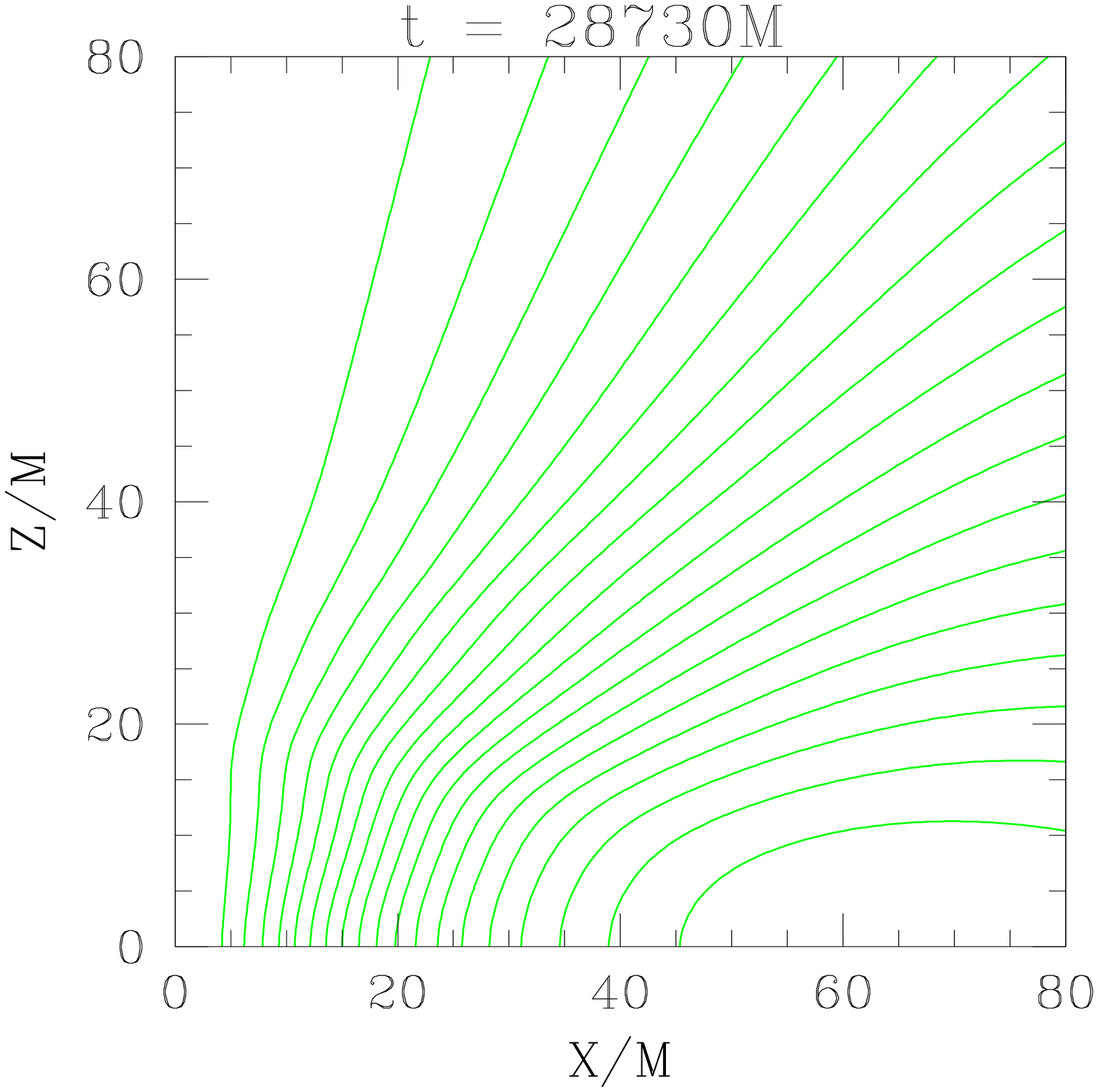}
\epsfxsize=2.15in
\leavevmode
\epsffile{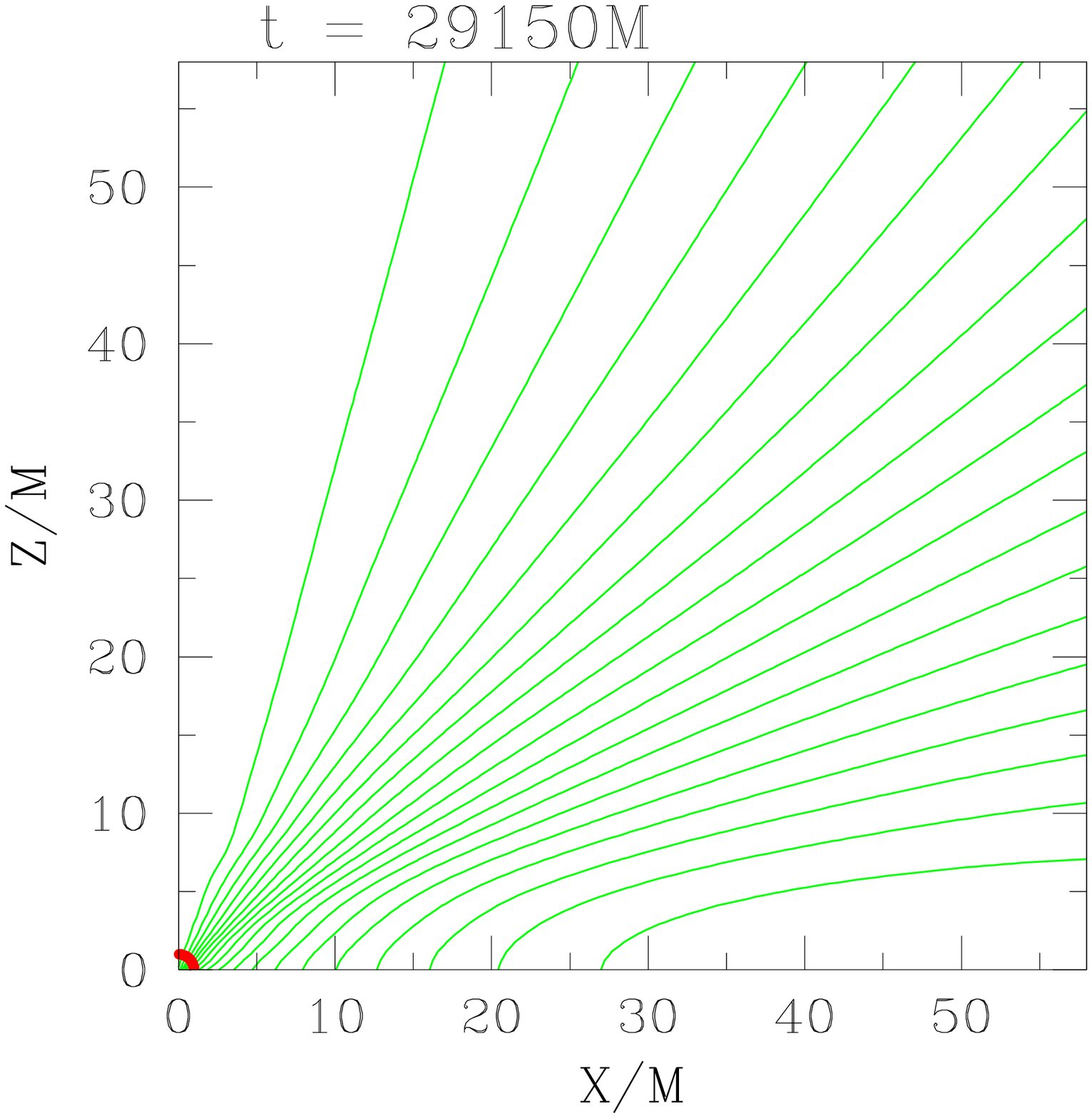}
\caption{Same as Fig.~\ref{fig:prexconS1} but for model S2.}
\label{fig:prexconS2}
\end{center}
\end{figure*}

Figure~\ref{fig:rho_alp} shows the evolution of central density and lapse 
for the three models. Figures~\ref{fig:prexconS0}--\ref{fig:prexconS2} 
show the density contours and velocity vectors during pre-excision 
evolution for models S0, S1 and S2 respectively. Poloidal 
magnetic field lines are also 
shown for models S1 and S2 in Figs.~\ref{fig:prexconS1} and
\ref{fig:prexconS2}. We see that magnetic fields slightly 
slow down the collapse. As mentioned in Sec.~\ref{sec:id}, 
the collapse proceeds in a homologous manner at the beginning. When  
the central lapse decreases to $\alpha_c \lesssim 0.9$, the central region 
collapses faster than the outer layers. The apparent horizon appears at 
$t=28280M$ for 
model S0, $t=28360M$ for S1 and $t=29149M$ for S2. Without excision,
the code becomes 
inaccurate soon after the formation of the apparent horizon because of the 
grid stretching. 

\begin{figure}
\begin{center}
\epsfxsize=3in
\leavevmode
\epsffile{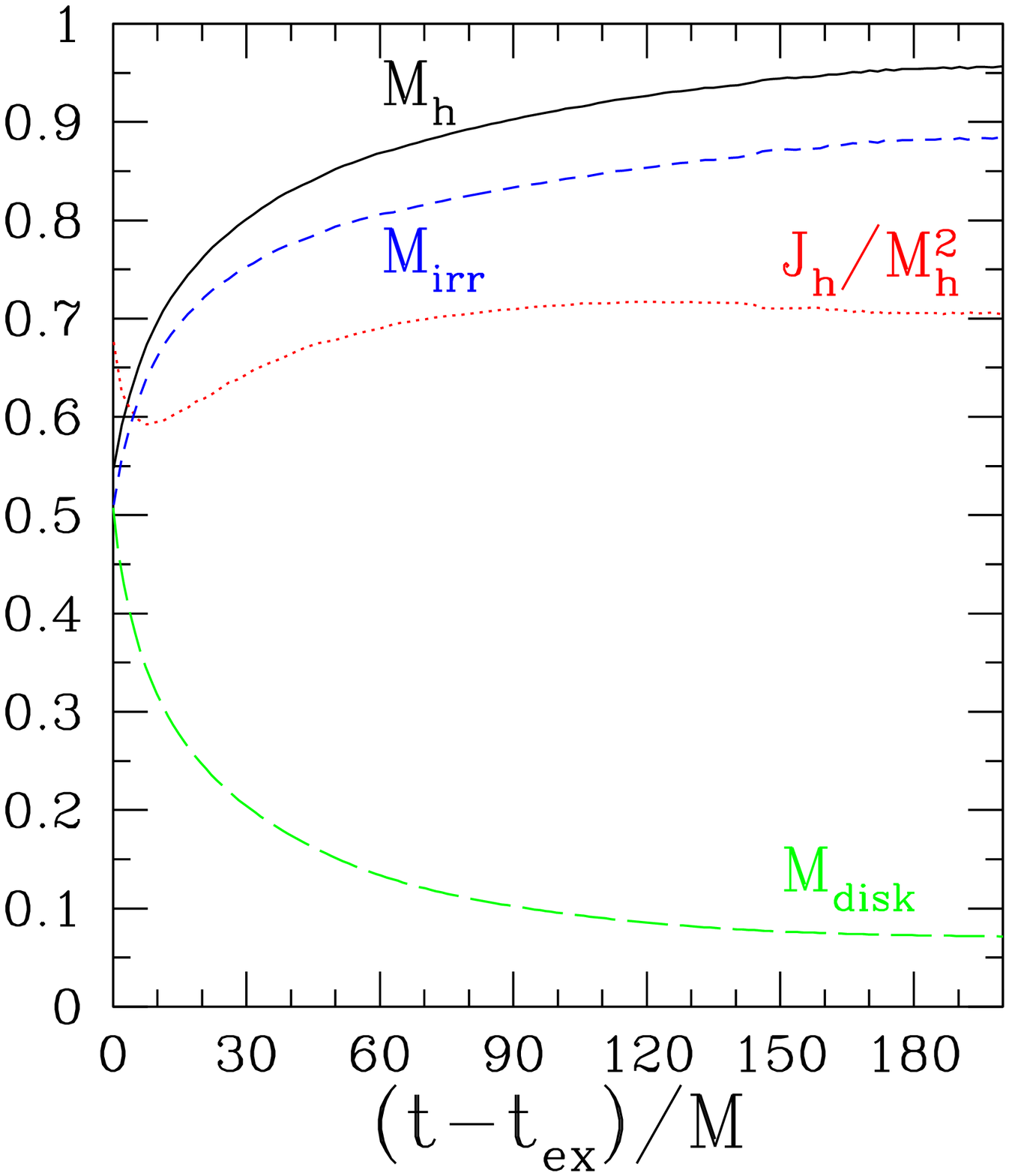}
\caption{Post-excision evolution of the mass $M_h$, spin parameter $J_h/M_h^2$, 
and the irreducible mass $M_{\rm irr}$ of the central black hole, 
and the rest mass of the disk $M_{\rm disk}$ outside the apparent 
horizon for model S0. Time is measured from the beginning
of excision ($t_{\rm ex} = 28284M$).}
\label{fig:hor_S0}
\end{center}
\end{figure}

\begin{figure}
\begin{center}
\epsfxsize=3in
\leavevmode
\epsffile{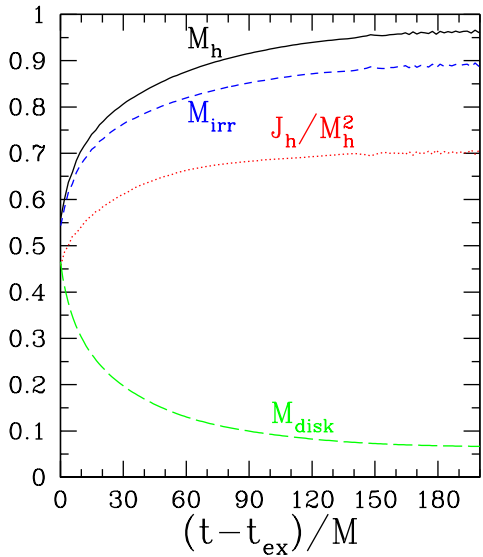}
\caption{Same as Fig.~\ref{fig:hor_S0} but for model S1.
Excision starts at $t_{\rm ex} = 28364M$.}
\label{fig:hor_S1}
\end{center}
\end{figure}

\begin{figure}
\begin{center}
\epsfxsize=3in
\leavevmode
\epsffile{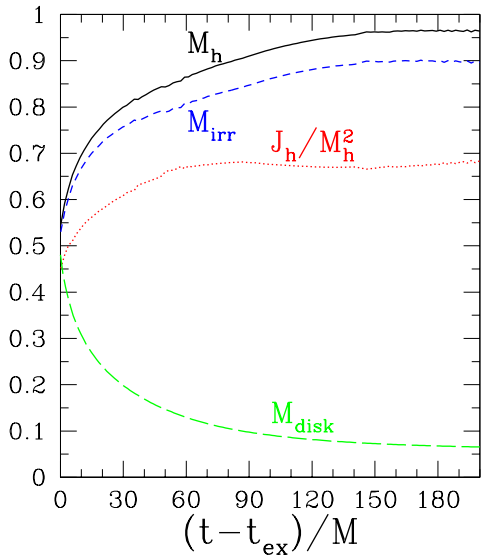}
\caption{Same as Fig.~\ref{fig:hor_S0} but for model S2. 
Excision starts at $t_{\rm ex} = 29150M$.}
\label{fig:hor_S2}
\end{center}
\end{figure}

\begin{figure*}
\vspace{-4mm}
\begin{center}
\epsfxsize=2.15in
\leavevmode
\epsffile{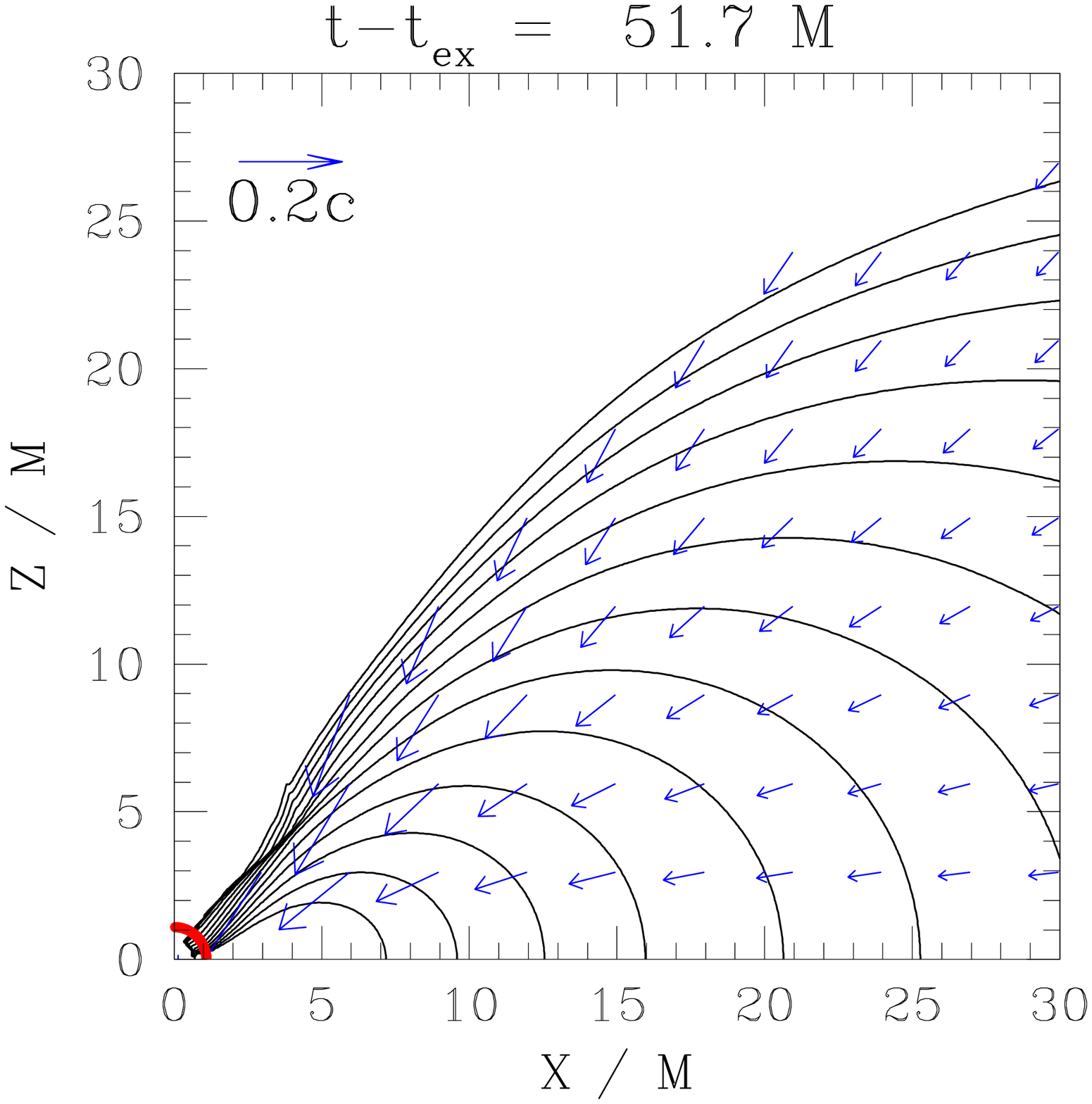}
\epsfxsize=2.15in
\leavevmode
\epsffile{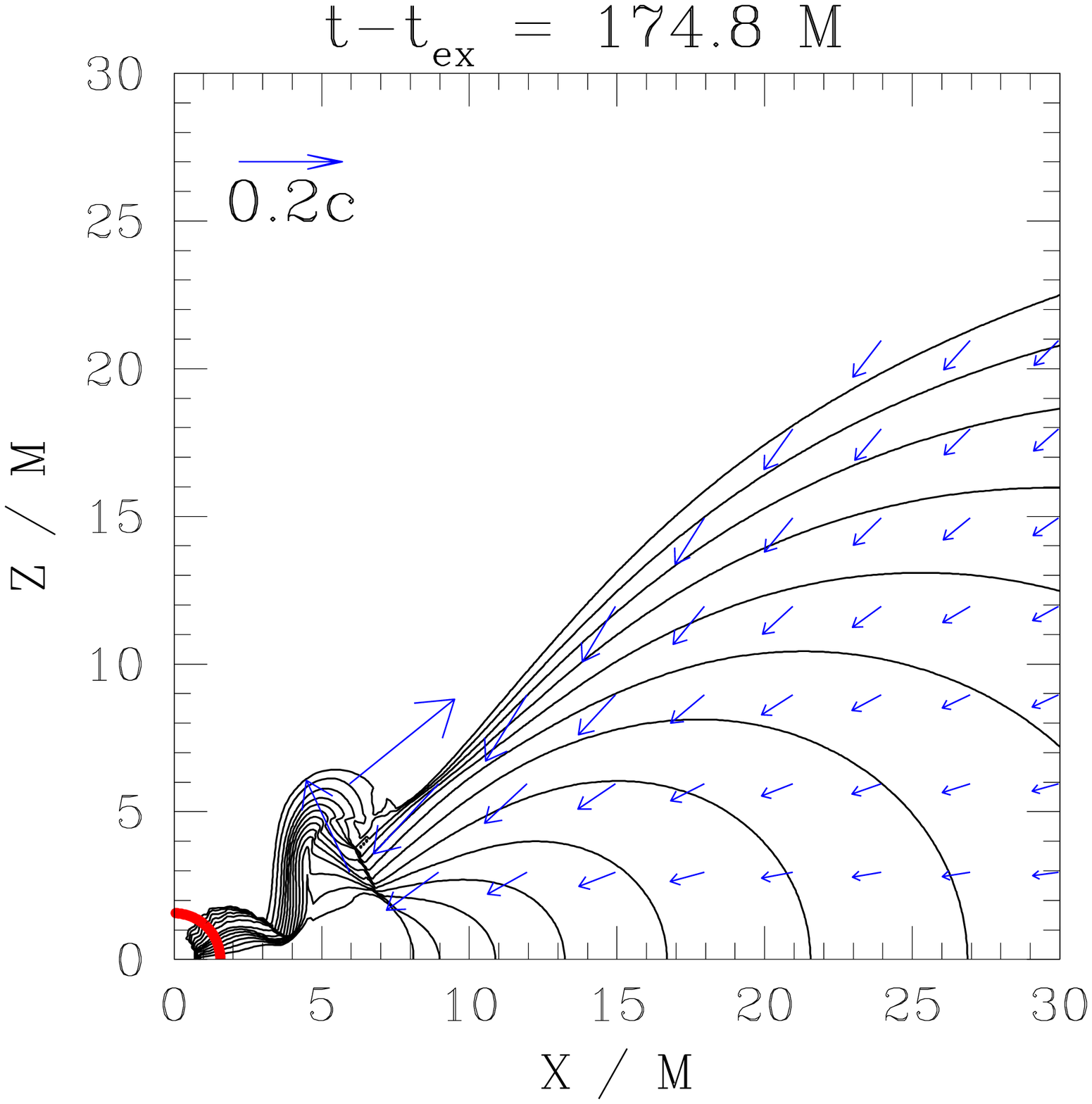}
\epsfxsize=2.15in
\leavevmode
\epsffile{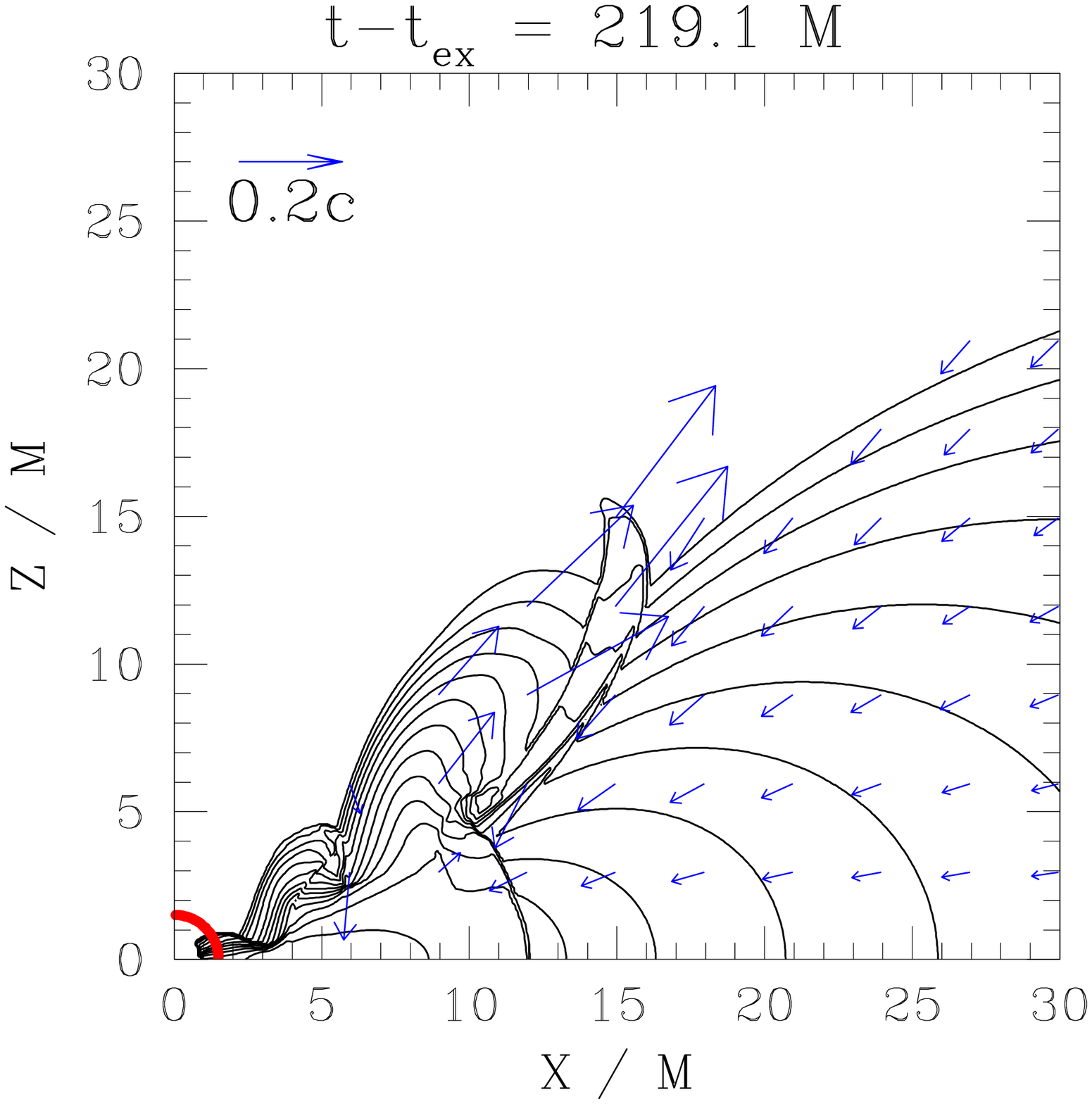}\\
\epsfxsize=2.15in
\leavevmode
\epsffile{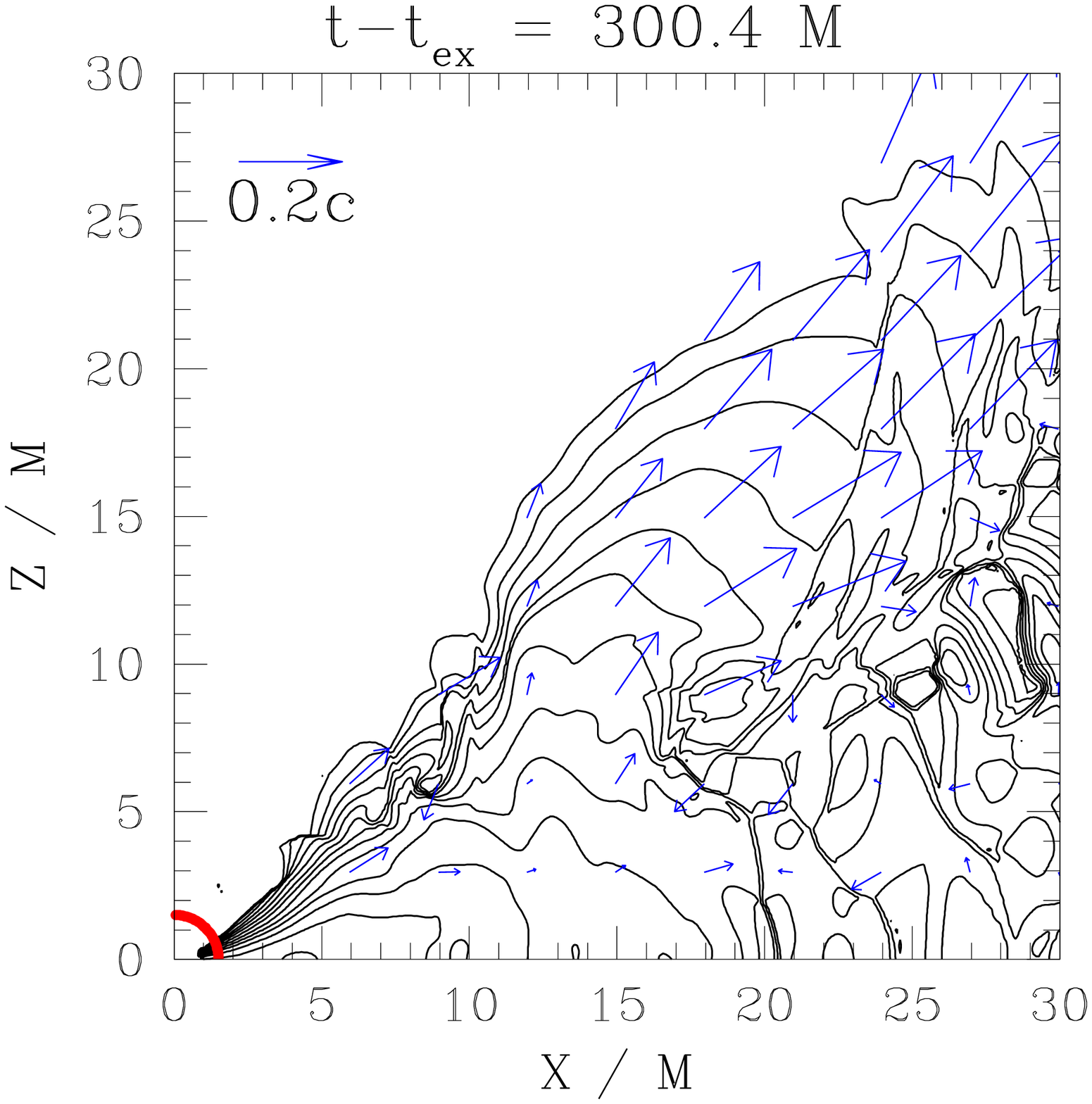}
\epsfxsize=2.15in
\leavevmode
\epsffile{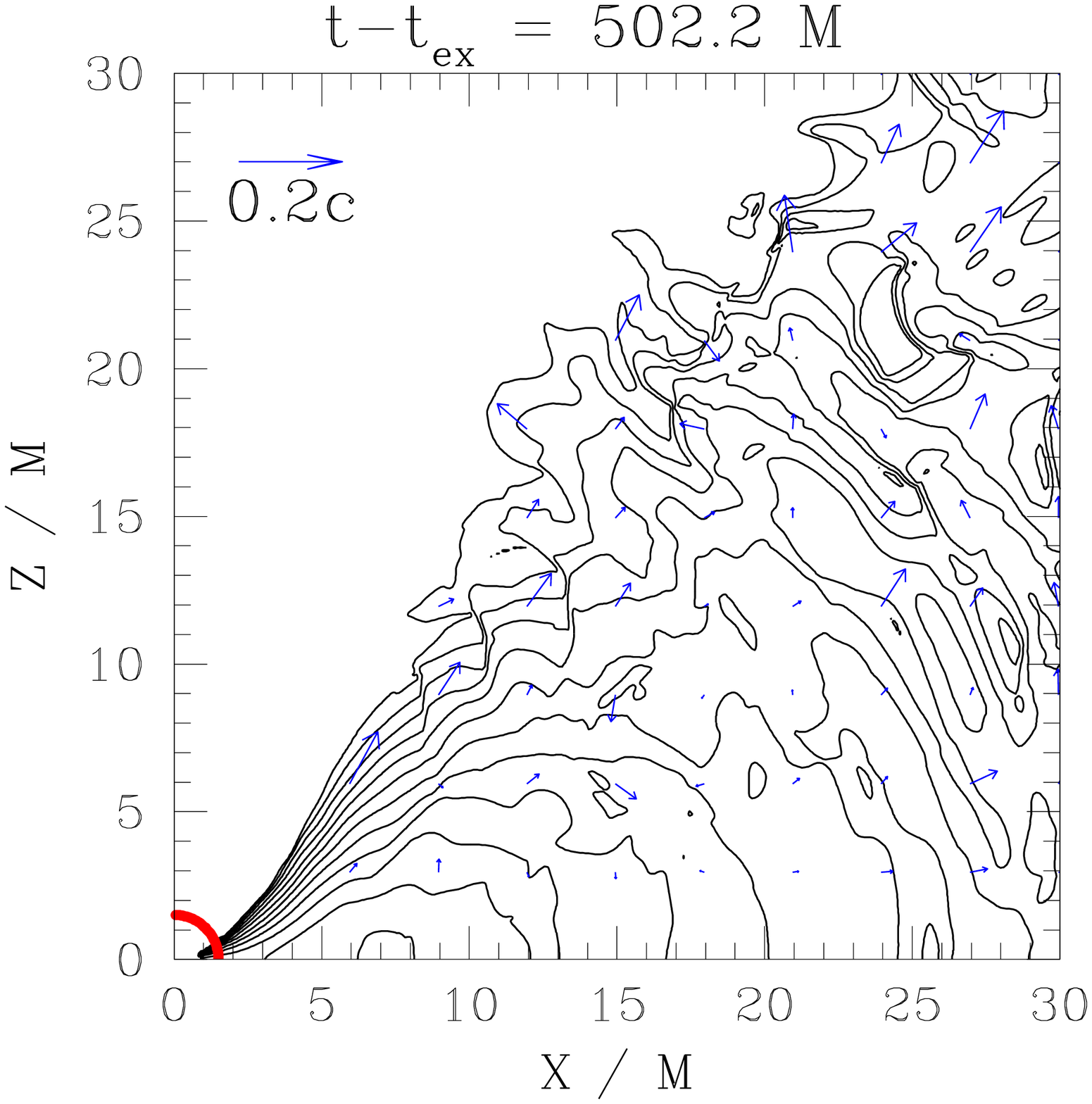}
\epsfxsize=2.15in
\leavevmode
\epsffile{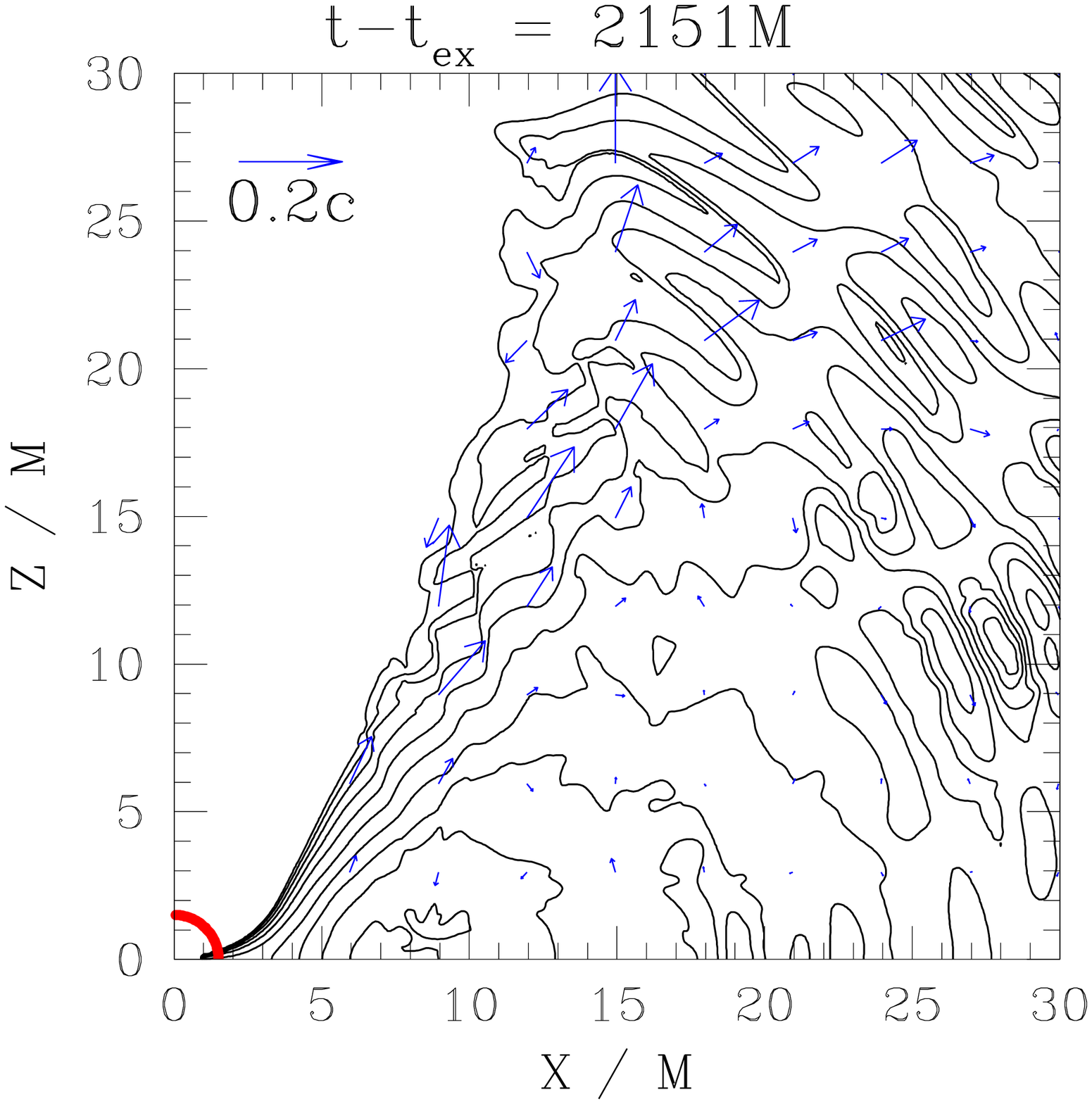}
\caption{Snapshots of density contour curves and velocity vectors 
in the post-excision
evolution for model S0. The contours are drawn for
$\rho_0 = 100 \rho_c(0) 10^{-0.3j}~(j=0$--10). The red
line denotes the apparent horizon.}
\label{fig:exconS0}
\end{center}
\end{figure*}

\begin{figure*}
\vspace{-4mm}
\begin{center}
\epsfxsize=2.15in
\leavevmode
\epsffile{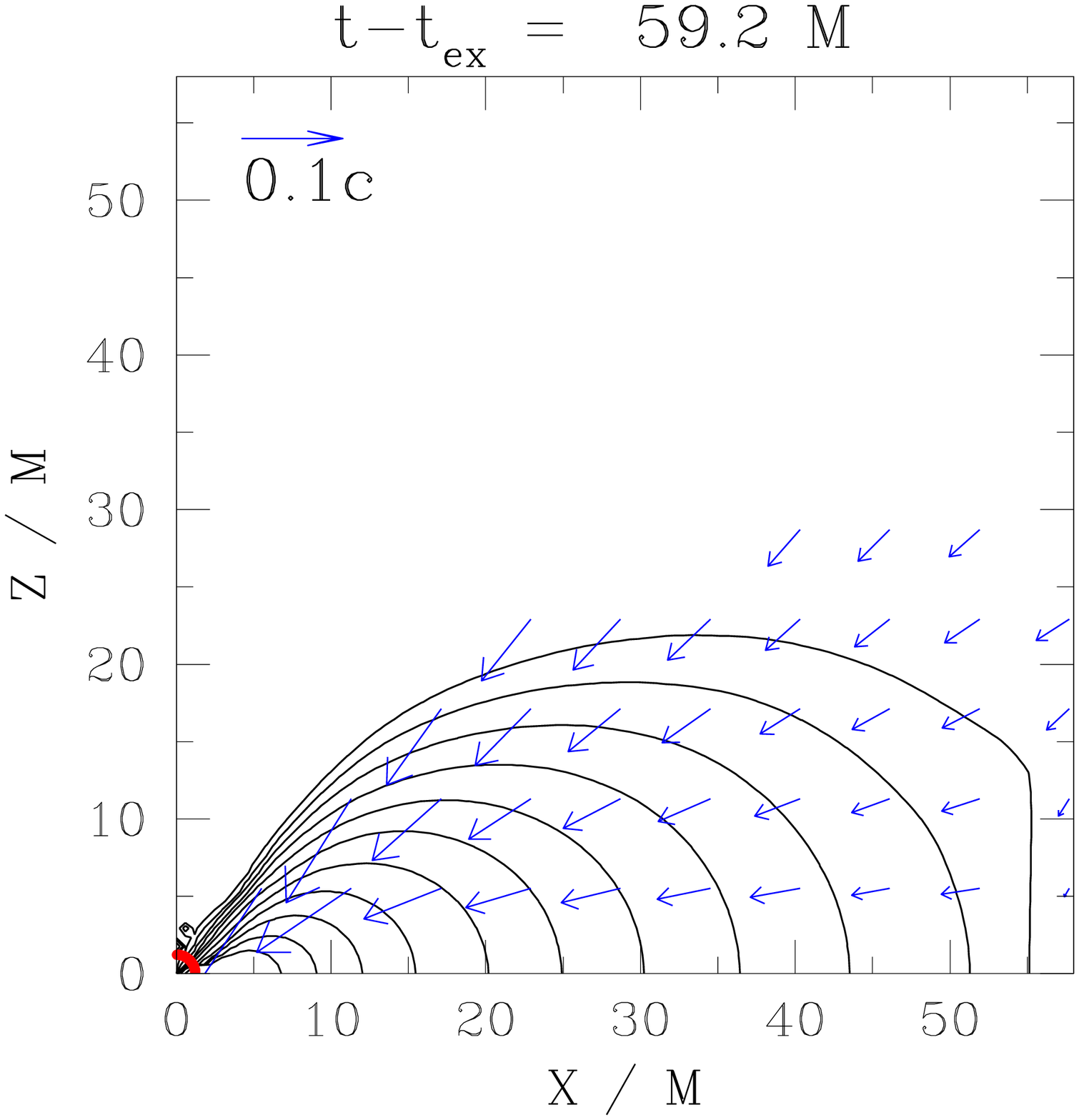}
\epsfxsize=2.15in
\leavevmode
\epsffile{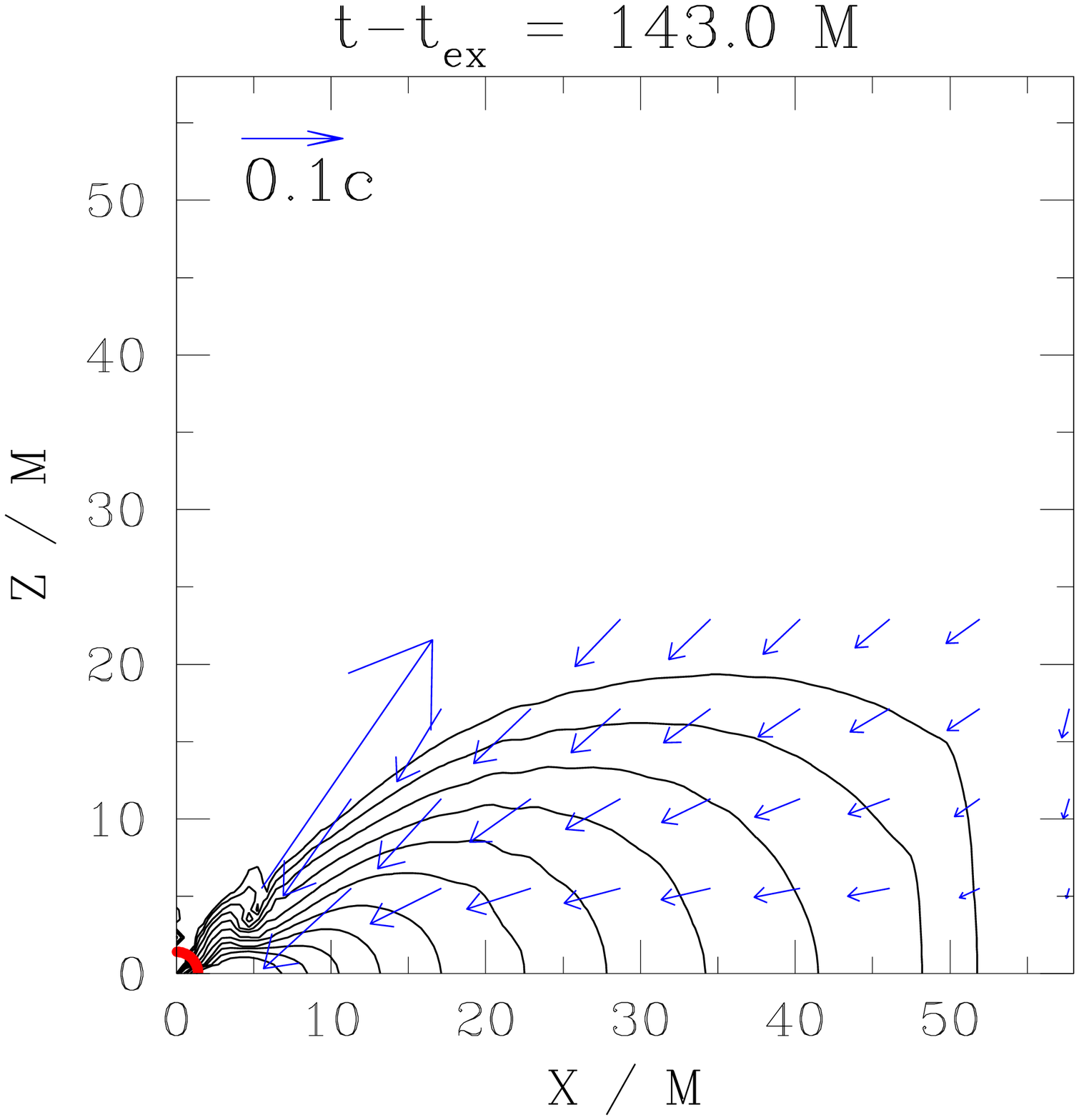}
\epsfxsize=2.15in
\leavevmode
\epsffile{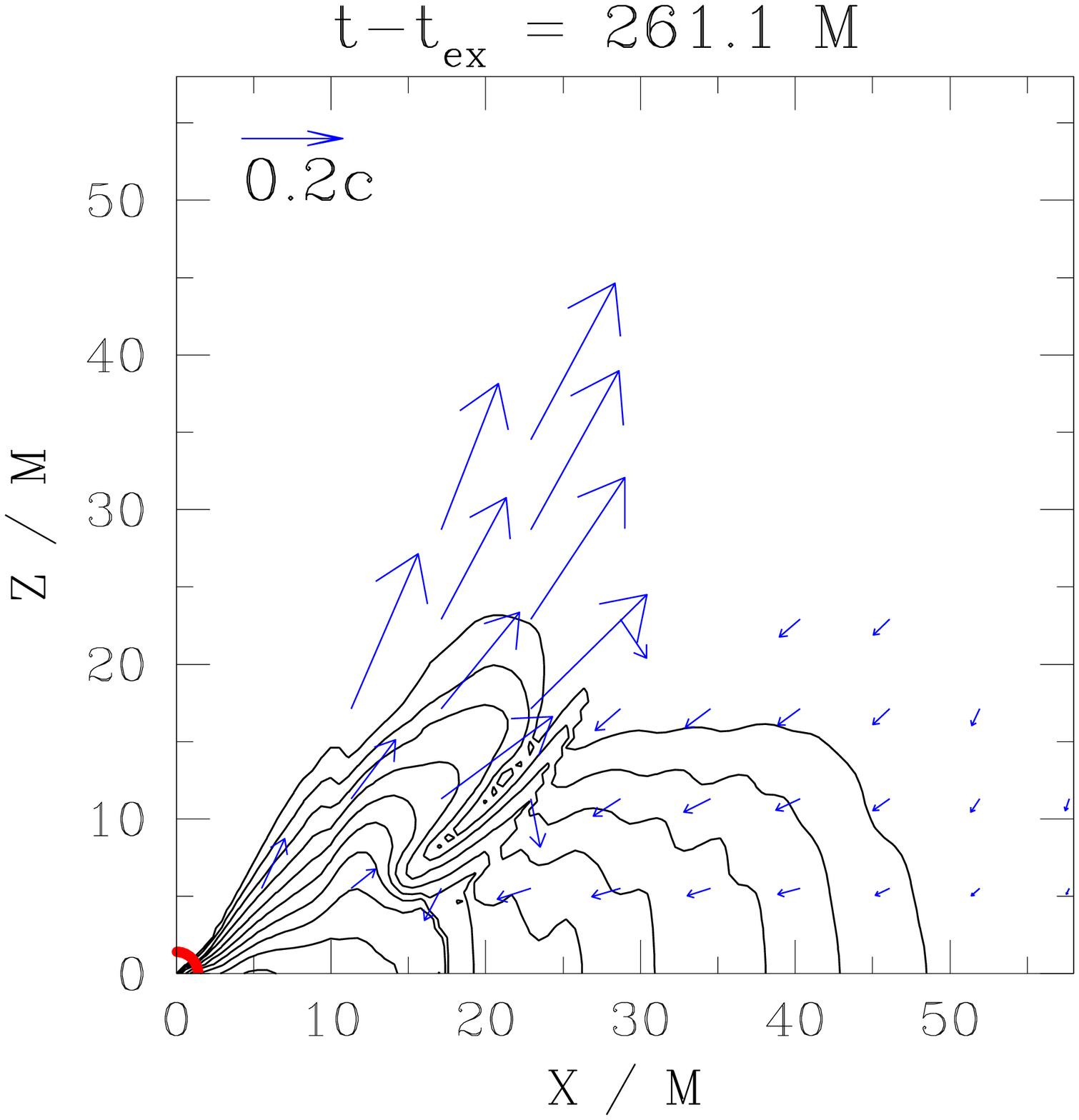}\\
\epsfxsize=2.15in
\leavevmode
\epsffile{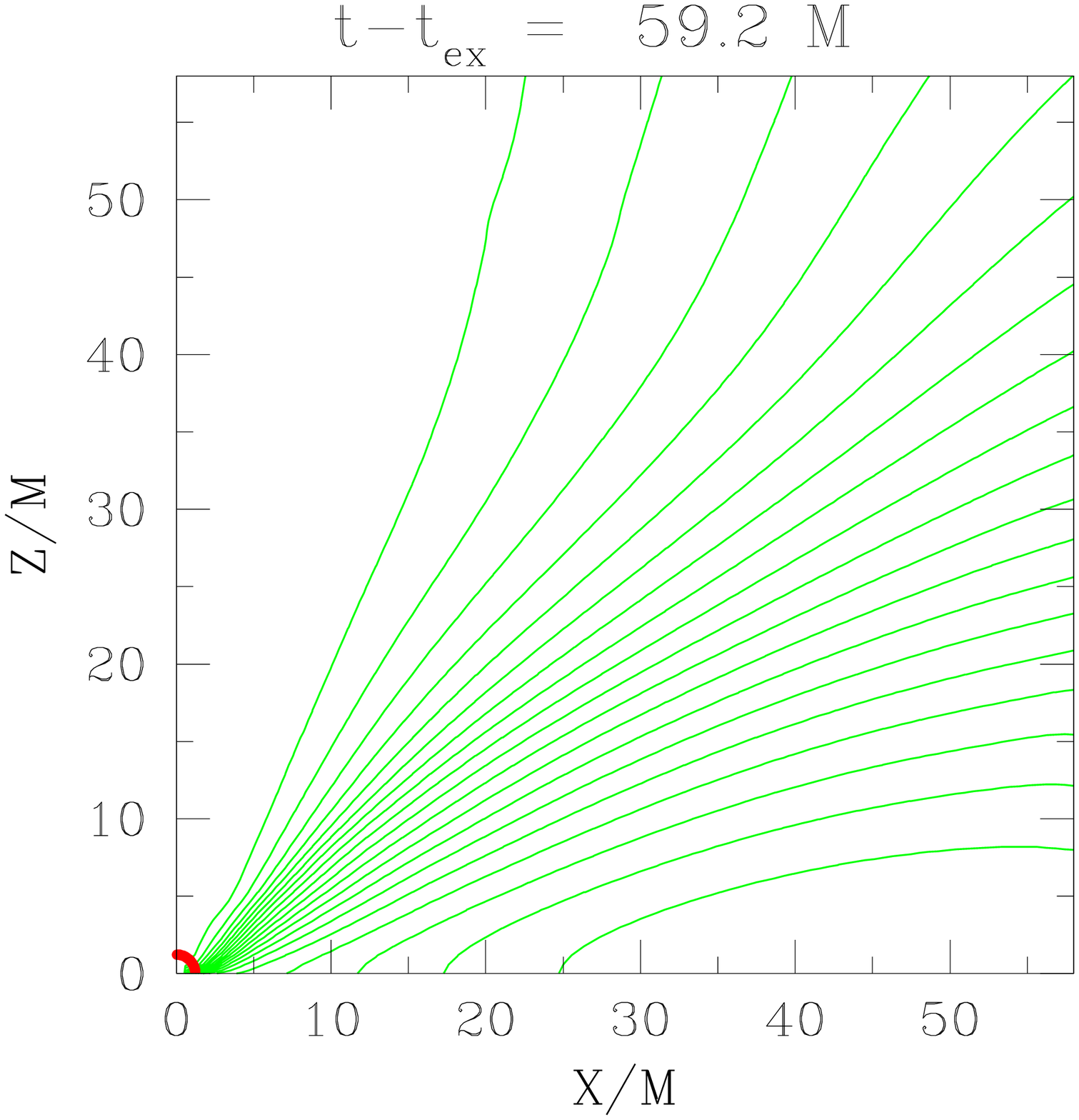}
\epsfxsize=2.15in
\leavevmode
\epsffile{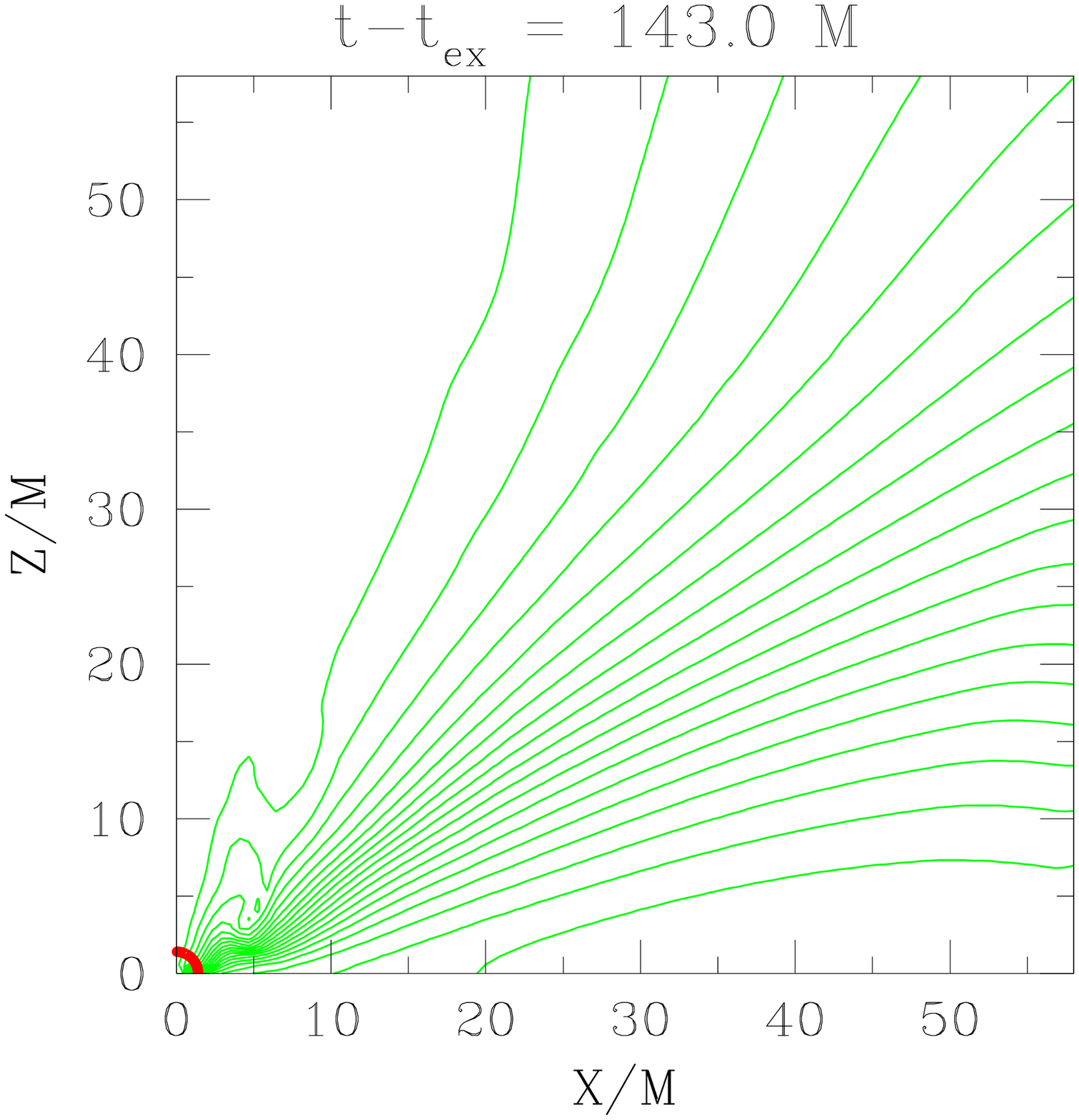}
\epsfxsize=2.15in
\leavevmode
\epsffile{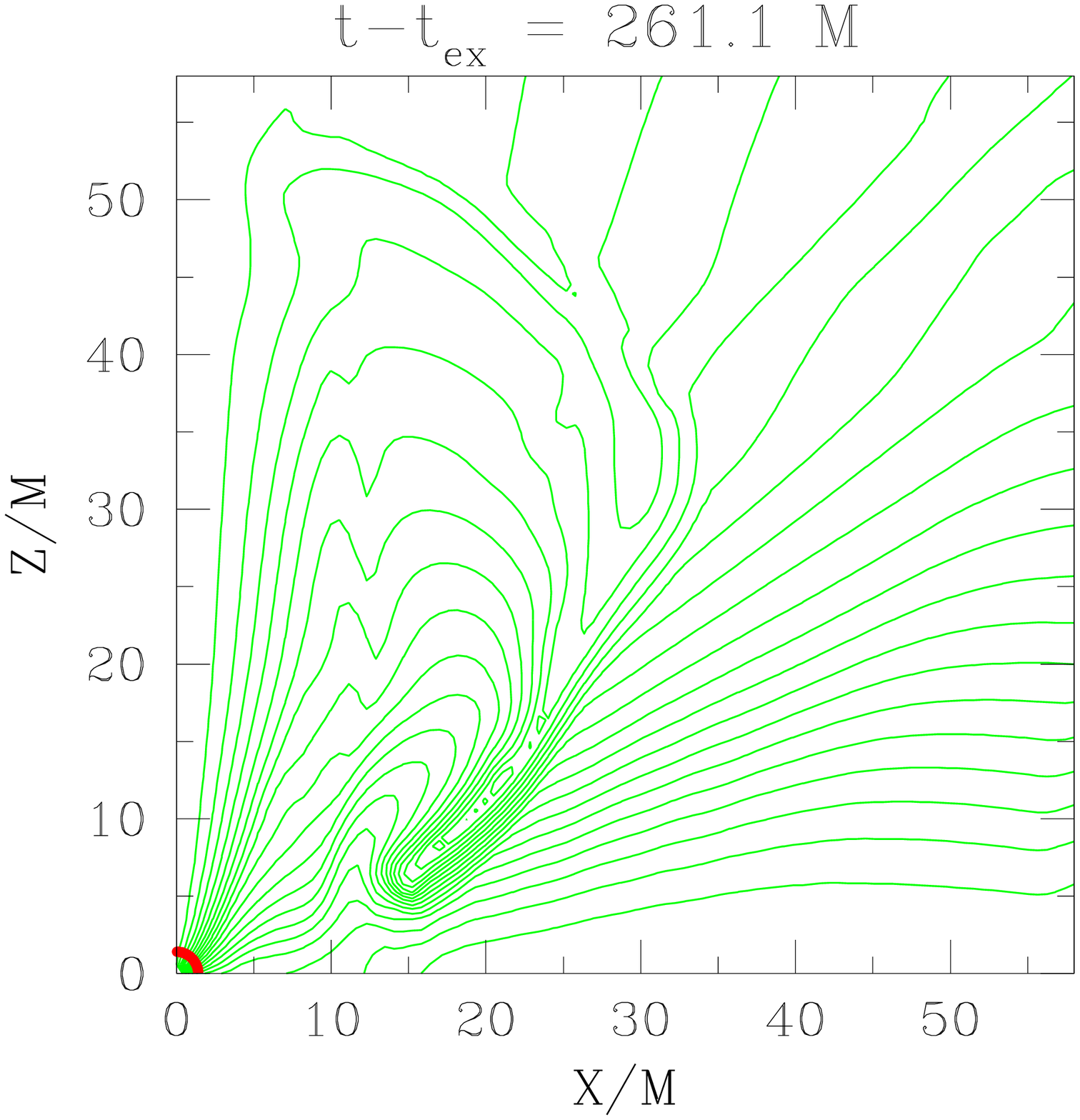}\\
\epsfxsize=2.15in
\leavevmode
\epsffile{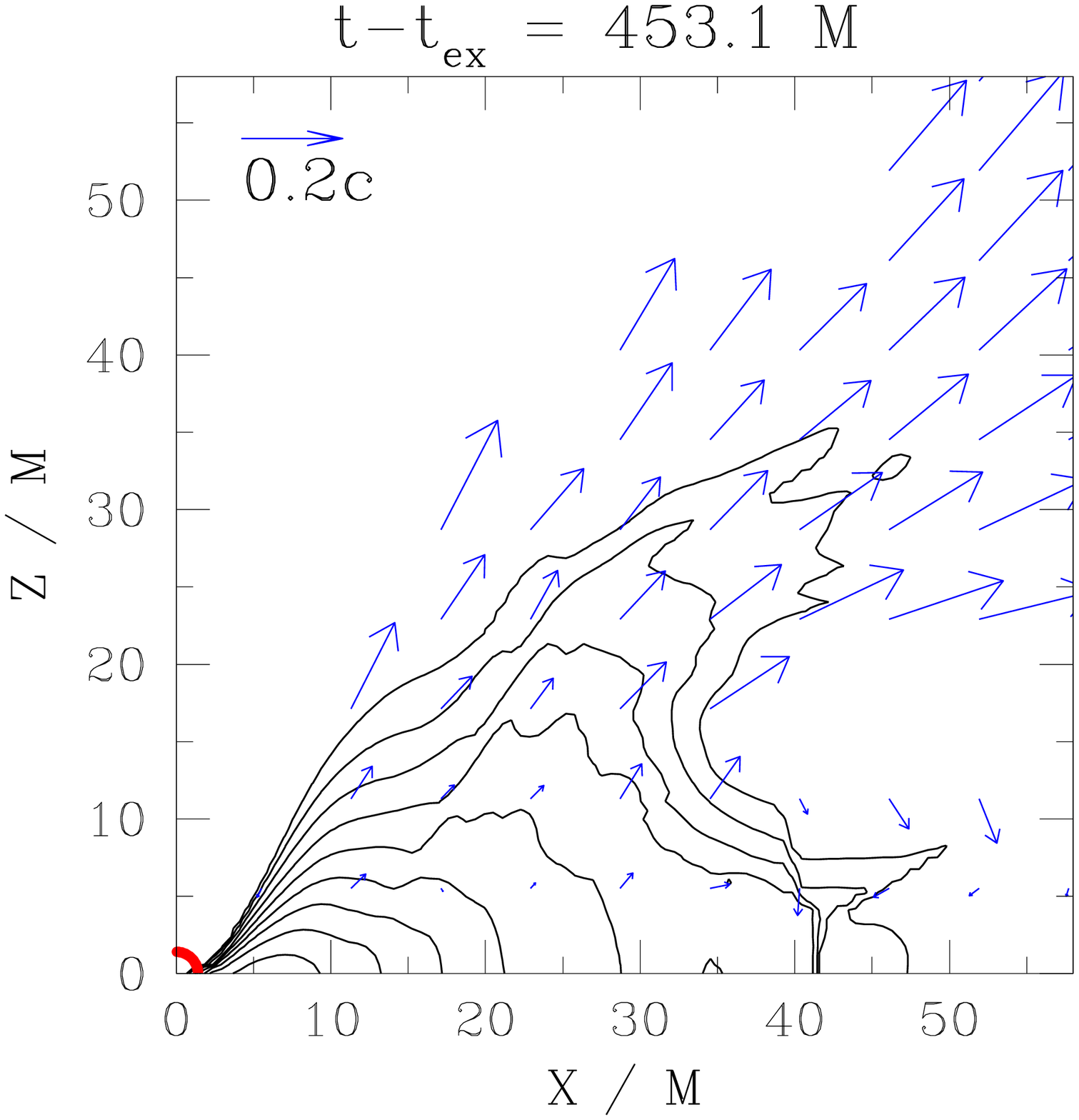}
\epsfxsize=2.15in
\leavevmode
\epsffile{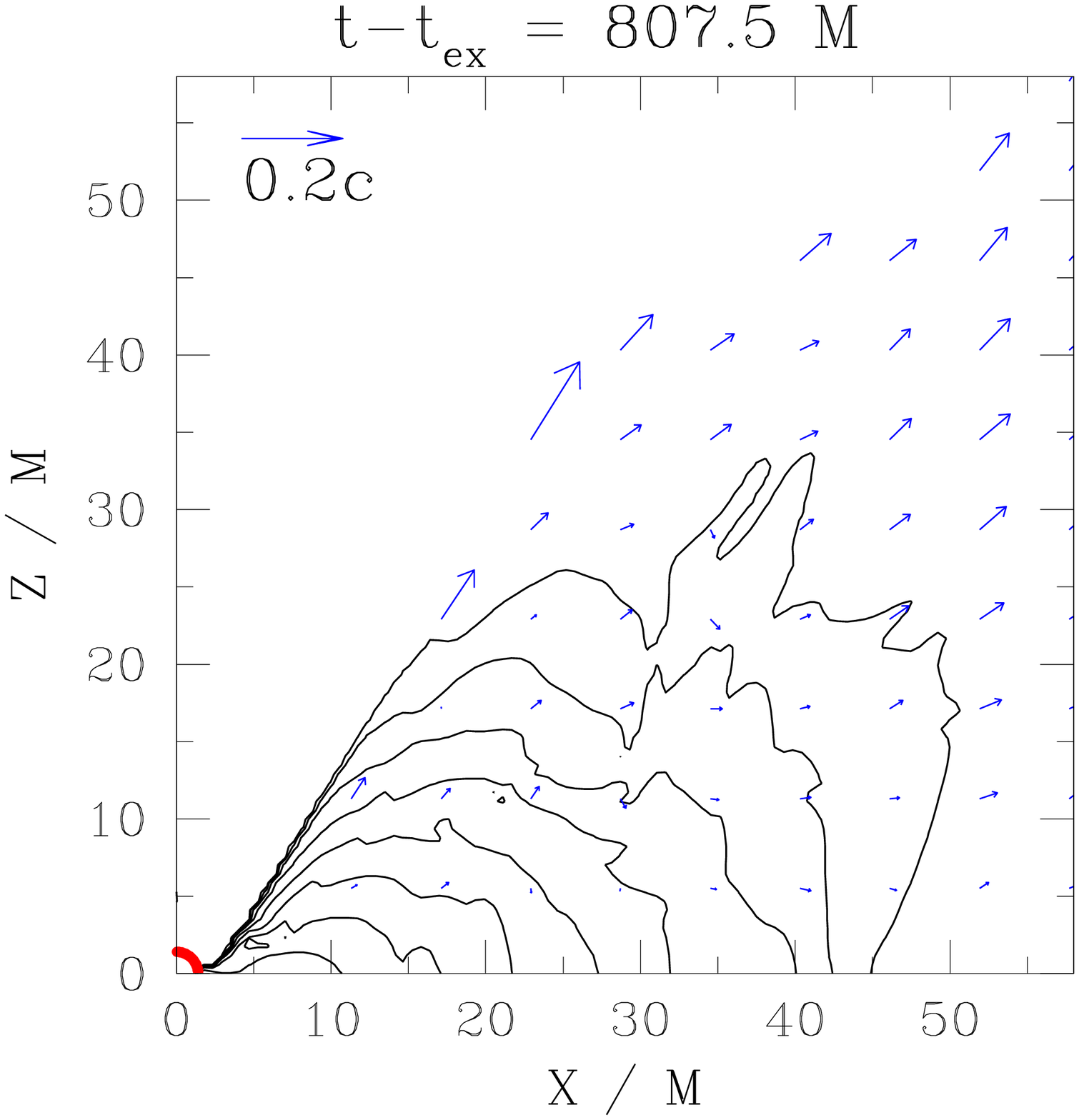}
\epsfxsize=2.15in
\leavevmode
\epsffile{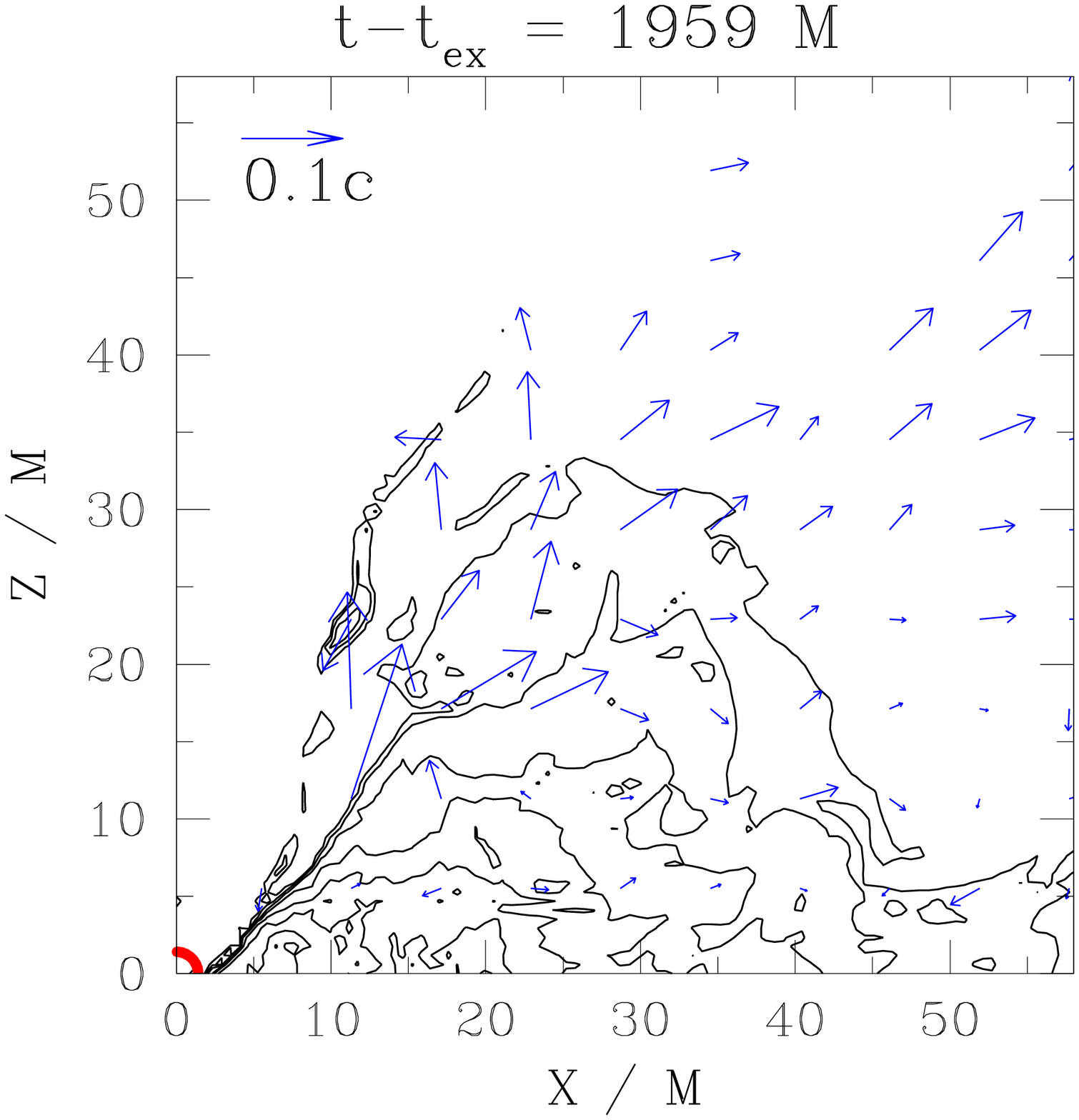}\\
\epsfxsize=2.15in
\leavevmode
\epsffile{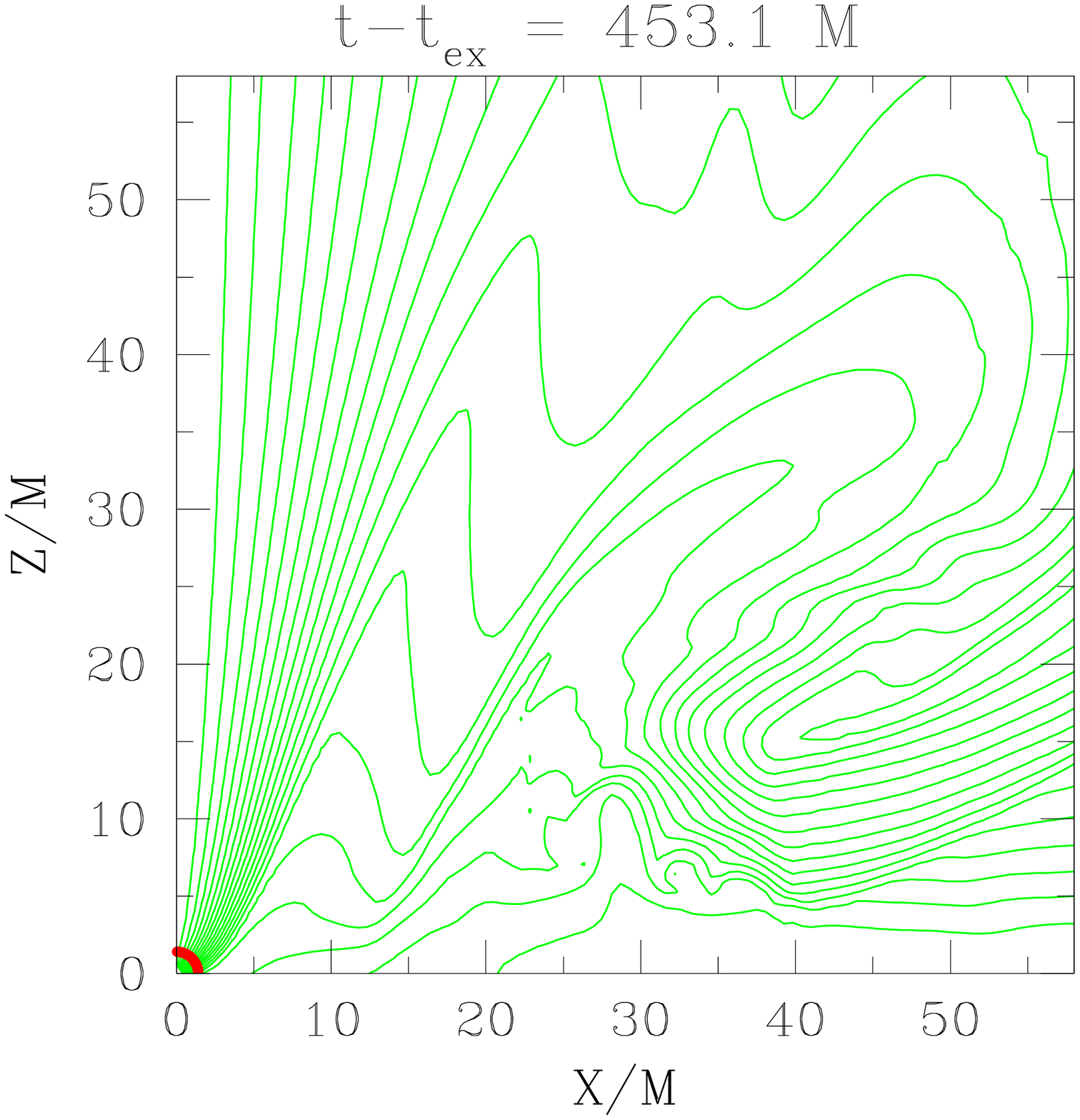}
\epsfxsize=2.15in
\leavevmode
\epsffile{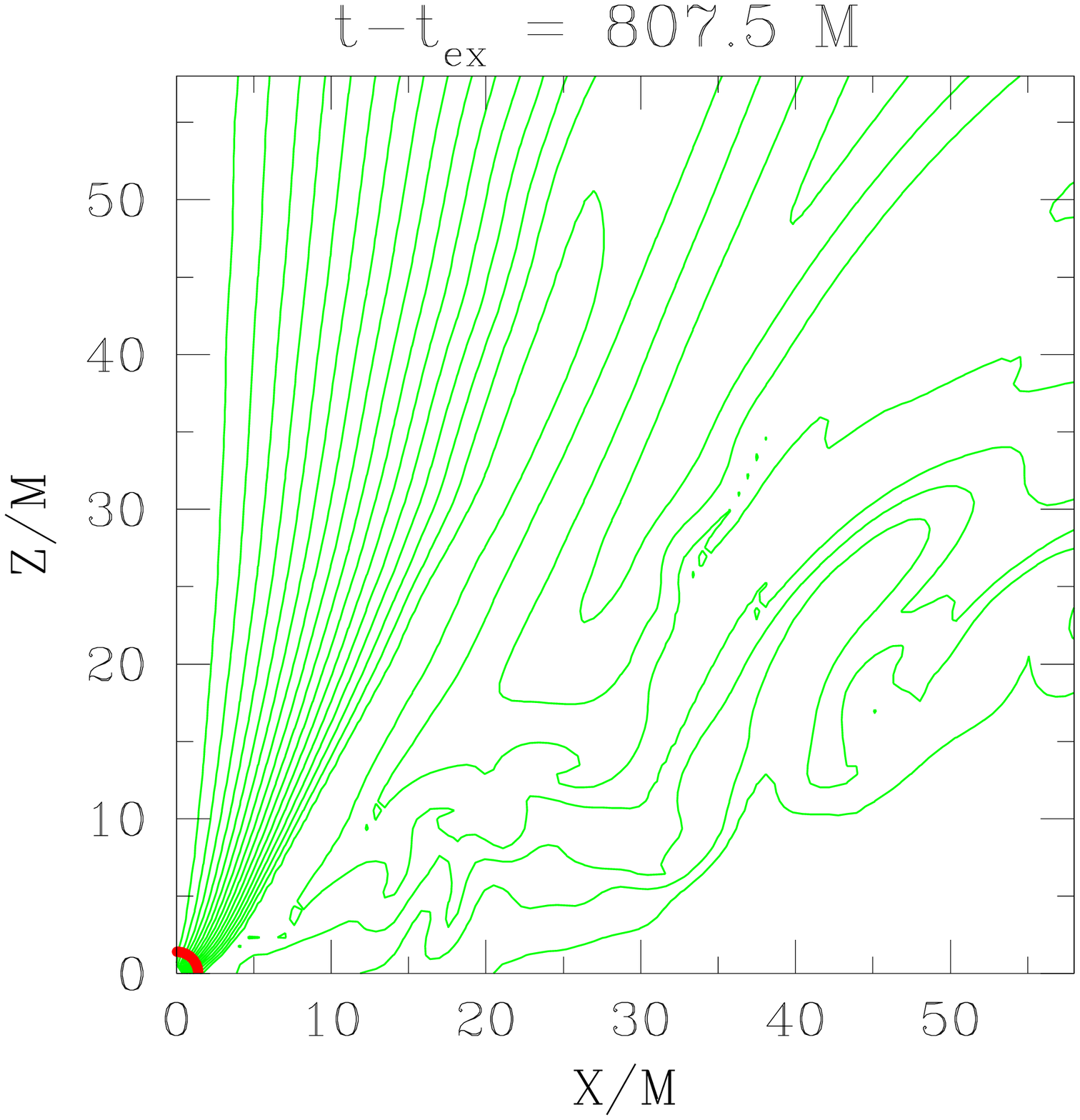}
\epsfxsize=2.15in
\leavevmode
\epsffile{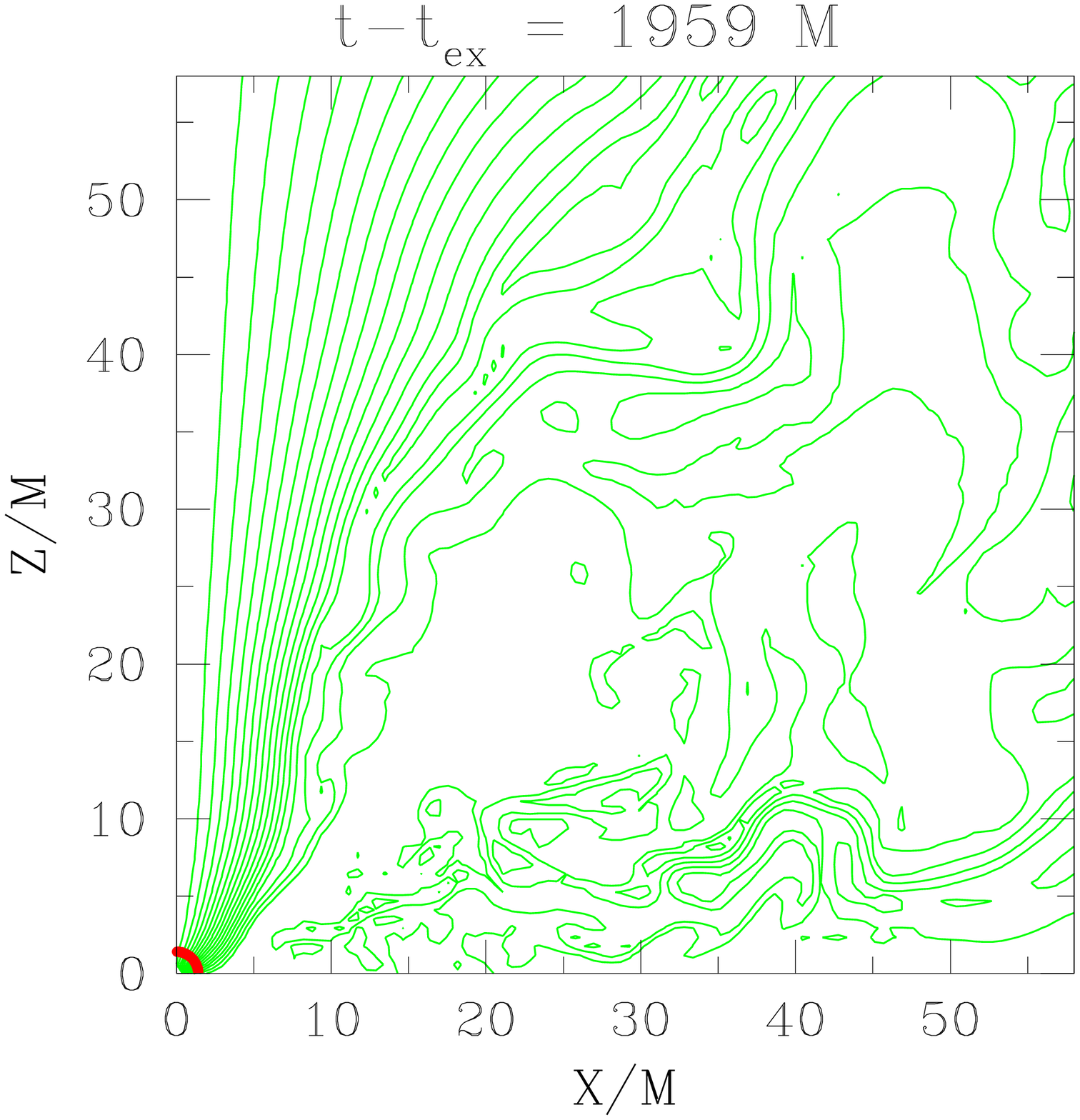}
\caption{Snapshots of density contour curves and velocity vectors 
(first and third rows), and poloidal magnetic field lines (second and 
fourth rows) in the post-excision evolution of model S1. The 
contours are drawn for
$\rho_0 = 100 \rho_c(0) 10^{-0.3j}~(j=0$--10). The thick (red)
line near the lower left corner denotes the apparent horizon. 
The poloidal magnetic field lines are drawn for 
$A_{\varphi}= (j/20)A_{\varphi,{\rm max}}$
with $j=1$--19.}
\label{fig:exconS1}
\end{center}
\end{figure*}

\begin{figure*}
\vspace{-4mm}
\begin{center}
\epsfxsize=2.15in
\leavevmode
\epsffile{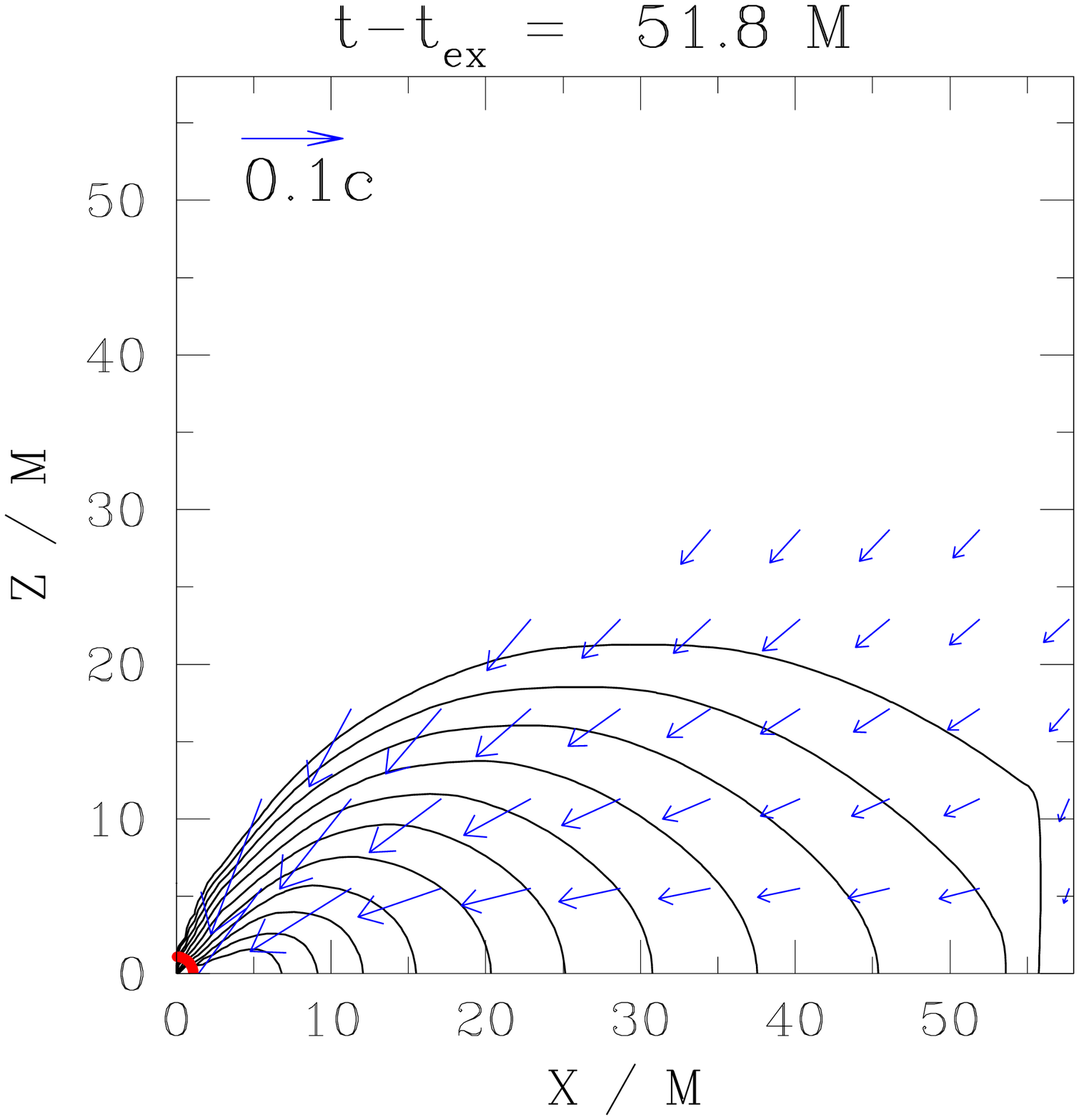}
\epsfxsize=2.15in
\leavevmode
\epsffile{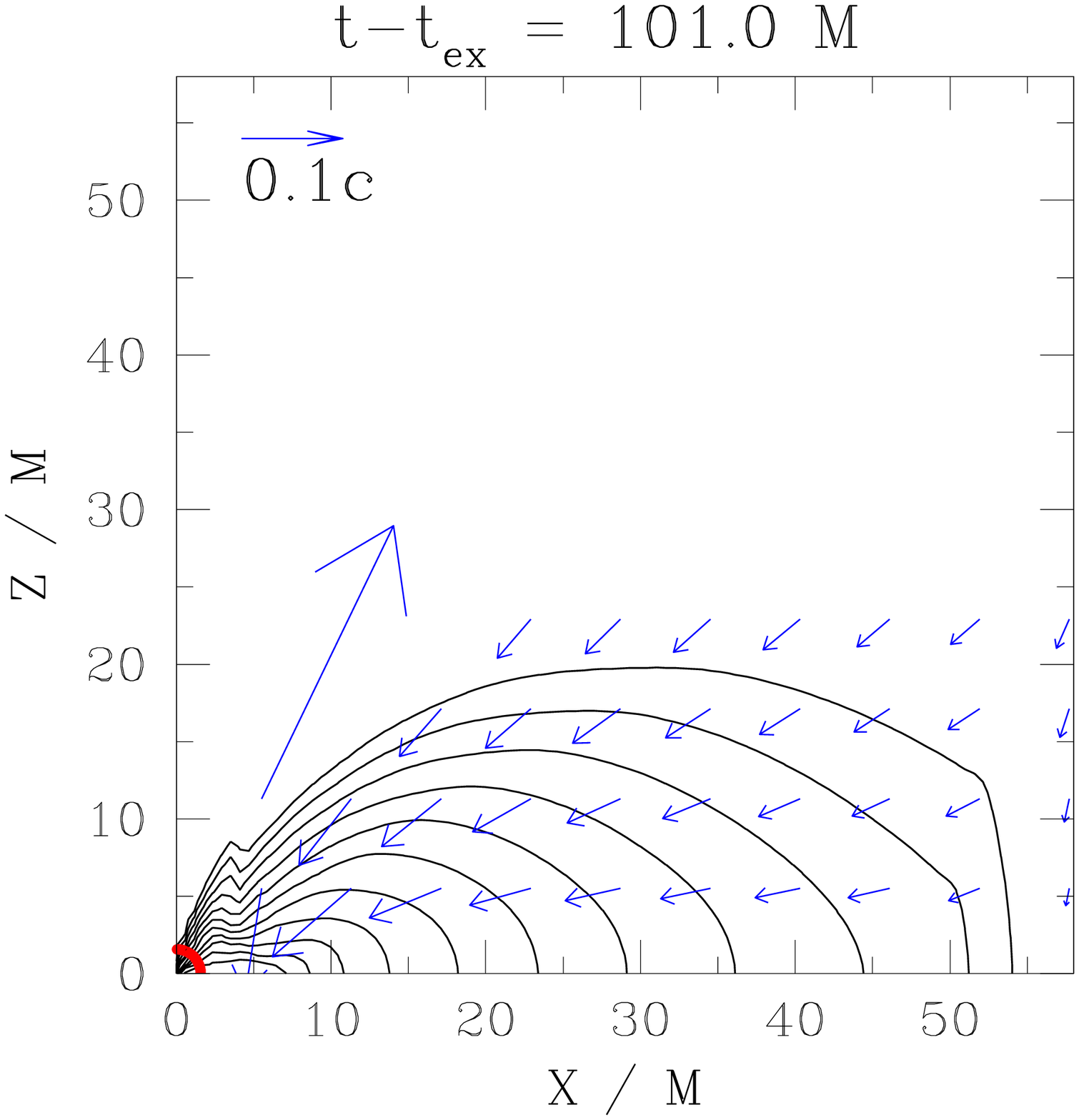}
\epsfxsize=2.15in
\leavevmode
\epsffile{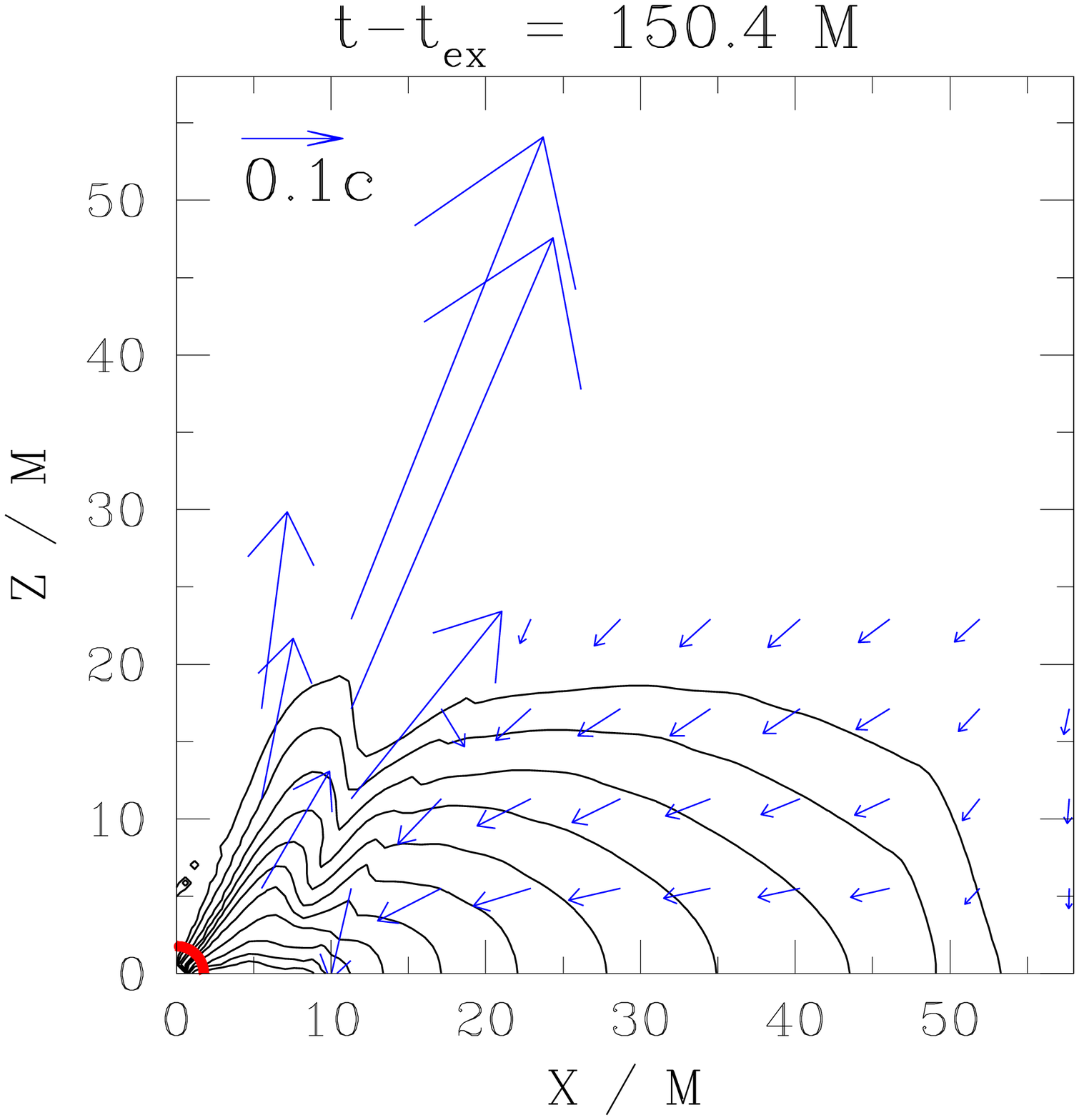}\\
\epsfxsize=2.15in
\leavevmode
\epsffile{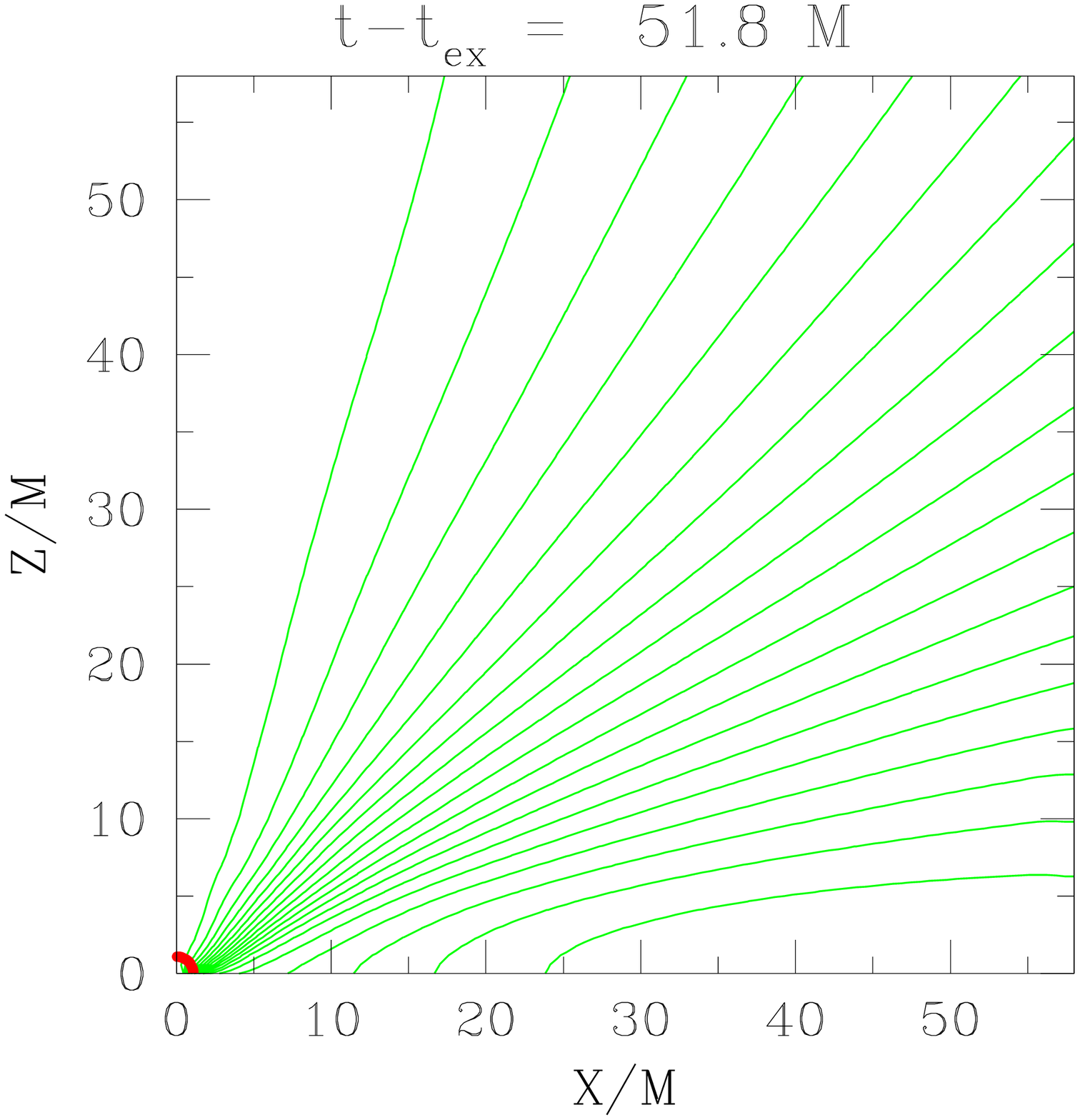}
\epsfxsize=2.15in
\leavevmode
\epsffile{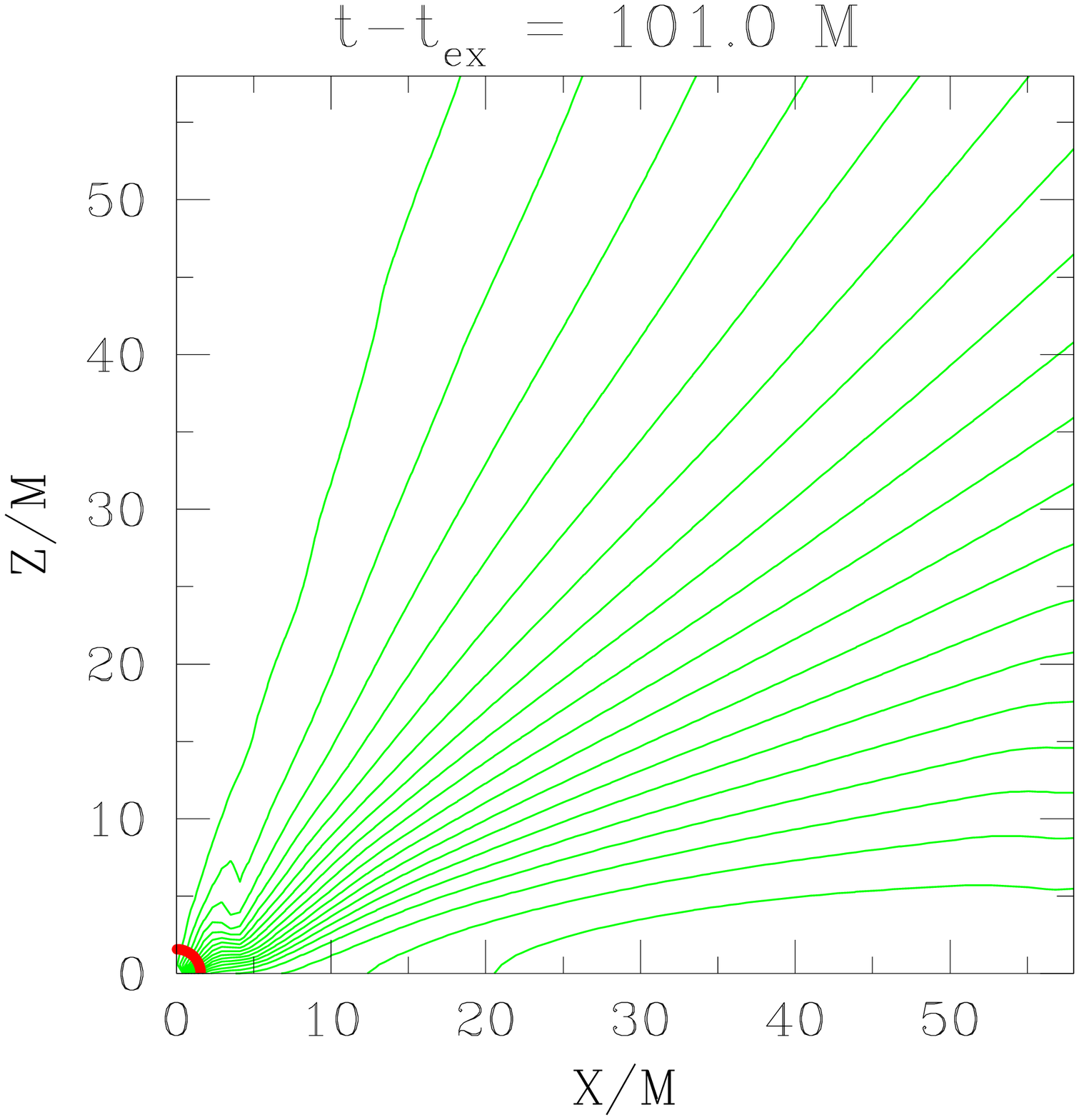}
\epsfxsize=2.15in
\leavevmode
\epsffile{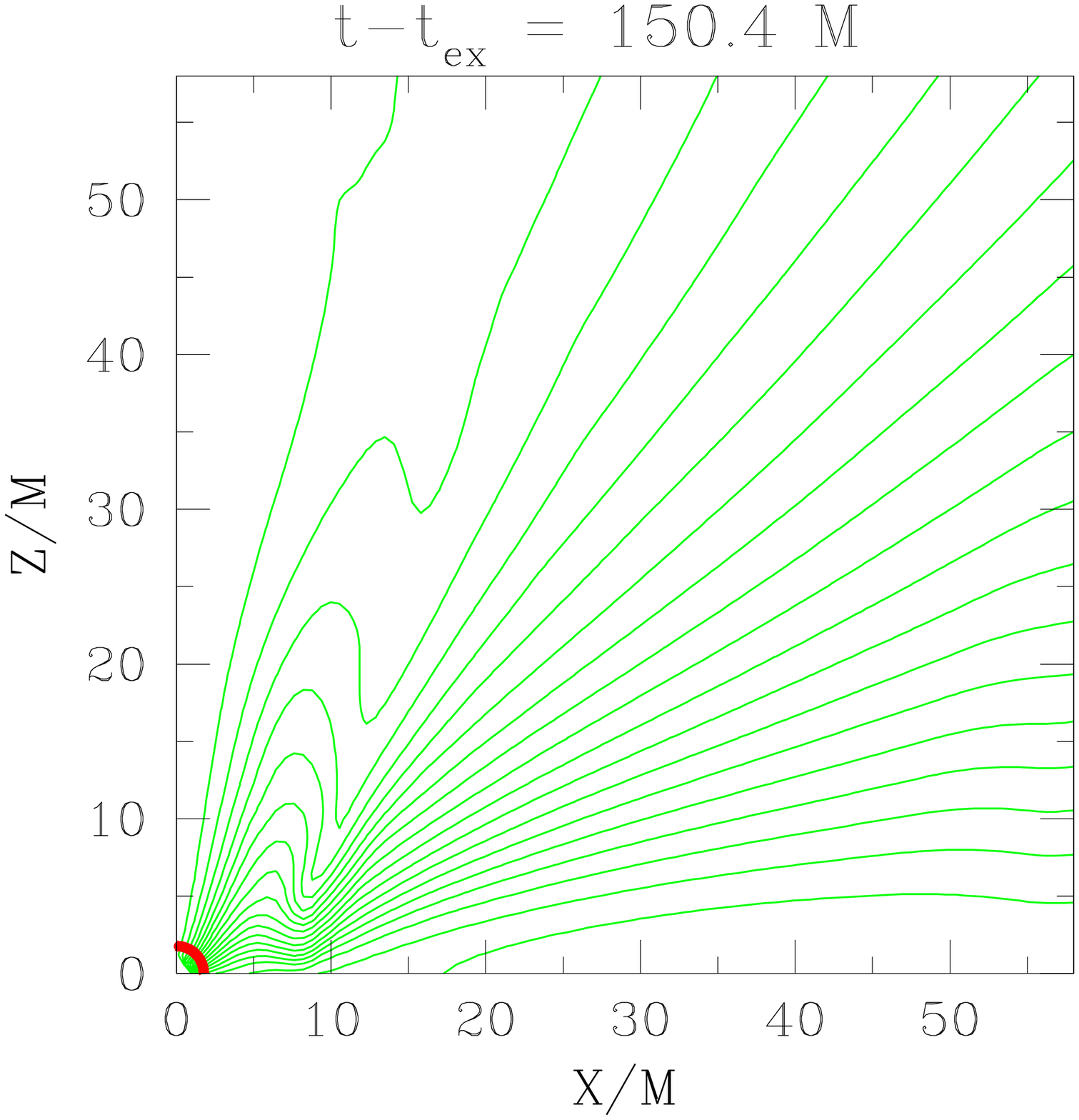}\\
\epsfxsize=2.15in
\leavevmode
\epsffile{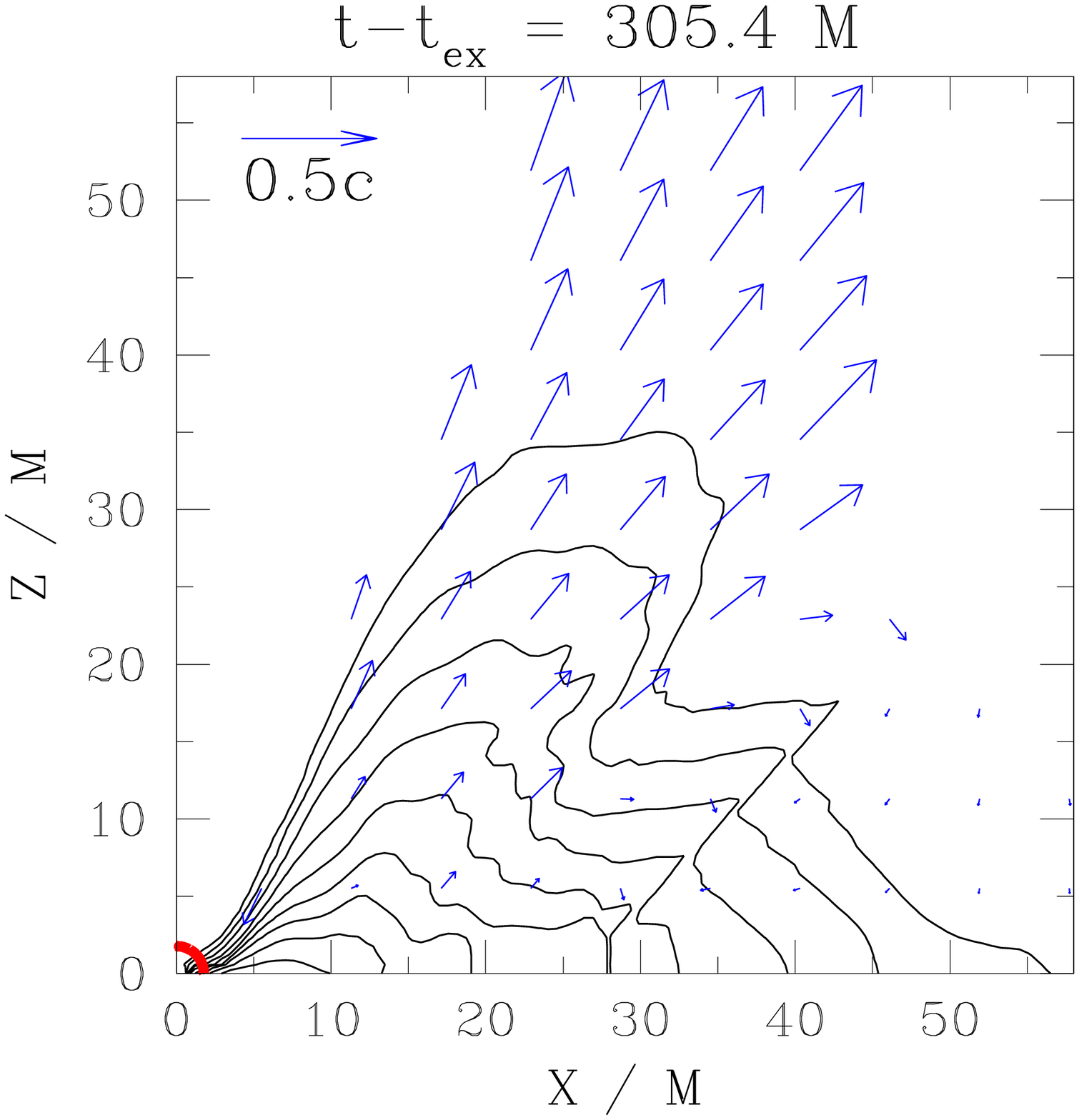}
\epsfxsize=2.15in
\leavevmode
\epsffile{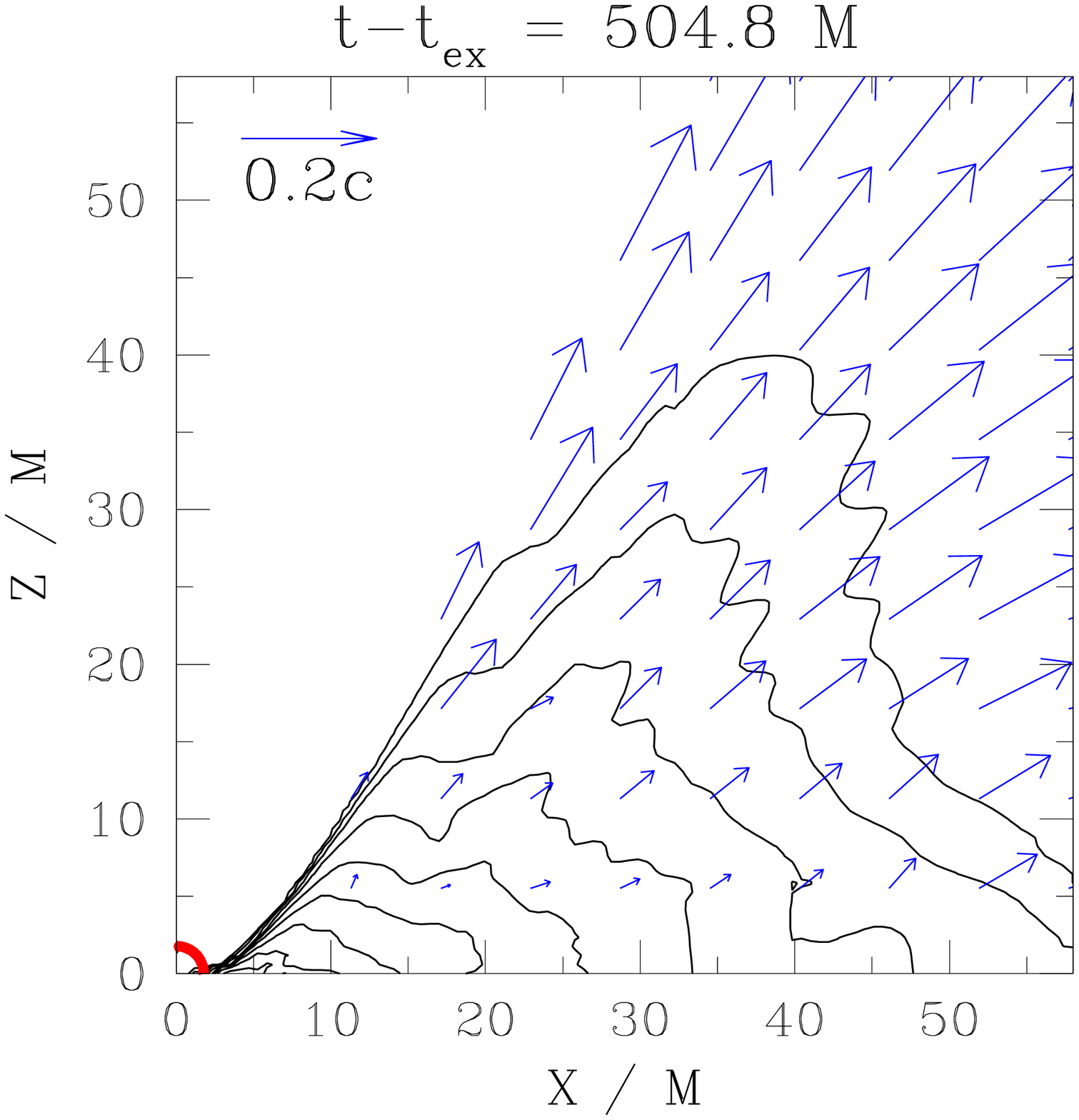}
\epsfxsize=2.15in
\leavevmode
\epsffile{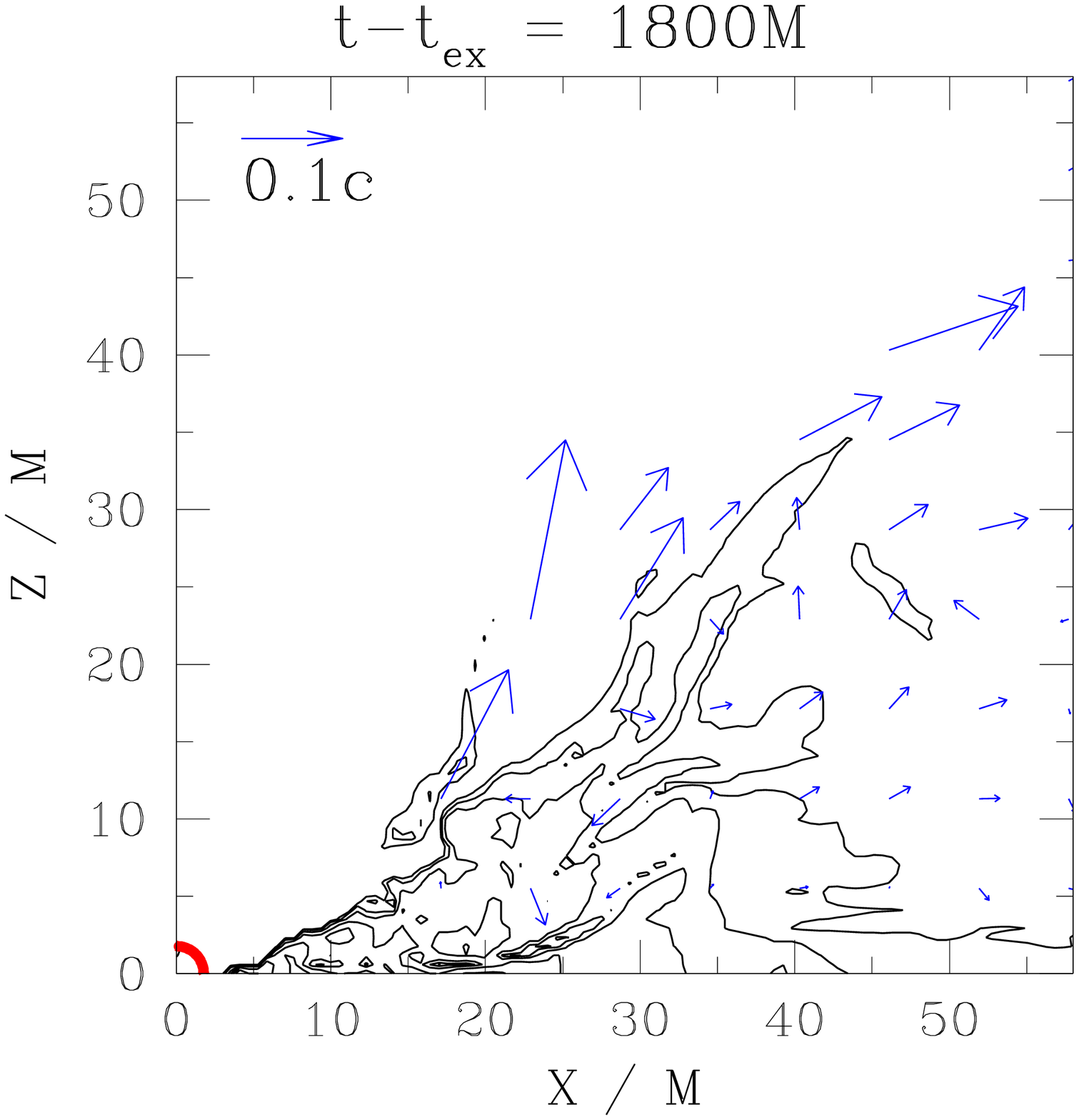}\\
\epsfxsize=2.15in
\leavevmode
\epsffile{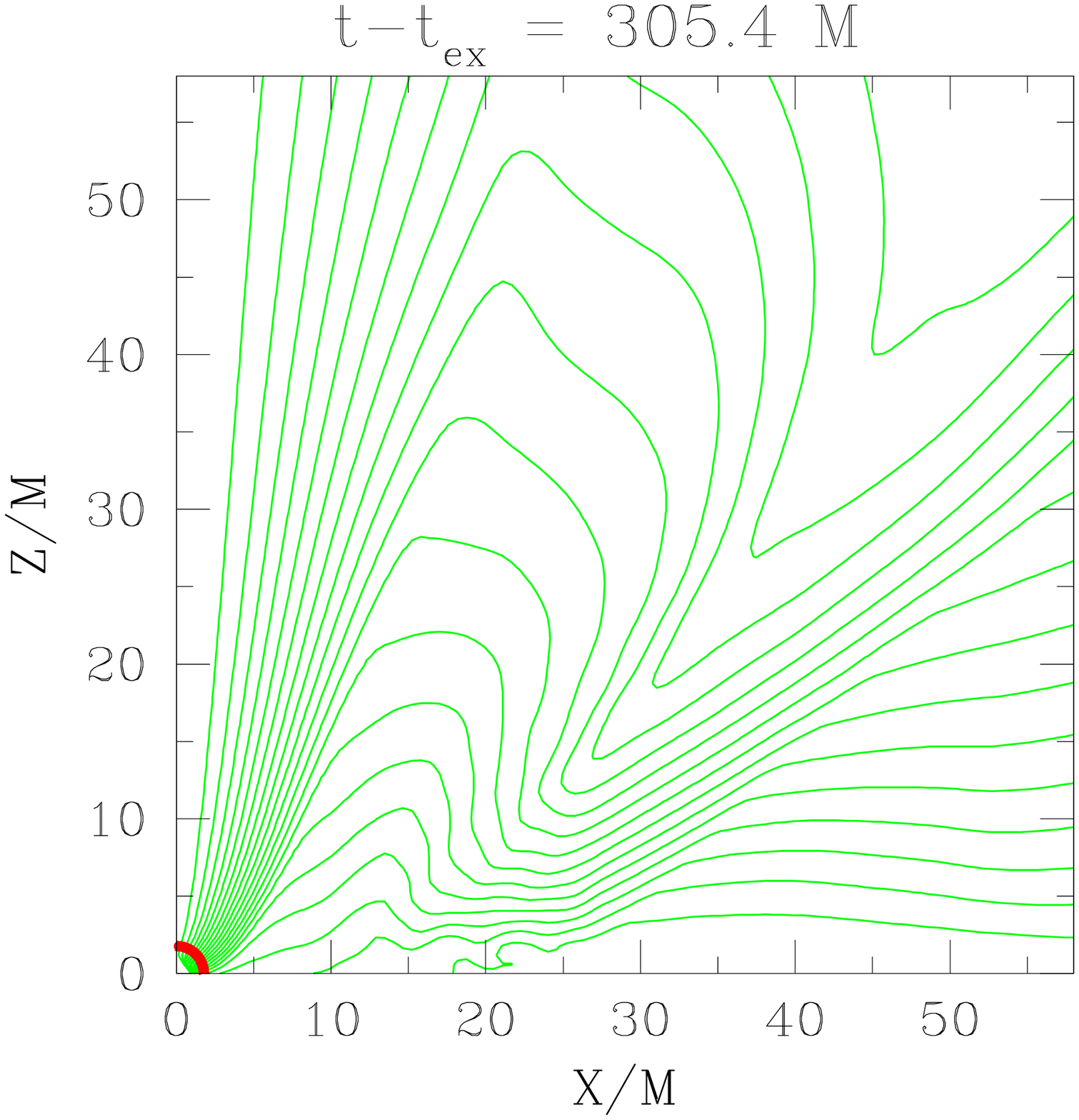}
\epsfxsize=2.15in
\leavevmode
\epsffile{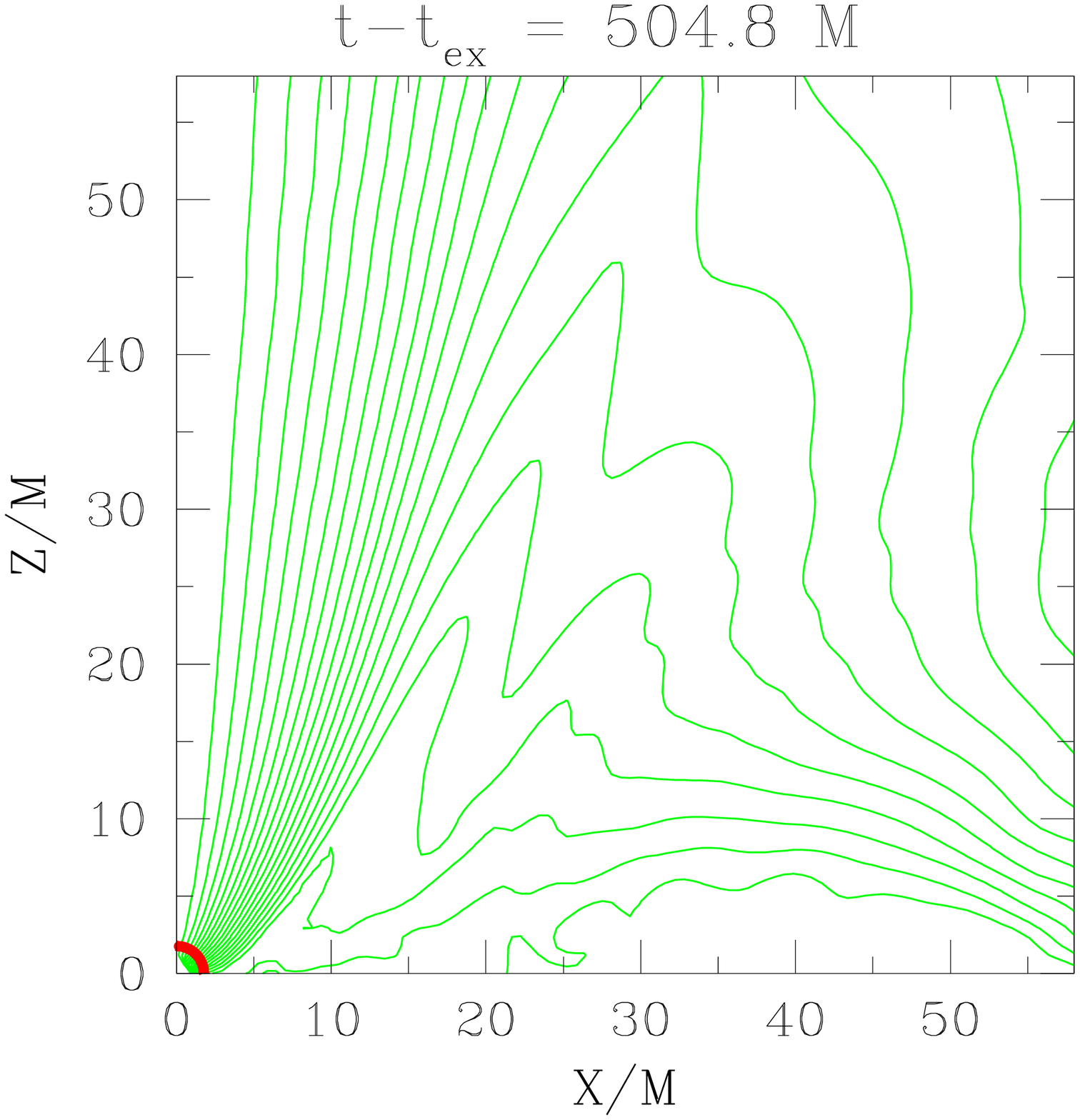}
\epsfxsize=2.15in
\leavevmode
\epsffile{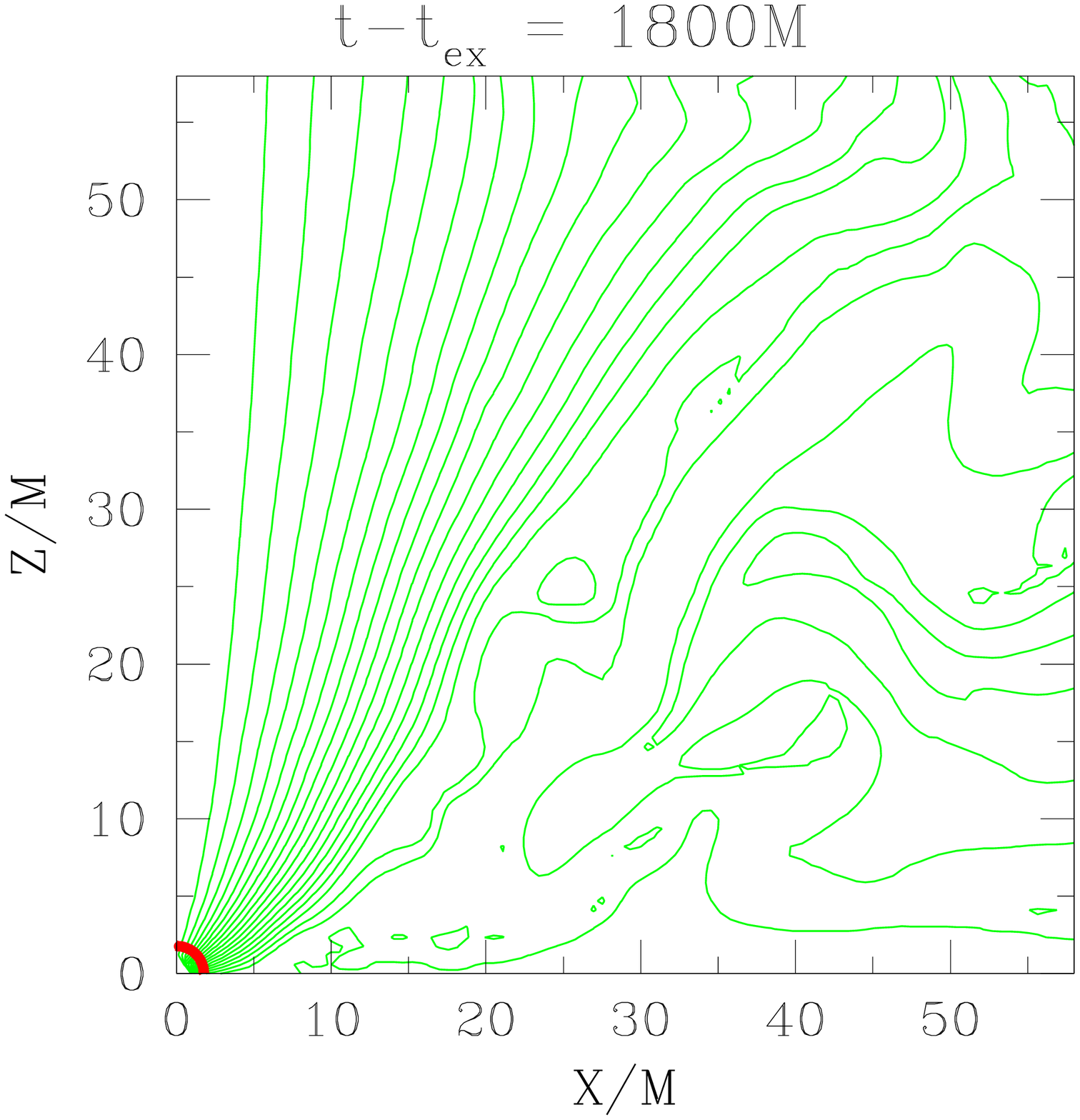}
\caption{Same as Fig.~\ref{fig:exconS1} but for model S2.}
\label{fig:exconS2}
\end{center}
\end{figure*}

To continue the evolution, we excise a spherical region
inside the apparent horizon. We start the excision evolution 
a few $\Delta t \sim M$ after the apparent horizon forms.
We are able to follow the evolution reliably
for another $\sim 200M$, after which the Hamiltonian and momentum 
constraints increase substantially. This eventual breakdown 
is probably because 
the metric inside the horizon, which is not computed accurately, 
slowly leaks out to the 
region outside due to superluminal gauge modes. We are currently 
investigating other gauge conditions, as well as other techniques 
to overcome the numerical difficulty.
As the collapse proceeds, the mass 
and angular momentum of the central black hole increase before 
settling down to quasi-stationary values.
Figures~\ref{fig:hor_S0}--\ref{fig:hor_S2} show the 
post-excision evolution of the black 
hole's irreducible mass $M_{\rm irr}$, mass $M_h$, spin parameter 
$J_h/M_h^2$, and the rest mass of the material outside the apparent 
horizon $M_{\rm disk}$, for the three models. 
We see that after $\Delta t\sim 150M$, 
the black hole settles down to a quasi-stationary state, with 
$M_h \approx 0.95M$ and $J_h/M_h^2 \approx 0.7$ for all the three models. 
This result agrees roughly with our earlier 
simulations and analytic estimates for unmagnetized collapse 
($M_h \approx 0.9M$ and $J_h/M_h^2 \approx 0.75$) in~\cite{ss02,ss02b,s04}. 
The remaining material, having too much angular momentum, forms a 
torus surrounding the black hole (see 
Figs.~\ref{fig:exconS0}--\ref{fig:exconS2}). Even though the central 
black hole has settled down after $\sim 150M$, the torus continues to 
evolve as material from the outer layers gradually reaches the central region. 
The dynamical timescale at radius $r$ is 
$t_{\rm dyn} \approx 2\pi \sqrt{r^3/M}$. 
Hence $t_{\rm dyn}\approx 1000M$ at $r=30M$, and $t_{\rm dyn}\approx 2000M$ 
at $r=50M$. Since the torus extends beyond $50M$, we need to follow 
the evolution for at least $2000M$.
To study the subsequent evolution, we adopt the Cowling 
approximation by freezing the metric at $t-t_{\rm ex}\sim 150M$, where 
$t_{\rm ex}$ is the time when excision starts.
This is a fairly good approximation since 
the material outside the horizon contributes only $\sim 5\%$ of the total 
mass and so the metric is dominated by the central black hole, 
which has settled down. We have compared the results of our 
Cowling (stationary metric) and non-Cowling (dynamic metric) runs during 
the transition interval  
$150M \lesssim t-t_{\rm ex} \lesssim 200M$ and find good agreement.

\begin{figure}
\vspace{-4mm}
\begin{center}
\epsfxsize=3.2in
\leavevmode
\epsffile{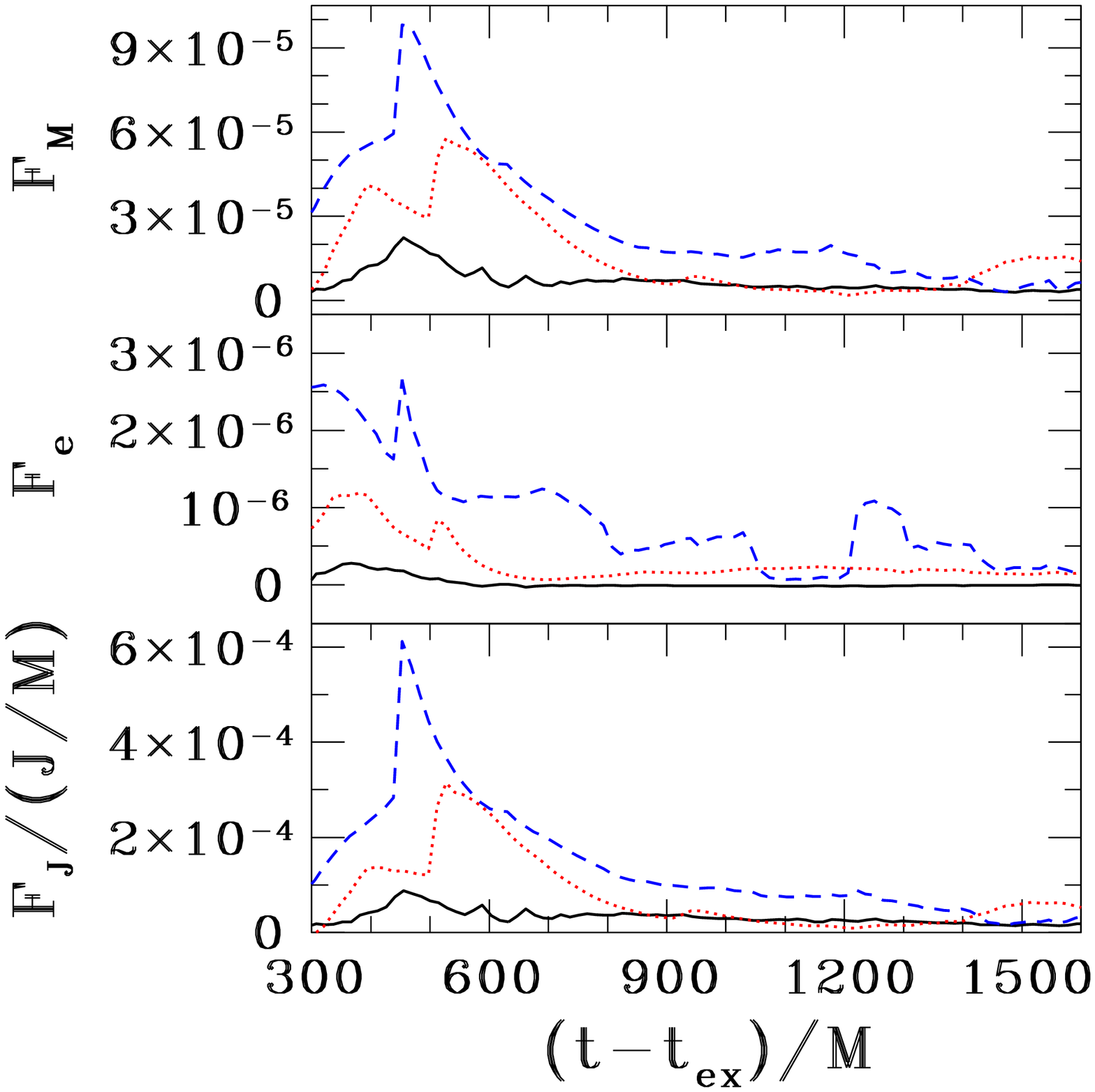}
\vspace{-2mm}
\caption{Rest-mass flux $F_M$, energy flux $F_e$ and angular momentum 
flux $F_J$ through 
a spherical surface of radius $50M$ for models S0 (black solid lines), 
S1 (red dotted lines) and S2 (blue dashed lines).} 
\label{fig:flux}
\end{center}
\end{figure}

Figure~\ref{fig:exconS0} shows snapshots of 
density contours and velocity fields for model S0 in the 
post-excision evolution. 
We terminate the simulation at $t-t_{\rm ex}=2200M$, 
where most of the dynamical processes have ended.
We find that an outflow develops at 
$t-t_{\rm ex} \sim 120M$ near the horizon and becomes prominent 
at $t-t_{\rm ex} \approx 170M$. The outflow is due to the fact 
that material from the outer layers 
arrives into the inner region 
with a substantial amount of angular
momentum. When it reaches the inner region,
the centrifugal barrier prevents it from falling into the black hole. 
The fluid particles move with ``zoom-whirl"-like 
trajectories~\cite{zoom-whirl} and accumulate 
near the black hole. As more fluid particles arrive and smash into  
interior layers, the fluid heats up and forms a shock, which propagates outward 
and creates an outflow along the surface of the torus. While this 
outflow is still expanding, we find 
that a secondary, weaker outflow forms at $t-t_{\rm ex} \sim 160M$, 
which can be seen in the second and third plots in Fig.~\ref{fig:exconS0}. 
A few more episodes of smaller outflow develop as more material from the 
outer layers 
arrive. However, the process damps by $t \gtrsim 250M$ by which time
most of the material has reached the central region and the residual 
infalling fluid 
does not have enough momentum to push on the torus and generate 
further outflow. To determine if the outflowing material is unbound, 
we calculate the quantity $-u_t$. As discussed in Sec.~\ref{sec:formalism}, 
any unbound fluid particle moving in a low density region (in which 
pressure and electromagnetic forces are negligible) has  
$-u_t >1$. We find that the outflow material is indeed unbound, but the 
total rest mass of the unbound fluid is only $10^{-3}M$. 
The outflow reaches the outer boundary of our grid 
($r \approx 60M$) after $t-t_{\rm ex} \gtrsim 500M$. 
Most of the unbound material leaves the grid after 
$t-t_{\rm ex} \gtrsim 700M$. During this same time, the infalling material 
in the outer 
region of the torus close to the equatorial plane also rebounds outward 
because of the centrifugal barrier. We have checked that this outward 
moving fluid remains bound ($-u_t <1$), but about $0.02M$ of rest mass 
leaves the grid by the time we terminate our simulation at 
$t-t_{\rm ex} = 2200M$. This outward moving fluid has too much angular 
momentum to be able to remain in the inner region. The torus in the 
inner region with 
$r \lesssim 30M$ settles down to quasi-equilibrium by $t \gtrsim 500M$.

Figures~\ref{fig:exconS1} and \ref{fig:exconS2} show snapshots of
density contours, velocity fields and poloidal magnetic field lines 
for models S1 and S2 in the post-excision evolution.
We find similar outflow as in the case S0, but the outflow in S2 
develops at time $t-t_{\rm ex} \approx 75M$, much earlier than that in S0. 
Unlike S0, the outflow in S1 and S2 is generated continuously rather than
intermittently. The outflow is stronger than S0 and about 
$4\times 10^{-3}M$ of the rest mass becomes unbound for S1 
and $8.7\times 10^{-3}M$ for S2, much larger than the case of S0. 
In the presence of magnetic fields, the outflow carries the 
frozen magnetic field and travels outward along the torus's surface. 
This causes the field lines near the boundary 
of the outflow and torus to bend (see 
Figs.~\ref{fig:exconS1} and \ref{fig:exconS2}). This bending amplifies the 
magnetic field in that region and hence the outflow is intensified by 
the extra magnetic pressure. A magnetic shock is also generated in 
the region, which leads to turbulence in the torus. 
The bending is more significant in S1 than in S2. This is because in S2, 
the magnetic field is strong enough to quickly counteract the bending 
and drives more fluid outward.
Figure~\ref{fig:flux} shows the rest-mass flux $F_M$, energy 
flux $F_e$ and angular momentum flux $F_J$ through a spherical surface of 
radius $50M$ for the three models. We see that the outflow 
is significantly stronger in the presence of magnetic fields. 
Figure~\ref{fig:flux} also indicates that a sustained flux  
is present in the time period $900M$--$1200M$ for model S2. We find that 
this flux is not due to outflow generated near the black hole, but due to  
a wind that arises in the middle of the torus. We find that in the wind 
the fluid moves along the magnetic field lines. The inclination angle 
between the field lines and $z$-axis is between $20^{\circ}$ and 
$40^{\circ}$. This suggests that the wind is driven by the 
magneto-centrifugal 
mechanism~\cite{magnetocen}. The outflows in models S1 and S2 
cause the field lines to collimate along the rotation 
axis of the black hole. For model S2, the outflow and the subsequent 
wind carry away a substantial amount of magnetic energy from 
the torus. At $t-t_{\rm ex}>1500M$, the wind subsides and 
the interior of the remaining torus has a weak 
magnetic field. The outflow and wind in model S2 are so strong that they perturb 
the equilibrium of the inner torus and causes it to oscillate radially.
As in the case of S0, there is bound fluid 
moving out of the grid as a result of centrifugal bounce in models S1 and 
S2. By the end of the simulation ($t-t_{\rm ex}=2000M$), only 0.04M of the rest 
mass remains in the inner torus in model S1 and $0.02M$ remains in 
model S2.

\begin{figure}
\vspace{-4mm}
\begin{center}
\epsfxsize=3.2in
\leavevmode
\epsffile{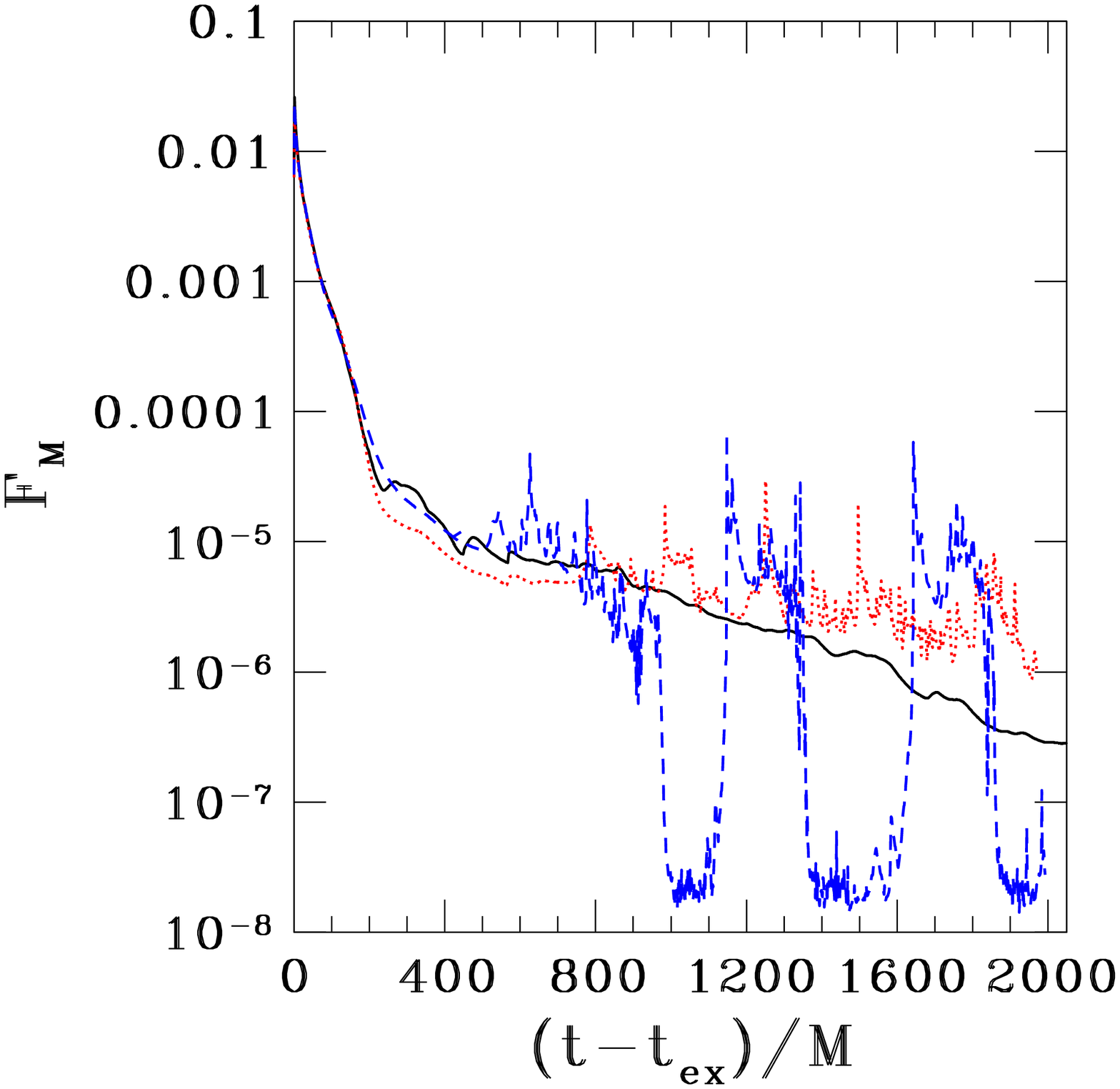}
\vspace{-2mm}
\caption{Rest-mass flux $F_M$ through the apparent horizon for models
S0 (black solid lines), S1 (red dotted lines) and S2 (blue dashed lines).}
\label{fig:flux_hor}
\end{center}
\end{figure}

Figure~\ref{fig:flux_hor} shows the rest-mass flux through the apparent 
horizon for the three models. For model S0, the inward flux decreases with time 
as the torus settles down to dynamical equilibrium. Without 
magnetic fields or viscosity, there is no dissipation to drive further 
accretion.  For model 
S1, we see that at late time ($t-t_{\rm ex} \gtrsim 800M$) material from 
the torus accretes into the central black hole in a stochastic manner. 
Stochastic accretion is often seen in simulations of magnetized 
accretion disks around stationary black holes 
(see e.g.~\cite{dvhk03,mhd-disk}).
This suggests that the accretion is due to magnetic-induced 
turbulence in the torus. 
The turbulence is generated initially by the magnetic shock as 
a result of the outflow, and is then sustained by the MRI. To verify 
that we are able to resolve the MRI, we compute the wavelength of the 
fastest growing MRI mode $\lambda_{\rm MRI}$ using Eq.~(\ref{eq:lmri}). 
We find that $\lambda_{\rm MRI}/\Delta >$15 in some region near the 
equatorial plane, where $\Delta$ is our grid spacing. This suggests 
that the MRI can be resolved in our simulation. For model S2, 
the radial oscillation of the inner torus causes episodic accretion 
into the central black hole. 
When the torus swings away from the black hole, no accretion 
occurs. Accretion resumes when the torus swings towards the black hole. 
This explains the episodic 
mass accretion pattern seen in Fig.~\ref{fig:flux_hor}. 
The small accretion rate in the figure is due to accretion from the 
atmosphere. 

When the magnetic field strength is much smaller than that in S1, 
we expect the dynamics of the fluid evolution to be very similar to 
S0 initially. As in the case of S1 
and S2, the outflow is expected to collimate the magnetic field lines 
and generate magnetic shocks which may create turbulence in the torus. 
Turbulence can also be generated by the MRI, 
which operates on the orbital timescale of the torus independent of the field 
strength. We should then expect 
to see the stochastic accretion similar to the case in S1. Both a collimated 
magnetic field and a massive, accretion torus surrounding a central 
black hole are essential ingredients for launching 
ultrarelativistic jets~\cite{jet-launching}. The black hole-torus system 
observed in our simulations provides a viable central engine for 
long-soft GRBs.

The radial oscillation observed in model S2 gives rise 
to gravitational radiation. The oscillation period of $\sim 500M$ 
corresponds to the gravitational wave frequency  
$f\sim 1/[500M (1+z)] \sim 0.04 (10^4M_{\odot}/M)/(1+z)$~Hz at 
redshift $z$. For a SMS with 
$M \gtrsim 10^4 M_{\odot}$, the signal is in the LISA frequency 
band. To estimate its amplitude, we apply the quadrupole 
formula $h\approx 2 \ddot{\mbox{\bI\, }}/D_L$, 
where $D_L$ is the source's luminosity 
distance, \bI \  is the tracefree quadrupole 
moment and $\ddot{\mbox{\bI\, }}\sim \omega^2 M_{\rm disk}\Delta R_c^2 
\sim 2\omega^2 M_{\rm disk} A R_c$. Here $R_c \sim 30M$ is 
the characteristic radius of the torus and $A\sim 5M$ is the 
amplitude of the oscillation. Setting $M_{\rm disk} \sim 0.04M$ and 
$\omega = 2\pi f$, we obtain 
\beq
  h \sim 4\times 10^{-23} \left( \frac{M}{10^4 M_{\odot}} \right) 
\left( \frac{48 {\rm Gpc}}{D_L} \right) \ ,
\eeq
where $D_L=48$Gpc corresponds to redshift $z=5$ in the concordance $\Lambda$CDM 
cosmology model with $H_0=71{\rm km}~{\rm s}^{-1}~{\rm Mpc}^{-1}$, 
$\Omega_M=0.27$ and $\Omega_{\Lambda}=0.73$~\cite{wmap}. 
We note that if the  
signal can be tracked for $n$ cycles, where $n$ is expected to be a few, 
the effective wave strength 
will be increased by a factor of $\sqrt{n}$.
Such a gravitational 
wave signal may be detectable by LISA [see~\cite{lisa-sen} for LISA's 
sensitivity curve].

Our simulations are adiabatic and do not take into account the heat loss due to 
neutrino cooling. To determine if this effect can be neglected during 
the phase in which the torus forms and evolves around the black hole, 
we estimate the neutrino cooling timescale. We first compute the 
temperature in the disk from the specific thermal energy density
$\epsilon_{\rm th} = \epsilon - \epsilon_{\rm cold}$, where 
$\epsilon_{\rm cold} = 3 \rho_0^{1/3}$ for our adopted $\Gamma=4/3$ 
EOS. We find from our data that the typical values of $\rho_0$ and 
$\epsilon_{\rm th}$ in the disk at late times are 
\beqn
  \rho_0 &\approx & 6000 \left( \frac{M}{10^4 M_{\odot}}\right)^{-2}
{\rm g}~{\rm cm}^{-3} \ , \label{rho0disk} \\ 
  \epsilon_{\rm th}/c^2 &\approx & 0.005 \ ,
\eeqn
where we have restored the speed of light $c$ in the above equation. 
To calculate the temperature $T$, we adopt the expression of 
$\epsilon_{\rm th}(\rho_0,T)$ in~\cite{pwf99}:
\beq
  \frac{\epsilon_{\rm th}}{c^2} = \frac{3kT}{2m_p c^2}\left( 
\frac{1+3X_{\rm nuc}}{4}\right) + f \frac{aT^4}{\rho_0 c^2} \ ,
\label{eq:thEOS}
\eeq
where $k$ is the Boltzmann constant, $a$ is radiation constant, 
$m_p$ is proton mass, and $X_{\rm nuc}$ is the mass fraction 
of free nucleons approximately given by~\cite{qw96}  
$X_{\rm nuc} \approx \min[ 34.8 \rho_{10}^{-3/4} T_{11}^{9/8} 
\exp(-0.61/T_{11}),1]$. Here $\rho_{10}=\rho_0/10^{10}{\rm g}~{\rm cm}^{-3}$ 
and $T_{11}=T/10^{11}K$. The first term in Eq.~(\ref{eq:thEOS}) is the 
specific thermal energy density of an ideal gas, and the second term is the 
contribution from thermal radiation. 
The quantity $f$ is a temperature-dependent numerical factor depending 
on the species of ultrarelativistic particles that contribute 
to thermal radiation.
When $T \gg 2m_ec^2/k \sim 10^{10}K$, photons, ultra-relativistic 
electrons and positrons are present (assuming thermal neutrinos 
are negligible) and $f=11/4$. On the other hand, 
when $T\ll 10^{10}K$, only photons contribute to thermal radiation and $f=1$. 
Combining Eqs.~(\ref{rho0disk})--(\ref{eq:thEOS}), we obtain 
\beq
  0.0345(1+3X_{\rm nuc})T_9 + 1.40 f 
\left( \frac{M}{10^4M_{\odot}}\right)^2 T_9^4 \approx 5 \ , 
\eeq
where $T_9=T/10^9K$. For $M=10^4M_{\odot}$, we find $T\approx 1.4\times 10^9K$ 
and, not surprisingly, $\epsilon_{\rm th}$ is dominated by thermal photon 
radiation. At this density and temperature, the torus is optically thin to 
neutrinos. The cooling rate $Q_{\nu}$ is dominated by the pair neutrino 
process and the value is 
$Q_{\nu}\approx 10^{16}{\rm erg}~{\rm cm}^{-3}~{\rm s}^{-1}$~\cite{itoh89}. 
The neutrino cooling timescale is 
$\tau_{\nu} \sim \rho_0 \epsilon_{\rm th}/Q_{\nu} \sim 3\times 10^6{\rm s} 
\sim 5\times 10^7M$, which is much longer than the timescale in our 
simulations ($\sim 2000M$). 
Even for $M=100M_{\odot}$, we find $\tau_{\nu} \sim 90{\rm s} \sim 2\times 
10^5M \gg 2000M$. The same conclusion (i.e.\ $\tau_{\nu} \gg 2000M$) holds 
for all $M \gtrsim 100M_{\odot}$. Hence neutrino cooling can be neglected 
in the torus evolution.

\section{Summary and conclusion}
\label{sec:conclusion}

In this paper, we study the magnetorotational collapse of 
very massive stars by performing full GRMHD simulations in
axisymmetry. We model the pre-collapse star by an $n=3$ 
polytrope uniformly rotating near the mass-shedding 
limit at the onset of radial collapse. We adopt an adiabatic 
$\Gamma = 4/3$ EOS for the fluid. We study three models, 
which we label S0, S1 and S2. The three models differ by the strength 
of the initial magnetic field (see Table~\ref{tab:models}). Model S0 
is unmagnetized (${\cal M}=0$), whereas the ratio of the initial magnetic to 
kinetic energies (${\cal M}/T$) are 1\% and 10\% for models S1 and S2, 
respectively. 

We find that these magnetic fields do not affect the 
initial collapse significantly. 
An apparent horizon forms at time $t \approx 29000M$. 
The black hole grows as the collapse proceeds, and settles down at
a time $\sim 150M$ after the formation of the apparent horizon. For all three 
models we study, we find that the mass $M_h$ and 
spin parameter $J_h/M_h^2$ of the black hole are approximately 
$0.95M$ and 0.7 respectively, 
where $M$ is the initial mass of the star. These values roughly agree with 
the semi-analytic estimates in~\cite{ss02,ss02b,s04}. 
The remaining material forms 
a torus around the central black hole. Although the central black 
hole has settled down to quasi-stationary equilibrium, 
the ambient torus continues to evolve as 
fluid from the outer layers of the star gradually reaches 
the central region. 
During this epoch, magnetic fields have substantial influence on the evolution 
of the torus. 
The infalling fluid particles have large angular 
momenta. They pile up near the black hole horizon, are heated by shocks and 
then get ejected along the surface of the torus, forming an unbound 
outflow. In the presence of magnetic fields, the outflow bends the 
magnetic field lines near the boundary of the outflow, which amplifies 
the field and induces magnetic shocks. The extra magnetic pressure 
makes the outflow stronger than in the unmagnetized case. The outflow 
also causes the magnetic fields to collimate along the black hole's 
rotation axis.
For model S0, when the outflow leaves the central region, the torus 
settles down to equilibrium. For model S1, MHD turbulence generated 
by magnetic shocks and MRI in the 
disk causes stochastic accretion of material into the 
black hole. For model S2, when the outflow leaves, strong magnetic 
fields in the torus create a magnetic wind, driving more material 
and magnetic field out of the torus. During this time, the torus 
acquires a quasiperiodic radial oscillation. The wind subsides as the magnetic 
field inside the torus decreases. The radial oscillations of the torus 
induce episodic accretion of material into the central black hole. The
oscillations also generate gravitational radiation, which might be 
detectable by LISA at redshift 
$z \sim 5$ if the mass of the star satisfies $M\gtrsim 10^4M_{\odot}$. 

If the initial magnetic field strength is smaller than that in model S1,
we expect the evolution to be similar to S1. In particular, the evolution 
in the collapse phase should remain unchanged. We also expect the outflow 
to collimate the magnetic field lines and generate magnetic shocks, which then 
leads to turbulence in the disk. Turbulence will be maintained as a result 
of the MRI. We thus expect stochastic accretion of the torus as in the 
case of S1.

In typical cases, the final stage of the magnetorotational 
collapse consists of a central 
black hole surrounded by a collimated magnetic field and a massive torus. 
These are the main ingredients for generating ultrarelativistic jets 
at large distance from the central source. The final system obtained in 
our simulations is thus capable 
of generating a long-soft GRB. In principle,
the collapse of a very massive star could result in the simultaneous detection
of gravitational waves and a GRB. The gravitational wave signal consists of
an initial burst signal due to collapse, a black-hole ring-down signal, and
a quasi-periodic signal due to the torus's oscillation if the magnetic
field is strong.

A few issues warrant further study. The first is the EOS. 
Our $\Gamma=4/3$ adiabatic EOS is 
a good approximation only for very massive stars.
But most of the observed long-soft GRBs are believed to
be triggered by the magnetorotational core collapse of smaller-mass Pop~I/II 
stars~\cite{mw99}.
The core mass of a Pop~I/II star 
is less than $2M_{\odot}$. A $\Gamma=4/3$ EOS 
describes the early phase of core collapse in such a star, when the pressure
is dominated by relativistic degenerate electrons. But 
the EOS stiffens when the core density exceeds nuclear density and this
happens before an apparent horizon forms. Also, a realistic 
EOS for this scenario must incorporate more detailed microphysics and 
neutrino transport.

A second issue concerns a search for a more 
robust singularity-avoiding algorithm once a black hole forms.
As mentioned in Sec.~\ref{sec:results}, 
we are only able to evolve the system for $\sim 200M$ after the black hole 
formation with our current excision technique. However, the evolution timescale 
of the torus is $> 2000M$. While this evolution could be
reliably tracked in the Cowling approximation, we are interested in
more general scenarios. We plan to explore this issue in two ways. 
The first will be to search for better lapse and shift conditions that can 
suppress troublesome superluminal gauge modes. The other will be
to identify a gauge that can drive the metric inside the horizon to a 
puncture-like solution, a technique which has been used with 
great success in binary black hole simulations~\cite{moving-punctures}. 
Simple experimentation 
with vacuum black holes and black holes immersed in hydrodynamic 
fluid suggest that there exist such gauge choices~\cite{fbest07}.

The third issue concerns our assumption of axisymmetry.
Nonaxisymmetric instabilities such as bar 
and/or one-armed spiral instabilities may develop during the 
collapse, which could affect the subsequent 
dynamics (but see~\cite{sbss02} for a treatment of
unmagnetized collapse in full $3+1$ post-Newtonian gravitation). 
Additionally, the MHD turbulence developed as a result of 
magnetic shocks and the MRI will be different. In particular, 
turbulence arises and persists 
more readily in $3+1$  due to the lack of symmetry. 
More specifically, according to
the axisymmetric anti-dynamo theorem~\cite{moffatt78}, sustained
growth of the magnetic field energy is not possible through axisymmetric
turbulence.
However, a full $3+1$ GRMHD simulation covering the required dynamic 
range for massive stellar collapse is computationally challenging 
and possibly beyond the resources currently available. 
This is because the torus extends to a large distance away from 
the central black hole, requiring vast dynamic range, 
and the dynamical timescale of the 
torus is very long. Though simulations in full $3+1$ dimensions will
eventually be necessary to capture the full behavior of the collapse, 
the $2+1$ results presented here likely provide a reasonable first
approximation.

\acknowledgments 

Numerical computations were performed at the National Center for 
Supercomputing Applications at the University of Illinois at 
Urbana-Champaign (UIUC). This work was supported in part by 
NSF Grants PHY-0205155, PHY-0345151 and PHY-0650377, NASA Grants NNG04GK54G, 
NNX07AG96G and NNG046N90H at UIUC.

\end{document}